%% file: main.tex
\documentclass[traditabstract,doublecolumn]{aa}
\usepackage{graphicx}
\usepackage{txfonts}
\usepackage{savesym}
\savesymbol{rotate}
\usepackage{deluxetable} 
\restoresymbol{DTBL}{rotate}

\usepackage[english]{babel}
\usepackage{times,epsfig}
\usepackage{longtable}
\usepackage{url}
\usepackage{hyperref}
\usepackage{mathrsfs}
\usepackage[normalem]{ulem}
\hypersetup{
    colorlinks = true,
    linkcolor = blue,
    anchorcolor = blue,
    citecolor = blue,
    filecolor = blue,
    urlcolor = blue
    }
    
\usepackage{placeins}

\newcommand{\orcid}[1]{\href{https://orcid.org/#1}{\includegraphics[width=10pt]{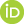}}}
\newcommand{\nosne}{35}
\newcommand{\numberspectra}{170}

\authorrunning{Holmbo, Stritzinger, Karamehmetoglu, et al.}
\titlerunning{Analysis of  CSP-I SE SN Spectroscopy.}
\begin{document} 

\title{The Carnegie Supernova Project~I. Spectroscopic analysis of stripped-envelope supernovae}

\author{
 S. Holmbo\inst{1}\orcid{0000-0002-3415-322X}
 \and
 M. D. Stritzinger\inst{1}\orcid{0000-0002-5571-1833}
 \and
 E. Karamehmetoglu\inst{1}\orcid{0000-0001-6209-838X}
 \and 
 C. R. Burns\inst{2}\orcid{0000-0003-4625-6629}
 \and
 N. Morrell\inst{3}\orcid{0000-0003-2535-3091}
 \and 
  C. Ashall\inst{4}\orcid{0000-0002-5221-7557}
  \and 
 E. Y. Hsiao\inst{5}\orcid{0000-0003-1039-2928}
 \and
 L. Galbany\inst{6,7}\orcid{0000-0002-1296-6887}
 \and 
 G. Folatelli\inst{8,9}\orcid{0000-0001-5247-1486}
 \and 
 M. M. Phillips\inst{3}\orcid{0000-0003-2734-0796}
 \and
 E. Baron\inst{10}\orcid{0000-0001-5393-1608}
 \and
 C.~P.~Guti\'errez\inst{11,12}\orcid{0000-0003-2375-2064}
 \and 
  G. Leloudas\inst{13}\orcid{0000-0002-8597-0756}
 \and 
 T. E. M\"uller-Bravo\inst{6}\orcid{0000-0003-3939-7167}
 \and
 P. Hoeflich\inst{5}\orcid{0000-0002-4338-6586}
 \and 
 F. Taddia\inst{1}\orcid{0000-0002-2387-6801}
 \and
 N.~B. Suntzeff\inst{14}\orcid{0000-0002-8102-181X} 
}

\institute{
 Department of Physics and Astronomy, Aarhus University, Ny Munkegade 120, DK-8000 Aarhus C, Denmark\\ (\email{max@phys.au.dk})
 \and
 Observatories of the Carnegie Institution for Science, 813 Santa Barbara St., Pasadena, CA 91101, USA
 \and
Carnegie Observatories, Las Campanas Observatory, Casilla 601, La Serena, Chile
 \and 
Department of Physics, Virginia Tech, Blacksburg, VA 24061, USA
\and
Department of Physics, Florida State University, 77 Chieftain Way, Tallahassee, FL, 32306, USA 
\and 
 Institute of Space Sciences (ICE, CSIC), Campus UAB, Carrer de Can Magrans, s/n, E-08193 Barcelona, Spain
 \and
 Institut d’Estudis Espacials de Catalunya (IEEC), E-08034 Barcelona, Spain
 \and
Instituto de Astrof\'{\i}sica de La Plata (IALP), CONICET, Paseo del Bosque S/N, 1900, Argentina
\and
Facultad de Ciencias Astron\'omicas y Geof\'{\i}sicas (FCAG), Universidad Nacional de La Plata (UNLP), Paseo del Bosque S/N, 1900, Argentina
 \and 
 Department of Physics and Astronomy, University of Oklahoma, 440 W. Brooks, Rm 100, Norman, OK 73019, USA
 \and 
 Finnish Centre for Astronomy with ESO (FINCA), FI-20014
University of Turku, Finland
\and 
Tuorla Observatory, Department of Physics and Astronomy,
FI-20014 University of Turku, Finland
\and
 DTU Space, National Space Institute, Technical University of Denmark, Elektrovej 327, DK-2800 Kgs. Lyngby, Denmark
 \and
 George P. and Cynthia Woods Mitchell Institute for Fundamental Physics and Astronomy, Department of Physics and Astronomy, Texas A\&M University, College Station, TX 77843, USA
}

 \date{Received 31 October, 2022; Accepted March 13, 2023.}
 
 \abstract{An analysis leveraging \numberspectra\  optical spectra of \nosne\ stripped-envelope (SE) core-collapse supernovae (SNe) observed by the Carnegie Supernova Project~I and published in a companion paper is presented. Mean template spectra were constructed for the SNe~IIb, Ib, and Ic subtypes, and parent ions associated with designated spectral features are identified with the aid of the spectral synthesis code \texttt{SYNAPPS}. Our modeled mean spectra suggest the $\sim 6150$~\AA\ feature in SNe~IIb may have an underlying contribution due to silicon, while the same feature in some SNe~Ib may have an underlying contribution due to hydrogen.  Standard spectral line diagnostics consisting of pseudo-equivalent widths (pEWs)  and blue-shifted Doppler velocity were measured for each of the spectral features. Correlation matrices and rolling mean values of both spectral diagnostics were constructed.  A principle component analysis (PCA) was applied to various wavelength ranges of the entire dataset and suggests clear separation among the different SE SN subtypes, which follows from  trends previously identified in the literature. In addition, our findings reveal the presence of two SNe~IIb subtypes, a select number of SNe~Ib displaying signatures of weak, high-velocity hydrogen, and  a single SN~Ic with evidence of weak helium features. Our PCA results can be leveraged to obtain robust subtyping of SE SNe based on a single spectrum taken during the so-called photospheric phase, separating SNe IIb from SNe Ib with $\sim80\%$ completion.}


\maketitle

\section{Introduction}

Between 2004 and 2009, the Carnegie Supernova Project~I (hereafter CSP-I; \citealt{hamuy2006}) obtained optical and near-infrared light curves \citep[][hereafter Paper~1]{stritzinger2018a} and visual-wavelength spectroscopy (Stritzinger et al., submitted; hereafter Paper~4) of nearly three dozen stripped envelope (SE) core-collapse supernovae (SNe). 
SE SNe are associated with the deaths of massive stars that have lost the majority of their hydrogen (and helium) envelopes prior to explosion. Within this context, stars with increasing amounts of mass stripping lead to characteristic spectra of different subtypes being either hydrogen (H) poor and helium (He) rich (SNe~IIb), hydrogen deficient with He features (SNe~Ib), or objects that are deficient of both H and He features (SNe~Ic). 

In a series of papers focusing on the CSP-I SE SN sample, we  summarize key  facets of our contemporary knowledge of SE SNe. This includes the topics of light curves and possible progenitor systems in Paper~1 \citep{stritzinger2018a}, which also presents the broadband photometry  of three dozen SE SNe. In Paper~2  \citep{stritzinger2018b}, the photometry is used to devise improved methods to estimate host-galaxy reddening, while Paper 3 \citep{taddia2018} presents a detailed analysis of the light curves and inferred explosion parameters. In Paper~4 a summary of the visual-wavelength spectroscopic properties and spectral classification of SE SNe is presented, along with the CSP-I SE SN spectroscopic sample consisting of 170 low-redshift (i.e., $z < 0.1$) spectra of 35 SE SNe \citep[see][]{Stritzinger2023}.

 In a classic paper on the spectroscopic studies of SE SNe, \citet{Matheson2001} studied 84 low-dispersion visual-wavelength spectra of 28 SE SNe extending from early to late phases, and in doing so were the first to characterize the heterogeneous nature of the different SE SN subtypes. 
Over the years, numerous single object case studies have been published \citep[for a review see][Chapters 15-17]{Branch2017}, while spectroscopic samples and associated analysis papers have been published by the Center for Astrophysics SN group \citep{Modjaz2014,Liu2016,Williamson2019}, as well as the Palomar Transient Factory (PTF) and intermediate PTF surveys \citep{Fremling2018}. 
The methods used in our analysis of the CSP-I SE SN spectroscopy dataset were inspired by these previous sample studies, as well as in part by a select number of papers analyzing the spectroscopic datasets of thermonuclear SNe \citep[e.g.,][]{Hsiao2007,2012AJ....143..126B,2012MNRAS.425.1819S,2013ApJ...773...53F}. 

The organization of this paper is as follows. 
First in Sect.~\ref{sec:lineIDs}, we focus on spectral line identification, including the construction of mean template spectra in Sect.~\ref{sec:meanspectra}, the calculation of synthetic spectra in Sect.~\ref{sec:SYNAPPSfits}, and the association of spectral line features with parent ion(s) in Sect.~\ref{sec:ions}. 
We then turn to line diagnostics measurements and correlation matrices for the line measurements of  pseudo-equivalent widths (pEWs)  in Sect.~\ref{sec:pEWs} and Doppler velocity line measurements in Sect.~\ref{sec:vel}. A principle component analysis (PCA) of the dataset is  presented in Sect.~\ref{sec:PCA}, which is then followed by the discussion in Sect.~\ref{sec:discussion} and a  summary of our key findings in Sect.~\ref{sec:summary}.

\section{Spectral line identification} 
\label{sec:lineIDs}

\subsection{Construction of SE SN mean spectra}
\label{sec:meanspectra}

Close inspection of the CSP-I SE SN spectral sequences presented in Paper~4 reveals the presence of numerous spectral features superposed on a pseudo-continuum with the shapes and strengths of the features being significantly time dependent. 
Before we make direct measurements of the various spectral features, we first set out to identify the key spectral features and determine their commonality among the different subtypes. 
Once the locations of key spectral features are identified we construct mean template spectra which are compared with synthetic spectra computed using \texttt{SYNAPPS}\footnote{\href{https://c3.lbl.gov/es/}{https://c3.lbl.gov/es/}} \citep{thomas2011}, which enables us to link the observed spectral features to parent ions (see Sect.~\ref{sec:SYNAPPSfits}).
Then, spectral line diagnostic measurements are made for the entire suite of identified features. 
 
\setcounter{figure}{0}
\begin{figure*}[t!]
 \resizebox{0.8\hsize}{!}
 {\includegraphics[]{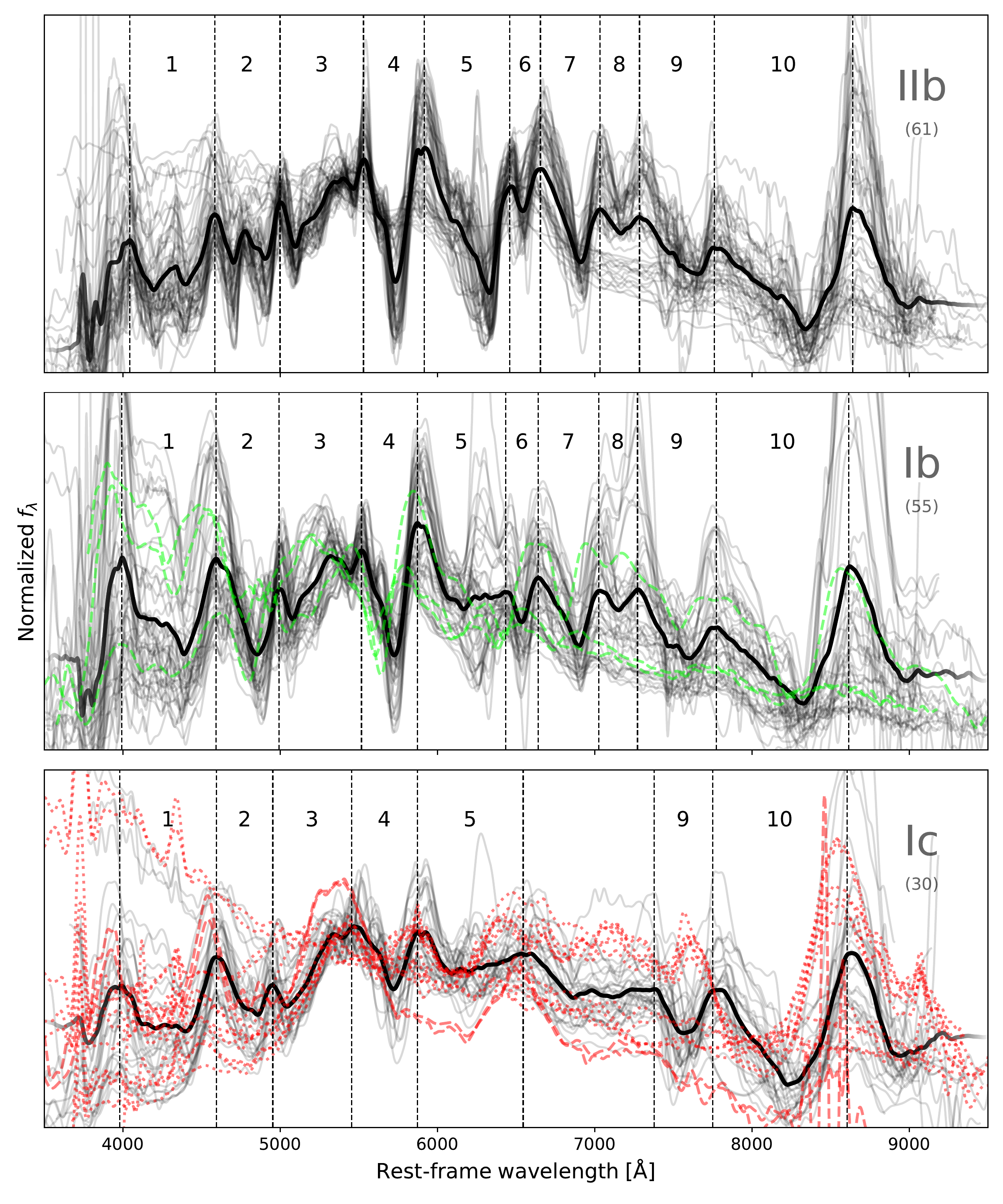}}
 \caption{Median spectrum for each SE SN subtype (black line) constructed from all of the spectra plotted in gray and listed in Table~\ref{tab:sample.tex}. The vertical dashed lines indicate 10 regions of interest. Regions 1--5, 9 and 10 are typically present in each SE SN subtype, while regions 6--8 are typically found in SNe~IIb and SNe~Ib and are usually most prevalent in the weeks after maximum light. Green lines in the middle panel are spectra of the high-velocity Type~Ib SN~2004gq, and red lines plotted in the bottom panel correspond to the spectra of the Type~Ic-BL SN~2009bb (dotted) and the unusual Type~Ic SN~2009ca (dashed). These spectra were excluded in making the median spectra. Finally, the number of spectra used to create each median spectrum plotted in each panel is indicated within parenthesis located just below the given spectral subtypes.}
 \label{fig:median}
\end{figure*}

 As a first step, a single median spectrum is constructed for each SE SN subtype using all of the spectra listed in Table~\ref{tab:sample.tex}, except those of SNe~2004qv, 2009bb, and 2009ca. The resulting median spectrum of each SE SN subtype are plotted in Fig.~\ref{fig:median}, along with the individual input spectra. In the middle panel, the spectra of the Type~Ib SN~2004gq that exhibit high-velocity features are plotted in green, while in the bottom panel the spectra of the broad-lined Type~Ic SNe~2009bb and 2009ca are plotted in red.  These spectra are not included in construction of the mean spectrum plotted in the figure of these two subtypes. 
 A comparison between the three median spectra reveals a number of common features as well as a handful of features that are typically only present in SN~IIb and/or SN~Ib spectra. As indicated in Fig.~\ref{fig:median}, 10 different complexes of spectral features are assigned a specific number (running from Feature~1 to Feature~10) and in some cases a feature may be attributed to multiple ions. Key characteristics of these features are highly dependent on the phase of the spectrum. An accurate study of the spectroscopic. properties of our sample requires a larger grid of mean template spectra. We therefore produce a coarse time-series of mean template spectra for each subtype to be modeled with {\tt SYNAPPS}.

\begin{figure*}
 \resizebox{\hsize}{!}
 {\includegraphics[]{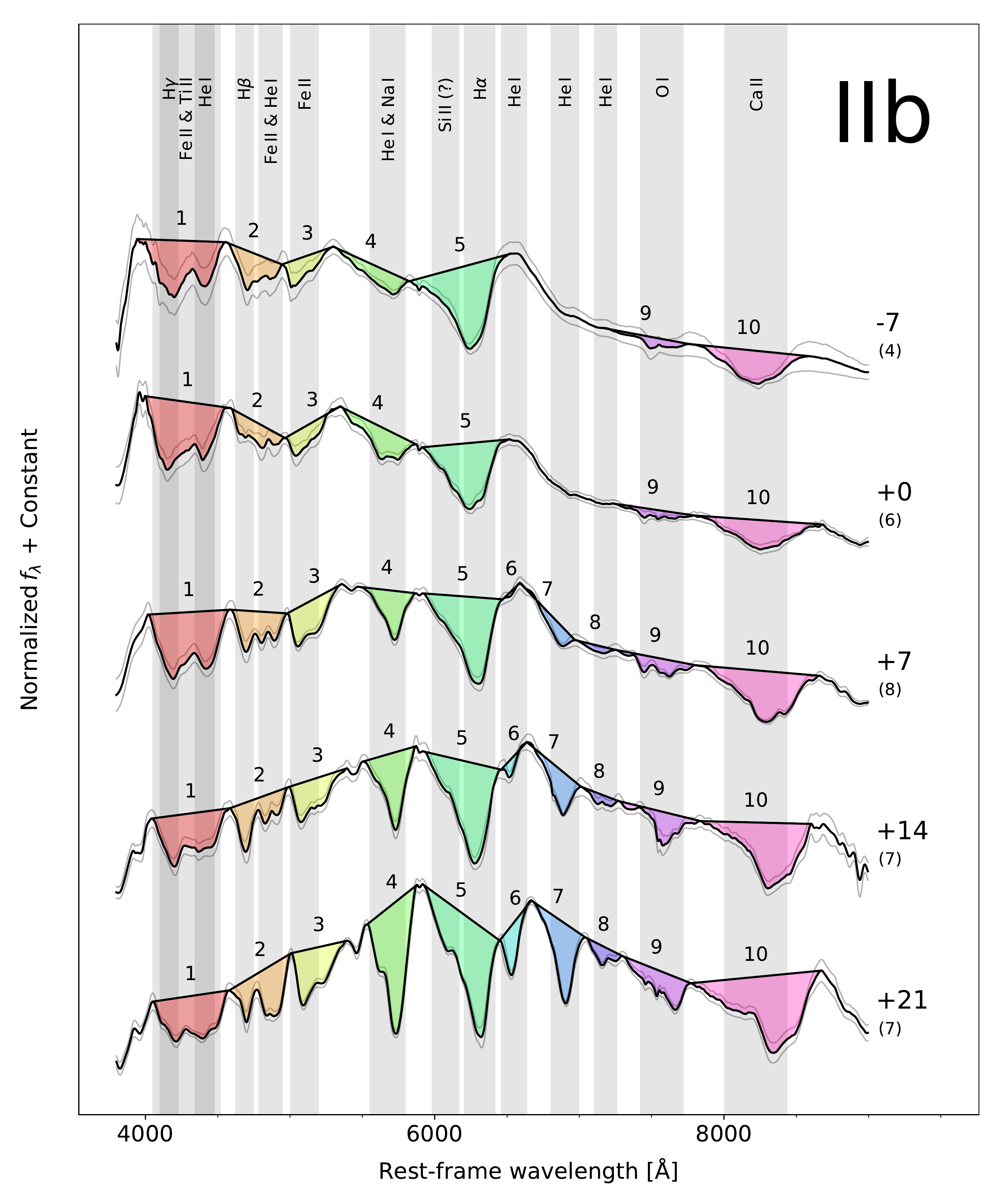}}
 \caption{Mean template spectra for the Type~IIb SE SN subtype at epochs of $-$7~d, 0~d, $+$7~d, $+$14~d, and $+$21~d.  The number of spectra used to construct each template  is indicated within parenthesis below the corresponding epoch. 
 Each numerically indicated spectral feature is labeled with its parent ion(s) as determined from {\tt SYNAPPS} fits.
 Thin gray lines indicate the 1-$\sigma$ error spectrum.}
 \label{fig:spectra_epoch1}
\end{figure*}

\begin{figure*}
 \resizebox{\hsize}{!}
 {\includegraphics[]{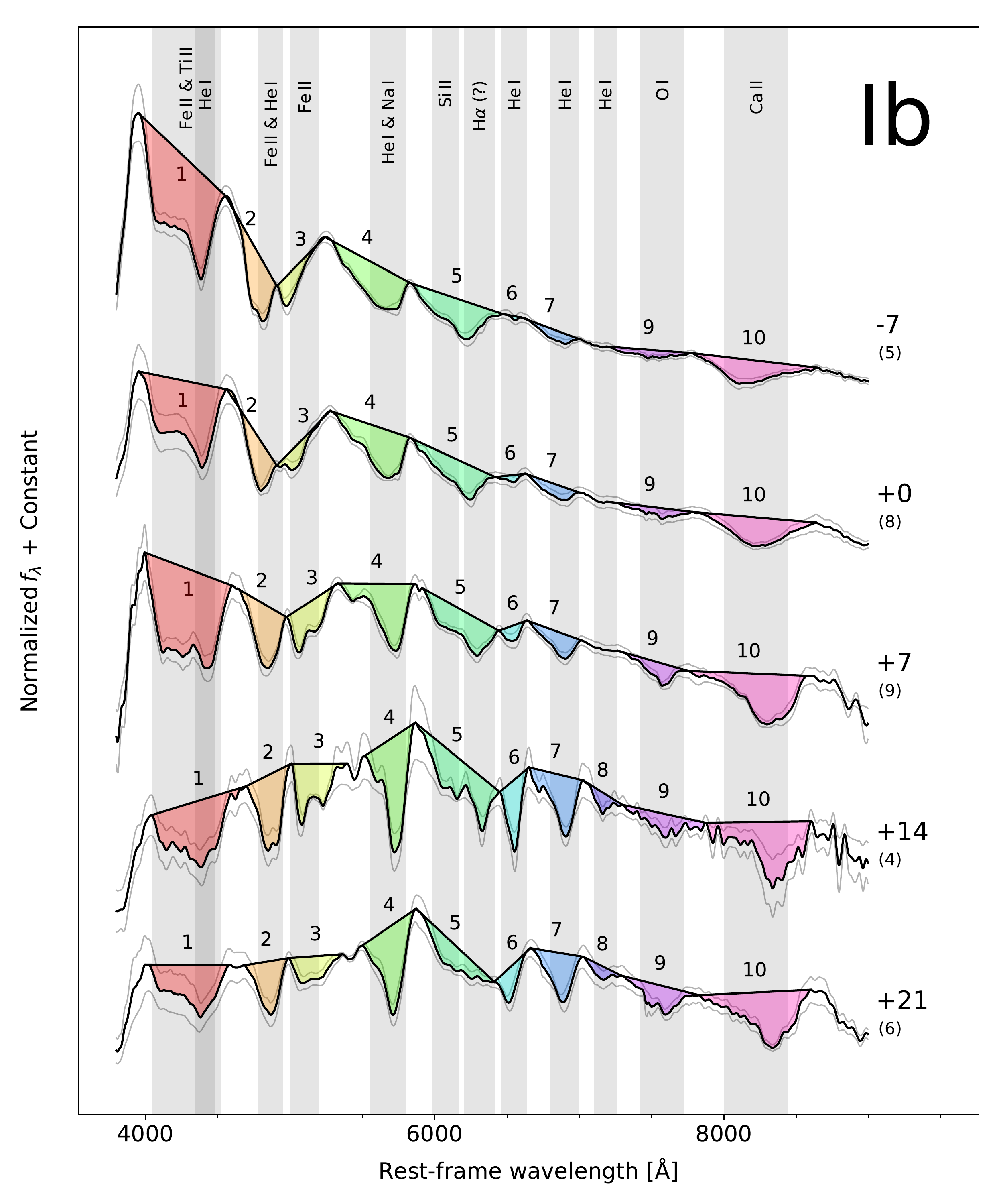}}
 \caption{Mean template spectra for the Type~Ib SE SN subtype at epochs of $-$7~d, 0~d, $+$7~d, $+$14~d, and $+$21d. The number of spectra used to construct each  template  is indicated within parenthesis below the corresponding epoch. Each numerically indicated spectral feature is labeled with its parent ion(s) as determined from {\tt SYNAPPS} fits. Thin gray lines indicate the 1-$\sigma$ error spectrum.}
  \label{fig:spectra_epoch2}
\end{figure*}

\begin{figure*}
 \resizebox{\hsize}{!}
 {\includegraphics[]{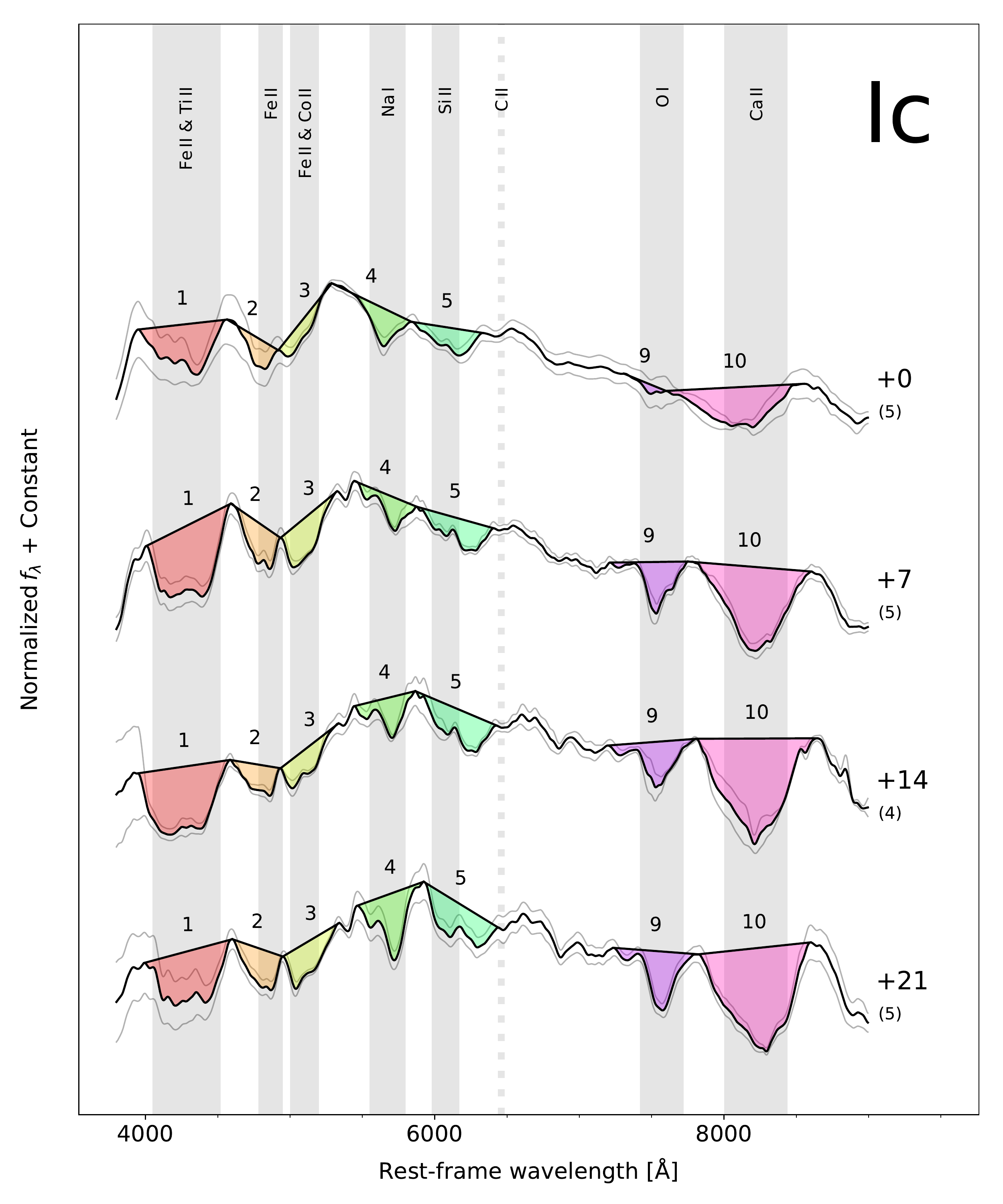}}
 \caption{Mean template spectra for the Type~Ic SE SN subtype at epochs of 0~d, $+$7~d, $+$14~d, and $+$21~d. The number of spectra used to construct each template  is indicated within parenthesis below the corresponding epoch. Each numerically indicated spectral feature is labeled with its parent ion(s) as determined from {\tt SYNAPPS} fits. Thin gray lines indicate the 1-$\sigma$ error spectrum. The vertical dashed line corresponds to the rest wavelength of \ion{C}{ii} $\lambda$6580.}
  \label{fig:spectra_epoch3}
\end{figure*}

 Mean spectra were constructed for the SN~IIb and SN~Ib subtypes for the epochs: $-7$~d, $+0$~d, $+$7~d, $+$14~d and $+$21~d. Due to a dearth of early data we did not construct a $-$7~d mean spectrum for SNe~Ic.
 The steps required to construct a mean spectrum are several fold. First we identified an input set of spectra that met the criteria of being obtained within 
$\pm$3.5~d of the epoch under consideration. The spectra of each subsample were then corrected for Milky Way and host reddening using values estimated by \citet{stritzinger2018b}. Next the spectra were smoothed using a Fourier Transform (FT) technique \citep[see][]{marion2009}, re-sampled, and then combined yielding a mean spectrum. 
An associated error spectrum was also computed using a semi-automated line fitting program written in \texttt{PYTHON}
named Measure Intricate Spectral Features In Transient Spectra (hereafter \texttt{misfits}\footnote{\href{https://github.com/sholmbo/misfits}{https://github.com/sholmbo/misfits}}).
 \texttt{misfits} enables robust measurements of standard line diagnostics, and in doing so estimates a realistic error snake. This is done by computing 10,000 realizations of each spectrum following a Monte Carlo approach. The resultant 1$\sigma$ standard deviation of the Monte Carlo distribution for each spectrum then serves as the error snake.

The resulting series of mean spectra and their 1-$\sigma$ error snakes of each SE SN subtype are plotted in Fig.~\ref{fig:spectra_epoch1}--\ref{fig:spectra_epoch3}. Within  Fig.~\ref{fig:spectra_epoch1} and Fig.~\ref{fig:spectra_epoch2} Features 1-10 are identified and labeled with their parent ion(s) as implied from {\tt SYNAPPS} modeling (see Sect.~\ref{sec:SYNAPPSfits}), while in Fig.~\ref{fig:spectra_epoch3} the features associated with \ion{He}{i} are excluded. 
Furthermore, a line is drawn connecting the red and blue edges of each feature to define a pseudo-continuum. 
The area contained within each feature is highlighted in color.

\subsection{Computing synthetic spectra with \texttt{SYNAPPS}}
\label{sec:SYNAPPSfits}

Synthetic spectra were computed with \texttt{SYNAPPS} \citep{thomas2011} which is an automated
implementation of the highly parameterized spectral synthesis code \texttt{SYNOW} (SYnthesis NOW; \citealt{2000PhDT.........6F}). \texttt{SYNOW} is a synthetic spectrum code based on a number of underlying assumptions, including spherical symmetry \citep[see][]{thomas2011}. 
Despite these shortcomings, \texttt{SYNOW} and \texttt{SYNAPPS} have proven to be effective tools to aid in spectroscopic studies of various flavors of stripped core-collapse and thermonuclear supernovae \citep[e.g.,][]{Deng2000,folatelli2006,branch2007,thomas2011,Hsiao2015,parrent2016}.  \citet{Holmbo2018} discusses in detail our  \texttt{SYNAPPS} analysis, which is summarized in  Appendix~\ref{appendixA} for the interested reader.

\subsection{Line identification}
\label{sec:ions}
 \subsubsection{SNe~IIb}

 \begin{itemize}
 
 \item Feature 1 is characterized by a W-shape line profile (see Fig.~\ref{fig:median}, panel a).  \texttt{SYNAPPS} fits  suggest this feature is produced from a blend of H$\gamma$ (forming the bluer of the two dips in the W), a forest of Fe lines including that of \ion{Fe}{II}~$\lambda$4550, a contribution from \ion{Ti}{II} $\lambda\lambda$4395, 4444, 4469, and also \ion{He}{i} $\lambda$4471 which mostly produces the red dip of the W-shape profile. 
 
 \item Feature 2 is produced by a blend of lines. At early phases the feature is primarily formed by H$\beta$ with a contribution from \ion{Fe}{ii} $\lambda\lambda4924,5018$, while at later phases a contribution due to \ion{He}{i} $\lambda$4922 emerges (see the $+$21~d mean SN~IIb spectrum in Fig.~\ref{fig:spectra_epoch1}).

 \item Feature 3 is largely produced by the third member of the \ion{Fe}{II} multiplet 42, $\lambda$5169. This feature is well defined in each of the SE SN subtypes, and is used as a proxy for the bulk velocity of the SN ejecta. 

 \item Feature 4 is produced by \ion{Na}{i} $\lambda\lambda$5890, 5896 and \ion{He}{i} $\lambda5876$.

 \item Feature 5 is ubiquitous to SNe~IIb and is formed by H$\alpha$. The absorption profile of the feature in the mean spectra show an extended blue wing, which as indicated by the \texttt{SYNAPPS} fits shown in Fig.~\ref{fig:synapps}(a), could be due to a contribution from the \ion{Si}{ii} $\lambda$6355 doublet. Indeed, the velocity would match that of \ion{Fe}{ii}, as we show for SNe Ib and Ic below. If Feature 5 was purely hydrogen, the pEW of this line would be expected to decrease with time. However, as can be seen in Fig. \ref{fig:pew_vs_t}, the pEW is constant with time. This is likely due to both the contribution from the blue feature (which is most likely \ion{Si}{ii}).
 
 \item Features 6, 7, 8 are attributed to \ion{He}{i} $\lambda$6678, \ion{He}{i} $\lambda$7065 and \ion{He}{i} $\lambda$7281, respectively. In general, the strength of
 these features grows over time becoming most prominent several weeks past maximum. 
 
 \item Feature 9 is attributed to \ion{O}{I} $\lambda7774$. This line is often contaminated by the telluric O$_{2}$ A-band absorption feature. As demonstrated in Paper~4, we were able to successfully correct for telluric absorption for many spectra in the sample, particularly those obtained with the \textit{du Pont} telescope equipped with the Wide Field re-imaging CCD (WFCCD) camera.
 
 \item Feature 10 is attributed to the \ion{Ca}{II} NIR triplet $\lambda\lambda8498,8542,8662$. In some objects, for example, the SNe~Ib/c (flat-velocity SNe~IIb) 2005bf and SN~2007Y, their early spectra exhibit a distinct high-velocity component that promptly vanishes over several days of evolution,  while a photospheric component emerges and grows in strength \citep[][]{folatelli2006,stritzinger2009}. 
 
 \end{itemize}

\subsubsection{SNe~Ib and SNe~Ic}
\label{sec:lineIDS-Ib-Ic}

 \begin{itemize}

 \item Feature 1 is produced (similar to SNe~IIb) by a forest of \ion{Fe}{II} and \ion{Ti}{II} lines, with an additional contribution from \ion{He}{i} $\lambda$4471 in SNe~Ib, which turns the characteristic SNe~IIb 
 W-shape profile to a Y-shape profile. 
 As SNe~Ic lack both H$\gamma$ and \ion{He}{i} the feature takes on a U-shape profile (see Fig.~\ref{fig:median}).

 \item Feature 2 is formed by \ion{Fe}{ii} $\lambda\lambda4924,5018$, with SNe~Ib having an additional contribution from \ion{He}{i} $\lambda$4921.

 \item Feature 3 is largely produced by \ion{Fe}{II} $\lambda$5169. In the case of SNe~Ic this feature may also contain a contribution from \ion{Co}{II} $\lambda5526$, which increases in strength during the post-maximum evolution. 
 
 \item Feature 4 is attributed to the \ion{Na}{i} $\lambda\lambda$5890,5896 doublet in SNe~Ic, while in in SNe~Ib a significant contribution comes from \ion{He}{i} $\lambda5876$.

 \item Feature 5, unlike in Type IIb SNe, is not hydrogen by the traditional spectral classification system. Although in the past it has been linked to the \ion{Si}{ii} $\lambda$6355 doublet \citep[e.g.,][]{1987ApJ...317..355H,2002ApJ...566.1005B}, various different ions have been proposed for this feature.  Other than \ion{Si}{ii}, these include high-velocity H$\alpha$ \citep{Liu2016, parrent2016}, as well as \ion{Fe}{ii}, \ion{Co}{ii}, \ion{C}{ii}, and  \ion{Ne}{i}  \citep[see][and references therein]{galyam2017}.
 
~~~Close inspection of the template spectra reveals evidence of a blend of at least two features contained within Feature~5, which we refer to as the red and blue portions. The red portion is stronger than the blue portion, especially up to +14 days in both SN subtypes. It is unlikely that the red portion, which is stronger, is produced by \ion{Si}{II}, as its position of maximum absorption would imply a red-shifted Doppler velocity. However, assuming the blue portion is due to \ion{Si}{II} brings the velocities in line with other ions, which is demonstrated in Fig. \ref{fig:vel_vs_t}.

 ~~~With the blue portion attributed to  \ion{Si}{II}, the synthetic spectra suggest that the red portion of the feature could be formed by a residual amount of H detached from the photosphere. In fact, it has been identified as H in so-called transitional SNe~Ib/c (e.g., SN~1999ex, \citealt{hamuy2002}), which \citet{folatelli2014} refer to as flat-velocity SNe~IIb. However, identifying Feature 5 as H possibly contradicts theoretical and observational evidence showing SNe~Ic to be deficient in H and He \citep{taddia2018,hachinger2012}. 
 
~~~As  previously mentioned, the red portion could instead be one of the other proposed lines (\ion{Fe}{ii}, \ion{Co}{ii}, \ion{C}{ii}, and  \ion{Ne}{i}), or a blend. \citet{Shahbandeh2022} found evidence of strong \ion{C}{i} features in NIR spectra of SNe~Ic, indicating the optical feature could at least partially be due to \ion{C}{ii}. In our analysis and figures, we assume the blue portion to be \ion{Si}{II} for SNe Ib and Ic when it can be reliably measured, while the red portion is not used.

 \item Features 6, 7, 8 are, as in the case of SNe~IIb, attributed to \ion{He}{i}
 $\lambda$6678, \ion{He}{i} $\lambda7065$, and \ion{He}{i} $\lambda$7281, respectively. By definition SNe~Ic contain no He features, but see Sect.~\ref{sec:interlopers}. 
 
 \item Feature 9 is attributed to \ion{O}{I} $\lambda7774$. 
 
 \item Feature 10 is attributed to the \ion{Ca}{II} NIR triplet.

 \end{itemize}

\section{Analysis of pseudo-equivalent width measurements}
\label{sec:pEWs}

In this section pEW measurements for Features 1--10 for the entire sample of spectra are computed and used to construct correlation matrices for various pairs of Features 1-10. 
The strength and evolution of pEWs (and Doppler velocities, see below) provides a wealth of information related to the progenitor stars \citep{Branch2017}. For example, the spectral features themselves provide a window to the ionizaton state and chemical content of the ejecta above the photosphere. As the SN ejecta expand and cool, the photosphere recedes into the inner ejecta enabling a direct view to the otherwise opaque inner regions of the progenitor stars.
In addition, the presence (or lack thereof), strength and time-evolution of features produced from certain ions (e.g., H and/or He) provides information on the 
spectral type of the progenitors, their mass-loss history and even their explosion physics, while the expansion velocities provides a measure of the explosion energy (see Paper~3, and references therein).

\subsection{Measuring pseudo-equivalent widths}
\label{sec:pEWdef}

Armed with the spectral line identifications of Features 1--10, we conduct a quantitative analysis of the line strength and evolution of the various features via pEW measurements. 
 This is a common line diagnostic having been utilized to study large spectroscopic samples of thermonuclear supernovae \citep[e.g.,][]{2006PASP..118..560B,2007A&A...470..411G,2012AJ....143..126B,2012MNRAS.425.1819S,2013ApJ...773...53F}. 
 \citet{Liu2016} and \citet{Fremling2018} have also followed suit, using both pEW and Doppler velocity measurements in their analysis of the CfA and PTF SE SN spectroscopic datasets, respectively. The use of pEW measurements serves as a flexible and accurate line diagnostic, particularly compared to fitting a Gaussian function, which is often not appropriate given the asymmetric and time-dependent spectral features inherent to SE SNe.

 The term pseudo in pEW highlights the difficulties faced when attempting to separate continuum flux from absorption and emission flux of time-dependent spectral features that also suffer significant Doppler broadening. Following \citet{2004NewAR..48..623F}, we define the pseudo-continuum to be a straight line connecting two maxima defining a spectral absorption feature: 
\begin{equation}
pEW = \sum_{i=0}^{N-1} \left( 1 - \frac{f(\lambda_i)}{f_c(\lambda_i)} \right) \Delta\lambda_i.
\label{eqn1}
\end{equation}
Here $\lambda_i$ corresponds to the wavelength of each pixel contained within the spectral range located between the blue and red edges of a specific feature. Furthermore, $N$ is the number of pixels contained between the red and blue edges, the parameter $\Delta\lambda_i$ represents the width of pixel $i$, $f(\lambda_i)$ corresponds to the observed flux at $\lambda_i$, $f_c(\lambda_i)$ is the pseudo-continuum at $\lambda_i$, and $\Delta\lambda_i$ is the sum over the defined wavelength interval. 

Given the number of spectra in the CSP-I sample and having up to 10 features, it quickly becomes cumbersome to measure the pEWs by hand. We therefore developed the \texttt{misfits} spectral analysis package, and in doing so, created a resource that minimizes user bias while enabling efficient measurements in a standard manner. 
To make pEW measurements of Features 1--10 for a given spectrum we adopted the following steps.
First \texttt{misfits} smooths an observed spectrum following the use of a Fourier Transform (FT) smoothing technique \citep{marion2009} and then identifies the highest peak located between the defined boundaries at the blue and red end of Features 1--10. The wavelength ranges for each blue/red boundary of each of the Features measured are listed in Table~\ref{tab:specregions}. 
With boundary end points identified, a pseudo-continuum is defined by connecting the boundary points with a straight line. This is demonstrated in Fig.~\ref{fig:spectra_epoch1} where the spectral regions of each feature are identified. These regions are used to infer pEW values using Eq.~(\ref{eqn1}).

Each pEW measurement also has an associated 1-$\sigma$ uncertainty estimated via the following recipe.
The error spectrum is first multiplied by a random number taken from a normal distribution and the resulting product is added to the FT smoothed observed spectrum. 
pEW measurements are computed from the altered spectrum, and following a Monte Carlo approach, this is done for 10,000 realizations. The 1-$\sigma$ value of the resulting pEW distribution then serves as the pEW 1-$\sigma$ measurement uncertainty. 
This method accounts for the uncertainty associated with the ability of the algorithms to accurately determine the heights of the blue/red boundaries of the spectral line features. 

\begin{figure*}[!t]
 \centering
 \includegraphics[width=14cm]{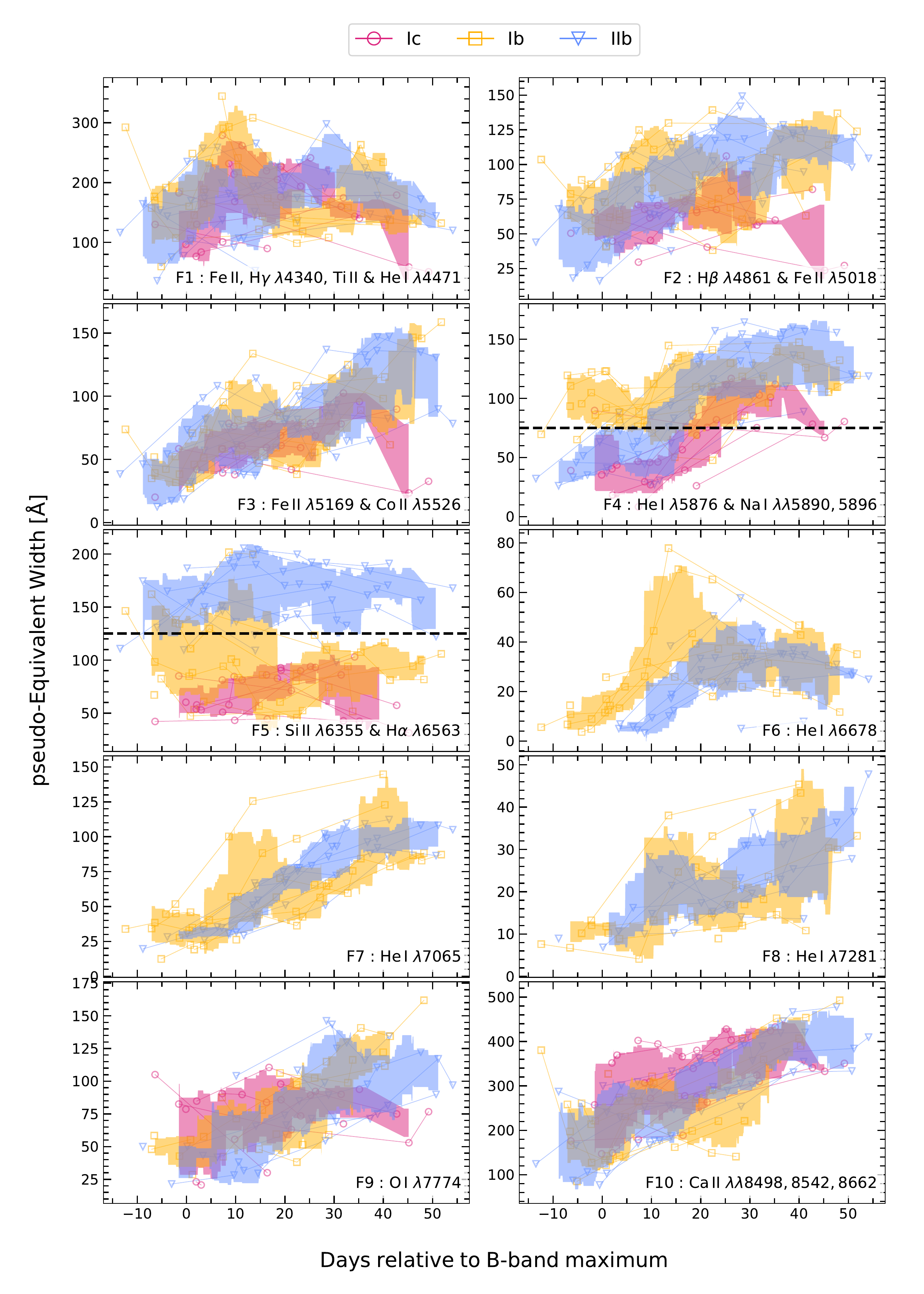}
 \caption{Pseudo-equivalent width values for Features 1--10 plotted vs days relative to the epoch of $T(B)_{max}$. The measurement symbols are encoded by SN subtype with SNe~IIb appearing as blue triangles, SNe~Ib as yellow squares, and SNe~Ic as magenta squares. Filled shaded regions correspond to the mean $\pm$1-$\sigma$ standard deviation of a rolling window characterized with a size of ten days. The rolling windows are evaluated on epochs containing at least  three distinct pEW measurements and  two pEW measurements in the preceding and subsequent epochs.}
 \label{fig:pew_vs_t}
\end{figure*}

\subsection{pEW measurement results for Features 1-10}
\label{sect:pEWs}

The definitive pEW measurements of Features 1--10 computed for all of the photospheric phase spectra are plotted in Fig.~\ref{fig:pew_vs_t} versus phase relative to the epoch of $B$-band maximum.
Over-plotted in each of the panels are filled shaded regions color-coded by spectral subtype representing the mean 1-$\sigma$ standard deviation computed using a rolling window with a size of ten days. The rolling windows were evaluated only on epochs containing a minimum of three measurements and have at least two preceding and subsequent measurements.

 Inspection of Fig.~\ref{fig:pew_vs_t} reveals that the pEWs for most of the features generally increase in strength over the week prior to maximum. Depending on the particular feature, this behavior continues, or they at least remain constant over the duration of $\approx 3-6$ weeks past maximum. The pEW values are found to range from as little as $\sim$ 10 \AA\ (e.g., $\ion{He}{i}$ features at early times) all the way up to $\gtrsim$600 \AA\ in the case of the $\ion{Ca}{II}$ NIR triplet.
We now briefly summarize the nature of the pEW measurement of Features 1--10 and in passing compare and contrast the results amongst the different subtypes.

\begin{itemize}

\item Feature 1 (\ion{Fe}{II}, \ion{Ti}{II} in each subtype with additional contributions from H$\gamma$ and \ion{He}{i} $\lambda4471$ in SNe~IIb and \ion{He}{i} in SNe~Ib).
 The pEW values typically increase within a few weeks from maximum and then remain relatively constant over the duration of our observations. pEW values range from $\sim 100$ ~\AA\ up to $\sim 250$~\AA. At early times SNe~Ic tend to exhibit smaller values as compared to SNe~IIb and SNe~Ib.

\item Feature 2 (\ion{Fe}{II} in each subtype with an additional contribution from H$\beta$ and \ion{He}{i} in SNe~IIb and only \ion{He}{i} in SNe~Ib) The pEW values generally increase over time for the SNe~IIb while the increase in SNe~Ib exhibits a shallower slope. pEW values typically reach peak values in excess of 100~\AA\ within 2 to 3 weeks past maximum and then remain constant and/or slowly decrease. 
The feature is less prevalent in SNe~Ic with pEW values reaching around $50~\AA$ at all epochs.

\item Feature 3 (\ion{Fe}{ii} and \ion{Co}{II} in SNe~Ic) consists of a forest of \ion{Fe}{ii} lines, often displaying a prominent P-Cygni \ion{Fe}{ii} emission profile, making it a somewhat troublesome feature for \texttt{misfits} to accurately identify the red boundary point automatically (see Fig.~\ref{fig:spectra_epoch1}). This can lead \texttt{misfits} to compute negative values at early phases for some of the objects. Nevertheless within a week past maximum the majority of objects exhibit pEW values between $\sim$~20~\AA\ to 70~\AA, and these continue to increase reaching $\sim$ 50 to 150~\AA\ by a month and half past maximum. 
In SNe~Ic there may be at later times a contribution from \ion{Co}{ii}, however, the mean pEW values are consistent with those inferred from the SNe~IIb and SNe~Ib, indicating this feature is largely dominated by \ion{Fe}{ii}.

\item Feature 4 (\ion{Na}{I} and/or \ion{He}{i}) in the majority of objects exhibits pEW values that increase with time. In the week leading to maximum the SNe~IIb and SNe~Ib exhibit very different pEW values with the former below 50~\AA, while the later are in excess of 75~\AA\ (indicated by a dashed line) with the rolling mean values around $\sim$100~\AA.  In the SNe~Ib at early times, the feature appears to slightly decrease and at around $+10$~d begins to increase and thereafter follow the same trend as the SNe~IIb. Over time the pEW values for both SNe~IIb and SNe~Ib continue to increase reaching  similar values out to $+$50~d. On the other hand, the SNe~Ic exhibit an average pEW value of $\sim 40\pm30$~\AA\ at maximum and generally exhibit lower pEW values as time goes on relative to both SNe~IIb and SNe~Ib.
Given the significant difference between the premaximum pEW measurements of this feature in SNe~Ib relative to the both SNe~IIb, the pEW value of this feature could serve as a useful aid in spectral classification  (see Sect.~\ref{sec:pEWlinediagnostics}). 

\item Feature 5 (H$\alpha$ and/or \ion{Si}{II}) pEW values show diversity dependent on spectral subtype and the phase of the spectrum. 
The majority of SNe~IIb exhibit pEW values in excess to those characterizing SNe~Ib and SNe~Ic and this trend appears to increase over time. 
As indicated by the dashed line in the Feature~5 panel of Fig.~\ref{fig:pew_vs_t} there is a clear separation at pEW of $\approx140$~\AA\ between the pEW values measured in SNe~IIb compared to the values measured in SNe~Ib and SNe~Ic. 
Quantitatively, the pEW values of SNe~IIb take on high values (pEW $\sim 140-200$ \AA) from early phases and onward. On the other hand, the SNe~Ib and SNe~Ic exhibit smaller pEW values ranging between $\approx 60-100$~\AA. In the earliest observations of SNe~Ib, Feature 5 shows higher values and this is likely due to absorption associated with a small shell of hydrogen material retained by their progenitor stars. As the photosphere recedes the pEW value in these cases drops to values similar to the SNe~Ic, indicating that the feature becomes dominated by \ion{Si}{ii} and has little to no contribution from H$\alpha$. 
In other words, the fact that SNe~Ib and SNe~Ic exhibit similar pEW values around maximum light and beyond, provides an argument that at these phases the red component of Feature~5 in SNe~Ib is largely produced by \ion{Si}{ii}. 
We quantify this separation in a PCA analysis presented below. 
The finding that SNe~IIb exhibit larger pEW values of Feature~5 at early times as compared to SNe~Ib is consistent with the idea that their progenitor stars retain higher amounts of hydrogen relative to the progenitors of SNe~Ib. 

\item Features 6-8 (\ion{He}{I}) increase in strength from early phases out to weeks past maximum. As expected by the atomic data of \ion{He}{i}, Feature~7 is significantly more prominent than Features~6 and 8 at all epochs. Quantitatively, Feature~7 reaches pEW values in excess of 100~\AA\ while Features 6 and 8 extend between 10-80 \AA. 

\item Feature~9 (\ion{O}{I}) in the SNe~IIb and SNe~Ib exhibits pEW values that generally increase from $\leq 50$~\AA\ around maximum to in excess of 100~\AA\ a month past maximum. SNe~Ic show more premaximum diversity with pEW values ranging from $\sim$ 25~\AA\ up to $\sim$100~\AA. The rolling means indicate that beyond +30~d the pEW values level off or even decrease to values $<100$~\AA. During the same phases, SNe~IIb and SNe~Ib exhibit mean pEW values $\gtrsim$100~\AA. 

\item Feature~10 (\ion{Ca}{II}) overall exhibits an increase in the pEW value for each SE SN subtype. Prior to maximum the SNe~IIb and SNe~Ib exhibit pEW values of around $160\pm50$~\AA\ that evolve up to $\sim$400~\AA\ by $+$40~d. Around maximum, SNe~Ic tend to reach somewhat larger pEW values compared to SNe~IIb and SNe~Ib, while by a month past maximum the three SE SN subtypes show similar values.

\end{itemize}

\begin{figure*}[!t]
\centering
\includegraphics[width=15cm]{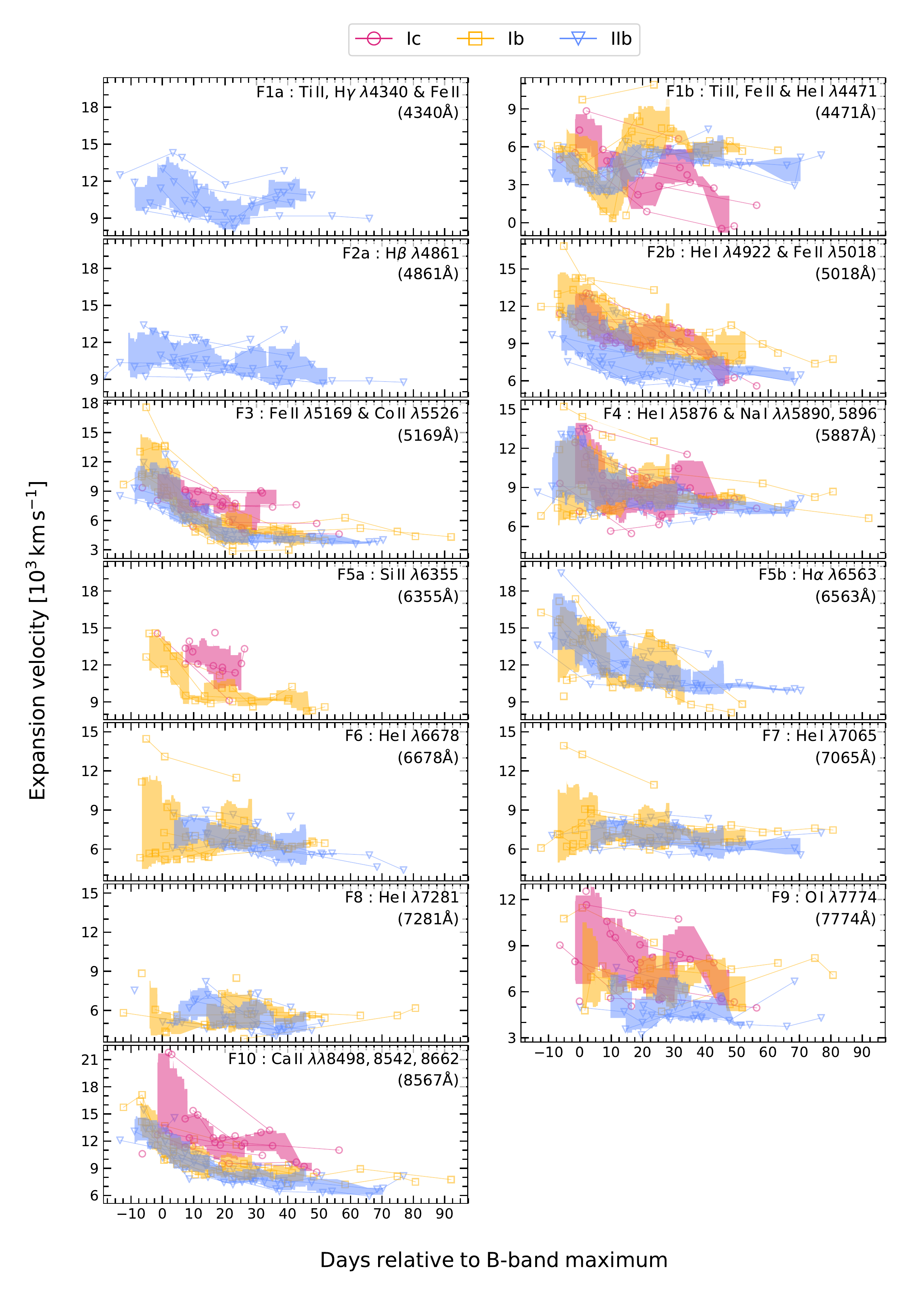}
 \caption{Time evolution of the Doppler velocity at maximum absorption ($-v_{abs}$) for Features 1--10 and color-coded by SE SN subtype. Shaded regions correspond to the standard deviation of the rolling mean computed using a window size of ten days and color-coded by spectroscopic subtype as indicated. The rolling windows are evaluated on epochs having a minimum of three measurements and at least two measurements before and after. We note that the peculiar SNe~Ic 2009bb and 2009ca are excluded.}
\label{fig:vel_vs_t}
\end{figure*}

\section{Analysis of Doppler velocity measurements}
\label{sec:vel}

The wavelength of maximum absorption for a given spectral feature provides an estimate on the bulk velocity of the line-forming material. 
Although prominent features may be produced from the blending of numerous lines, in general, the Doppler velocity at maximum absorption ($-v_{abs}$) provides a useful measure of the kinematics of the line-forming region, and depending on the spectral features used, a constraint on the explosion energy of the supernova \citep[see, e.g.,][]{2002ApJ...566.1005B,Fremling2018,taddia2018}.

\subsection{Measuring Doppler velocity}

Determining a value of $-v_{abs}$ based on a spectral feature in a 1D supernova spectrum is done using the relativistic Doppler approximation \citep[see][their Eq.~6]{Blondin2006}.
Measuring the observed wavelength ($\lambda_{obs}$) of a feature from a 1D spectrum is relatively straightforward, however, the use of an automatic detection of the exact position of a feature's minimum becomes increasingly difficult in low signal-to-noise spectra due to an accompanying increase in the number of local minima.
To overcome this problem an algorithm was developed to detect groupings of minima within a certain threshold and treat each of the groupings as a single minimum.

\subsection{Doppler velocities and evolution}
\label{sec:pDVresults}

Plotted in Fig.~\ref{fig:vel_vs_t} are the $-v_{abs}$ values measured for Features 1--10 using all of our spectra obtained prior to $+$100~days relative to the epoch of $B$-band maximum. We note that in SNe~IIb Feature~1 and Feature~2 have an additional absorption component attributed to H$\gamma$ and H$\beta$, respectively. We therefore separate each feature into two features denoted as Feature (a) and Feature (b). 
We now summarize the overall trends for Features 1--10, and then examine the correlation matrices computed for the various pairs of the $-v_{abs}$. 

 \begin{itemize}

\item Feature 1a (mostly H$\gamma$ in SNe~IIb) can be problematic to measure given the spectral range of this feature is often noisy. Feature~1a plotted in the top left panel of Fig.~\ref{fig:vel_vs_t} exhibits, as do the majority of other features, $-v_{abs}$ values that decrease over the first two weeks of evolution. The feature exhibits $-v_{abs} \sim 10,000-14,000$~km~s$^{-1}$ in the week prior to maximum, and by $+$21~d decreases to $\sim 8000-10,000$~km~s$^{-1}$. 
 
 \item Feature~1b (\ion{Fe}{ii}, \ion{Ti}{ii} \& \ion{He}{i}) is similar in SNe~IIb and SNe~Ib with $-v_{abs}$ values between 3,000-6,000 km~s$^{-1}$ from $-10$~d to $+10$~d. Upon reaching a minima $-v_{abs}$ at around $+10$~d, it experiences an upturn, reaching within three weeks, similar (in SNe~IIb) or even higher values (SNe~Ib) compared to those inferred from spectra in the week leading up to maximum. By $+$40~d the velocity evolution decreases in SNe~Ib leveling off to mean $-v_{abs}$ values of $\sim 6,000$ km~s$^{-1}$, which is similar to the SNe~IIb mean value of $\sim 4,000$ km~s$^{-1}$.
 Coverage of the SNe~Ic begins a week after the SNe~IIb/Ib with a mean $-v_{abs}$ value between $+0$~d to $+$10~d of $\sim 7,000$~km~s$^{-1}$. Between $+$10~d to $+$25~d the mean $-v_{abs}$ value drops to $\sim 2,500$ km~s$^{-1}$, similar to the mean values exhibited by the SNe~IIb and SNe~Ib when they reached their initial minima occurring 10-15 days earlier. 
 The (delayed) upturn in the SNe~Ic extends through $+$30~d, and then again, turns over declining to values around a factor of 2 or more less than inferred from the SNe~IIb and SNe~Ib at similar epochs (i.e., around $+$45~d). 

\item Feature 2a (H$\beta$) rolling mean $-v_{abs}$ values extend between 
$\sim 11,000\pm2,000$ km~s$^{-1}$ down to $\sim 9,000\pm1,000$ km~s$^{-1}$
from $-10$~d to $+50$~d. The velocities and evolution of this feature are quite consistent with that of Feature~1a.  

\item Feature 2b (\ion{Fe}{II} $\lambda$5018 \& \ion{He}{i}) in SNe~Ib exhibits rolling mean $-v_{abs}$ values that are $\sim 2,000$ km~s$^{-1}$ higher than the SNe~IIb. At the earliest phases covered, SNe~Ib reach rolling mean $-v_{abs}$ values of $\sim 13,000$ km~s$^{-1}$, while SNe~IIb never exceed $12,000$ km~s$^{-1}$. SNe~Ic mean values lie between the SNe~IIb/Ib until $+20$~d, and then subsequently drop to $\sim 6,000$ km~s$^{-1}$, slightly below the SNe~Ib mean values. Close inspection of panel F2b also reveals an SN~Ib at early times exhibiting a $-v_{abs}$ value $\sim 17,000$ kms$^{-1}$. This object is SN~2004gv and also exhibits high $-v_{abs}$ values in its other Features, particularly those attributed to \ion{He}{i} (see F3, F4, F6, F7). SN~2004gv is not included in the computation of the various rolling mean $-v_{abs}$ values plotted in the figure.

\item Feature 3 (\ion{Fe}{II} \& possibly \ion{Co}{ii} in SNe~Ic) exhibits at most epochs similar rolling mean $-v_{abs}$ values in both SNe~Ib and SNe~IIb. At the earliest phases of coverage SNe~Ib show $-v_{abs}$ values extending $\gtrsim 4,000$ km~s$^{-1}$ higher than that of the 
SNe~Ib rolling mean 1-$\sigma$ dispersion. Quantitatively, in the week leading to maximum, SNe~Ib exhibit mean $-v_{abs} \sim 13,000\pm2,000$~km~s$^{-1}$ while SNe~IIb exhibit mean $-v_{abs} \sim 10,000\pm2,000$~km~s$^{-1}$.
Between $\sim +0$~d to $+10$~d the SNe~Ic exhibit mean $-v_{abs}$ values fully consistent with the SNe~IIb, while beyond $+10$~d their values are typically a few 1,000~km~s$^{-1}$ higher than the mean values of both SNe~IIb and SNe~Ib. 
 
\item Feature~4 (\ion{Na}{i} \& \ion{He}{i} in SNe~IIb and SN~Ib) exhibits at all phases similar mean $-v_{abs}$ values and 1-$\sigma$ error snakes. At early phases a large range of velocities are measured extending from $\sim 6,000$ km~s$^{-1}$ up to and in excess of $\sim 14,000$~km~s$^{-1}$, with SNe~Ib exhibiting somewhat more dispersion than SNe~IIb. Turning to SNe~Ic, the rolling mean $-v_{abs}$ values are consistent with the SNe~IIb/Ib though their error snake is characterized by larger dispersion. 

\item Feature 5a (\ion{Si}{ii} in SNe~Ib and SNe~Ic) $-v_{abs}$ values were measured for the objects in the sample having a clear feature likely produced by \ion{Si}{ii}. Between maximum and $+$14~d the rolling mean $-v_{abs}$ values of the SNe~Ib decrease from $\sim 13,000\pm1,000$ km~s$^{-1}$ to $\sim 9,000$ km~s$^{-1}$ , whereupon they slow to just above 8,000 km~s$^{-1}$ out to $+45$~d. Our temporal coverage of this feature is limited in the SNe~Ic sample, with mean $-v_{abs}$ values ranging between 13,000~km~s$^{-1}$ to 9,000 km~s$^{-1}$ between $+$0~d to $+$21~d. 

\item Feature 5b (H$\alpha$) exhibits a smoothly evolving rolling mean $-v_{abs}$ values in the SNe~IIb sample ranging from $\sim 16,000\pm2,000$ km~s$^{-1}$ down to a value of $\sim 10,000\pm1,000$~km~s$^{-1}$ at $+30$~d. Subsequently, the feature remains constant for weeks, consistent with the idea H gas is not mixed with the heavier, lower velocity elements. 
The purported H$\alpha$ feature measured in our sample of SNe~Ib has rolling mean values and evolution similar to that of SNe~IIb, though its 1-$\sigma$ error snake is larger and our sample does not extend in phase as far. 
The feature vanishes from the majority of the SNe~Ib by $+$21~d (see Fig.~\ref{fig:spectra_epoch2}). This accounts for the rolling mean of the SNe~Ib ending around $+$30~d, while that of the SNe~IIb extends to later phases, and is consistent with the idea that the former retain smaller amounts of hydrogen compared with the latter. 

\item Features 6, 7 and 8 (\ion{He}{I}) slowly evolve over time. The $-v_{abs}$ values and evolution of Features 6 and 7 are completely consistent with one another. In the case of SNe~Ib both features exhibit rolling mean $-v_{abs}$ values that extend from $\sim 8,000\pm3,000$ km~s$^{-1}$ a week before maximum to $\sim 6,000\pm1,000$ km~s$^{-1}$ by $+40$~d. We note that the rolling mean $-v_{abs}$ values for the SNe~IIb sample begin several days post maximum.  
Feature 8 consistently shows lower $-v_{abs}$ values compared to Features 6 and 7 with both subtypes exhibiting rolling mean $-v_{abs}$ values of $\sim 6,000\pm2,000$ km~s$^{-1}$ between $+0$~d to $+$40~d.

\item Feature 9 (\ion{O}{I}) in general emerges first in SNe~Ic followed soon after by SNe~Ib, while for at least our small sample, this feature typically emerges in SNe~IIb more than a week past maximum. SNe~Ic exhibit high rolling mean $-v_{abs}$ values with significant dispersion at all epochs. Around maximum the SNe~Ic mean values are around $\sim 9,500\pm2000$~km~s$^{-1}$ and then subsequent slowly evolve over a period of a month. Rolling mean $-v_{abs}$ values of  SNe~IIb are consistently $\sim 2,000$~km~s$^{-1}$ less than the SNe~Ic mean values, and typically $\sim 2,000$ km~s$^{-1}$ higher compared with the SNe~IIb. 
 
\item Feature 10 (\ion{Ca}{II})
shows an exponentially declining rolling mean $-v_{abs}$ evolution for each SE SN subtype. Already by $-7$~d the SNe~IIb and SNe~Ib exhibit a high degree of similarity with rolling mean values of $\sim 14,000\pm1,000$~km~s$^{-1}$. SNe~Ic exhibit higher rolling mean $-v_{abs}$ values compared to the He-rich subtypes at all phases though the associated mean error snake is large at early phases mostly due to SN~2009dp. This is a noteworthy object as it is as bright as SN~2009bb and shows high $-v_{abs}$ values yet no broad-line features. 

\end{itemize}

\subsection{Doppler velocity correlation coefficients}\label{sec:vel_relations}

Spearman's rank correlation coefficients were computed for different pairs of $-v_{abs}$ values for Features 1-10, and examined and shown in the Appendix~\ref{appendixB}. SNe~Ic and SNe~Ib are found to only have a handful of pairs that are correlated, and those that are correlated typically show a low-to-moderate degree of correlation. In the case of the SNe~Ib, the \ion{He}{i} features only show correlations of low statistical significance as their $-v_{abs}$ values evolve very little, and do not monotonically change as required to produce a statistically significant correlation. On the other hand, SNe~IIb show a larger number of moderately to highly correlated pairs, particularly for the photospheric phase subset of the data. This is due to the spectral features appearing more prominent at earlier phases compared to in the SNe~Ib and SNe~Ic, and as a result of the rapid early evolution, a number of moderately to highly correlated pairs were computed. 

\section{Principal component analysis}
\label{sec:PCA}

 Principal Component Analysis (PCA; \citealt{Pearson1901}) provides a means to reconstruct the multidimensional information contained within the CSP-I SE SN spectral library using just a few variables.
This can be achieved through the use of the Singular Value Decomposition (SVD) of a data matrix (see \citealt{Hsiao2007,Cormier2011} and \citealt{Holmbo2018,Williamson2019,Shahbandeh2022} for the applications of PCA to SNe~Ia and SE SN spectral datasets, respectively). PCA is applied to summarize the large amount of information contained within an extended dataset by reducing its dimensionality while using only the most informative explanatory variables that can be derived from the dataset.

PCA is essentially a linear decomposition of a collection of data by a change of basis defined by the principal components (PCs; also known as eigenvectors) of the covariance matrix and the amplitudes (also known as projections)  defined by the inner product between the data and the new basis.
PCs are sorted/ranked by the degree to which they contribute to the variance within the data. This effectively means that the first basis vector, that is PC$_1$, accounts for the largest variation within the dataset, PC$_2$ the second largest, and so on and so forth as the dimensionality increases.
Since PCA requires little human intervention and is algorithmically performed by a computer in a matter of seconds, it provides a means to explore data in a much less labor intensive manner as compared to the line diagnostics visited in the previous sections. 

In the following we examine the PCs contributing to the largest variations within various  segments of our  spectral data library. Our analysis makes use of the \texttt{scikit-learn} PCA decomposition toolbox \citep{scikit-learn} and adheres to standard procedures as described in detail by \citet{Holmbo2018,Holmbo2020}. First, each observed spectrum is normalized to a common scale such that its  mean flux is equal to zero, and a common spectral range is used. A mean spectrum is then determined  for these input spectra, from which PCs and the amount of variance they account for are calculated using SVD. In practice this means that, for each input spectrum$_j$ used to estimate the mean spectrum, a set of PC$_i$ are obtained, along with their corresponding Amplitudes$_{ji}$ that reflect the degree to which PC$_i$ contributes to the variations of the data used to determine the mean spectrum. This is represented by the formalism:

\begin{equation}
{\rm Spectrum}_j = {\rm Mean~Spectrum} + \sum_{i=1}^{N} {\rm Amplitude}_{j,i} \cdot {\rm PC}_i.
\end{equation}

We  now examine our PCA results obtained using a large subset of the CSP-I SE SN spectral library before turning our focus toward more nuanced aspects of the data by inspecting sets of PCs determined from particular phase and spectral wavelengths ranges that effectively trace the strength and temporal evolution of the spectral features associated with  \ion{H}{i} and \ion{He}{i}.  

\begin{figure}[t!]
 {\includegraphics[width=8cm]{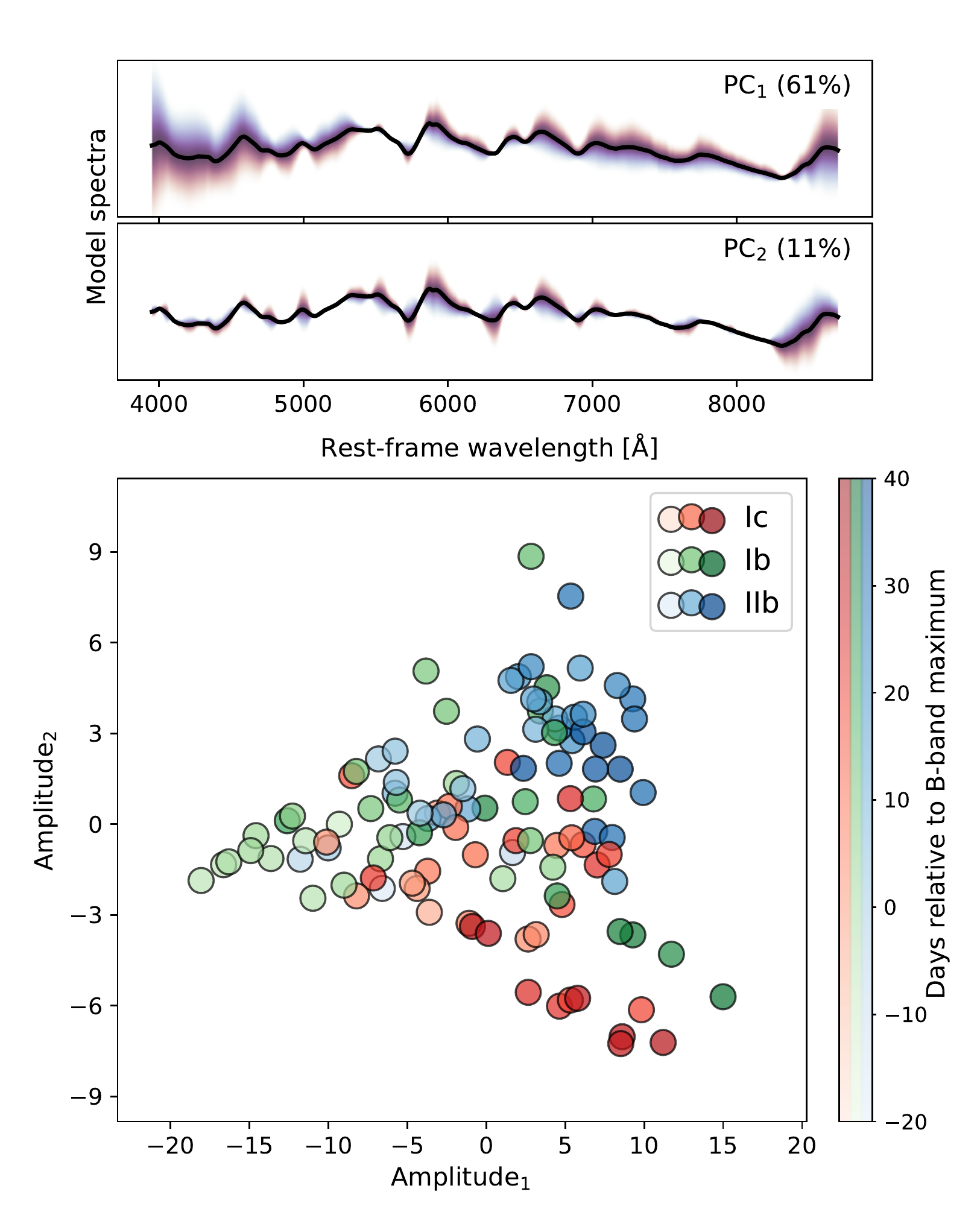}}
 \caption{Results from PCA. \textit{Top:} Solid black line is the mean spectrum computed using all observed (i.e., 111) spectra obtained out to $+40$~d and over the spectral range of 3950--8700 \AA. The shaded regions correspond to Mean spectrum + Amplitude$_i$ $\cdot$ PC$_i$, yielding the range of amplitudes shown in the panel below. Here the blue  shaded regions correspond to negative values, and the red regions correspond to positive values. 
 \textit{Bottom:} Amplitude$_1$ versus Amplitude$_2$, color-coded by spectral subtype and with the intensity of colors corresponding to the phase of the observed spectra following the multicolumn colorbar located to the right.}
 \label{fig:time_vs_int}
\end{figure}

\subsection{PCs associated with color and spectral line strength}
\label{sec:color}

 Plotted in the top of Fig.~\ref{fig:time_vs_int} is the  mean  spectrum (solid black line) computed using all CSP-I SE SNe spectra, which was used for this portion of our PCA. This includes 111 spectra covering the temporal phase out to $+40$~d and over the spectral wavelength range 3950--8700~\AA.\footnote{The phase range was limited to $+40$~d since there are only a few SNe~Ic spectra afterwards.}
 Over-plotted on the mean spectrum with shaded red and blue coloring is the full range of Amplitude$_i$ $\cdot$ PC$_i$ covered by the input data, with  PC$_{1}$ shown in the upper half-panel and  PC$_{2}$ in the lower half-panel. Here the blue  shading indicates a negative while the red indicates a positive contribution. 
 PC$_1$ accounts for 61\% of the total spectral variations of the data, PC$_2$ accounts for 11\%, and PC$_3$ (not shown) accounts for 7\%. The significant amount of variation associated with PC$_1$, particularly at the blue  end of the spectral wavelength range, is consistent with the observed broadband photometric colors of SE SNe (see Paper 2), while the positions and amplitudes of PC$_2$ suggests a connection to the depth/height of the spectral features. As shown in Paper~4, the spectral colors are directly correlated to the photometric colors, with a typical rms uncertainty of $\lesssim$ 0.1 mag.
 
Turning to the  bottom of Fig.~\ref{fig:time_vs_int},  we plot the  values of Amplitude$_{1}$ and Amplitude$_{2}$ as determined from our application of PCA. Each point in the figure is computed from a single spectrum with the color-coding of the points differentiating the subtype, and with the intensity of the coloring providing an indication of its temporal phase relative to the epoch of $B$-band maximum. 
Here, negative  Amplitude$_{1}$ values correspond to a subtraction from the mean spectrum and are coded blue. On the other hand, points with positive Amplitude$_{1}$ values correspond to adding a larger contribution of PC$_1$, and are coded red. This explains why the cluster of SNe~IIb and SNe~Ib  points located within the left-half of the plot are associated with early phase spectra when the spectral energy distributions of their associated SNe are hot and bluer. Subsequently,  as the SN ejecta expands and cools, their broadband colors evolve to longer (red) wavelengths. This therefore explains why the right-half side of the plot is populated predominately with points associated with  post-maximum phase spectra. Unfortunately, there is a dearth of  SNe~Ic spectra in the days leading up to maximum light, preventing a more rigorous comparison between their early phase spectra with those of the He-rich SE SNe.

\begin{figure}[!t]
 \includegraphics[width=8cm]{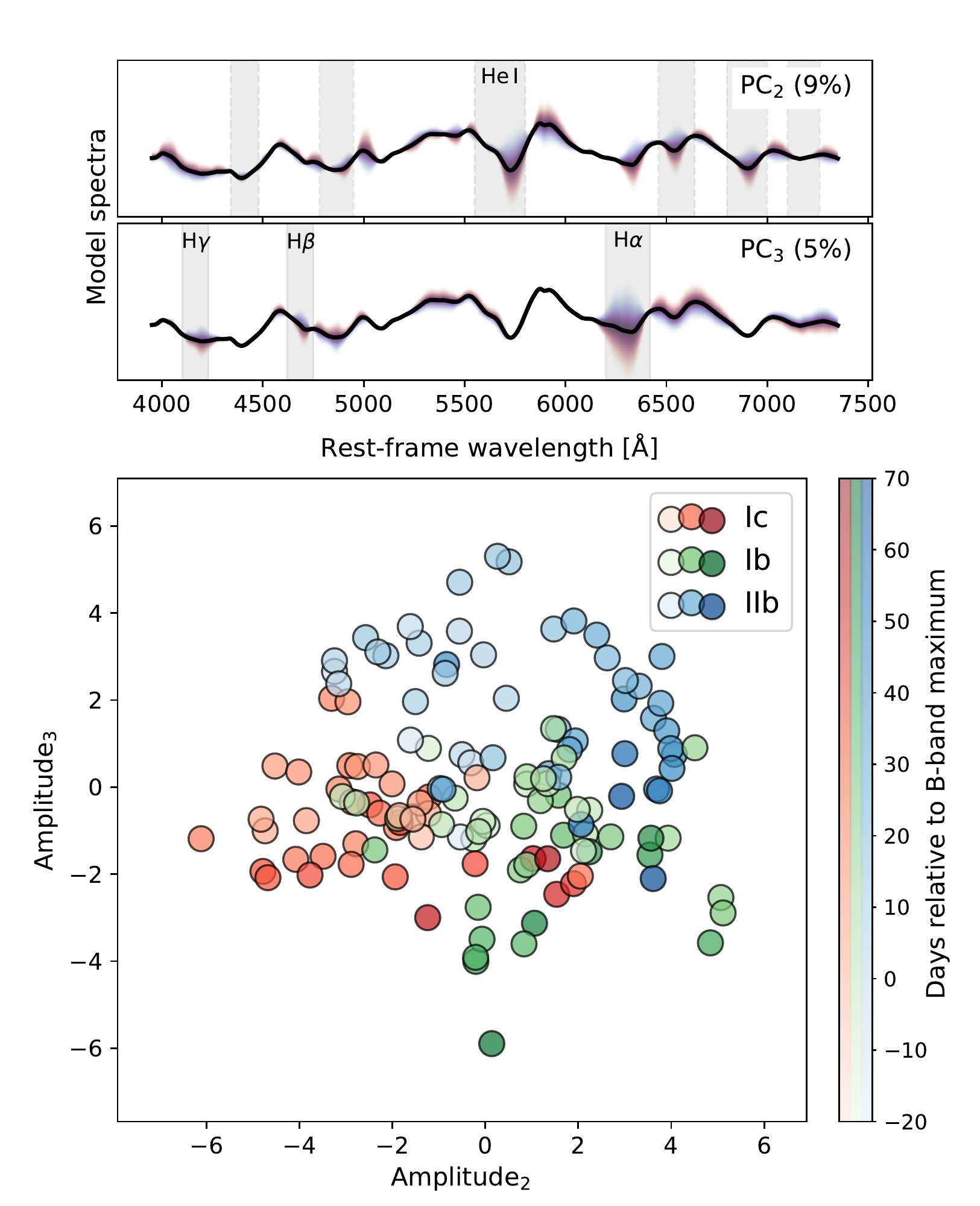}
 \caption{Results from PCA. \textit{Top:} Solid black line is the mean PCA spectrum computing using all observed spectra obtained out to $+$70~d and covering the wavelength region 3950--7350~\AA. Blue and red shaded regions correspond to the full range of Amplitude$_{i}$ with blue (red) corresponding to negative (positive) values. Vertical gray regions correspond to \ion{He}{i} (PC$_{2}$ panel) and \ion{H}{i} (PC$_{3}$ panel) related features (same as in Fig.~\ref{fig:spectra_epoch1}), indicating that variation in PC$_{2}$ corresponds strongly to Helium and in PC$_{3}$ to Hydrogen. \textit{Bottom:} Amplitude$_2$ versus Amplitude$_{3}$, color-coded by spectral subtype  with the  intensity of colors  corresponding to the phase of the observed spectra following the multicolumn colorbar located to the right.}
 \label{fig:hel_vs_hyd}
\end{figure}

Further inspection of the distribution of points indicates significant diversity among the Amplitude$_{2}$ values inferred from the post-maximum phase spectra of the different SE SN subtypes. 
Interestingly, the vast majority of post-maximum phase SNe~IIb are found to preferentially cluster within the upper-right quadrant of the figure contained within the parameter space of  Amplitude$_{1} > 0$ and   Amplitude$_{2} \gtrsim 1$. This is fully consistent with the temporal evolution of the pEWs of SNe~IIb previously documented  in Sect.~\ref{sect:pEWs}, which indicated a  strengthening  of most spectral features (e.g., those associated with \ion{He}{i}) over time. 
On the other hand, the SNe~Ib (and to a lesser extent the SNe~Ic) exhibit  a range of Amplitude$_{2}$ values that produces the triangular distribution of points within the bottom of  Fig.~\ref{fig:time_vs_int}.
For example, some maximum and post-maximum spectra of both subtypes exhibit little change in their  Amplitude$_{2}$ relative to premaximum epochs. However, other points associated with SN~Ib exhibit Amplitude$_{2}$ values ranging between $-$6 all the way up to 9, while in the case of SNe~Ic the  Amplitude$_{2}$ parameter space ranges between $\sim -9$ up to $2$. In short, the wide diversity among the Amplitude$_{2}$ parameter space is largely inherent to the difference in the suite of lines present in the various SE SN subtypes and the phase of the spectra. In order to assess the level of variations within the PCs that are largely associated with the various Balmer and \ion{He}{i} features, we now turn to PCA results obtained by examining more limited spectral wavelength regions.

\subsection{PCs associated with  \ion{H}{i} and \ion{He}{i}}
\label{section:PCAHeIHI}

\subsubsection{PC$_{2}$ versus PC$_{3}$}

Figure~\ref{fig:hel_vs_hyd} presents the PCA results obtained using all (128) of the spectra contained within the CSP-I SE SN library extending up to $+$70~d and over the spectral wavelength range of 3950-7350~\AA. 
The temporal range ensures that the line strength evolution of the features are well sampled, while the  spectral wavelength range excludes features associated with \ion{O}{i} $\lambda7773$, \ion{Ca}{ii} H\&K and the \ion{Ca}{ii} NIR triplet.\footnote{While $+$40~d was used in Sect.~\ref{sec:color}, we now extend it to cover line evolution, since the PCA analysis worked well and the mean epoch of the template is slightly after peak. However, the trade-off is that the number of SNe Ic spectra from late epochs included in the template is minimal.} By using a more limited wavelength range that excludes these common SE SN lines, the top PCs will contain more of the variance in H and He features, thereby avoiding diluting our results.
As in Fig.~\ref{fig:time_vs_int}, the  black solid line in the top of Fig.~\ref{fig:hel_vs_hyd} corresponds to the mean spectrum and the over-plotted blue/red  (negative/positive) shaded regions display the full range of  Amplitude$_{i} \cdot$ PC$_{i}$ covered by the data, where $i=2$ is shown in the upper panel and $i=3$ in the lower panel. 
PC$_1$ (not plotted)  accounts for 65\% of the spectral variations and is linked to color, as previously discussed. Meanwhile, PC$_2$  and  PC$_3$  account for 9\% and 6\% of the overall spectral variations, respectively. The majority of the blue/red shaded regions of  PC$_{2}$  are located  bluewards of the \ion{He}{i} rest-wavelengths (i.e., F1b, F2b, F4, F6, F7, F8). Similarly, PC$_{3}$ exhibits clear variations from the mean spectra at locations just bluewards of the rest wavelengths of Balmer features (i.e., F1a, F2a, F5b). However, both PCs also likely have contamination from other (non-He or non-H) features (interested readers are referred to Sect.~\ref{sec:lineIDS-Ib-Ic}).

We first consider the comparison between Amplitude$_{2}$  versus Amplitude$_{3}$, which is presented in the bottom of Fig.~\ref{fig:hel_vs_hyd}. As before, the points are color-coded based on SE SN subtype with the intensity of the color corresponding to  the temporal phase. 
Interestingly, the comparison of PC$_{2}$ and PC$_{3}$ exhibits groupings  separating  SNe~IIb from SNe~Ib and SNe~Ic, and also SNe~Ib from SNe~Ic, although the latter is less well differentiated. We now examine some specifics of the groupings.

 Relative to the coordinate origin (0,0) located at the center  of the plot, the SNe~IIb points overwhelmingly populate the Amplitude$_{3} \gtrsim 0$ parameter space, that is, they essentially cover the entire top-half region of the figure. On the other hand, the points associated with SNe~Ib and SNe~Ic tend to populate the entire bottom-half region (i.e., Amplitude$_{3} \lesssim 0$) with SNe~Ib mostly located in the bottom-right quadrant (i.e.,  Amplitude$_{2} \gtrsim  0$) and the SNe~Ic in the bottom-left quadrant (i.e.,  Amplitude$_{2} \lesssim 0$). These groupings are essentially dictated by the presence (or lack thereof), strength and time-dependence of the  Balmer and \ion{He}{i} features, as traced by $\text{PC}_2$ and $\text{PC}_3$.  The temporal dependence of the Balmer features explains why the SNe~IIb points associated with premaximum and around maximum spectra generally exhibit high Amplitudes$_{3}$ values as this is when the Balmer features are most prevalent. Similarly, the points associated with the post-maximum SNe~IIb points (when Balmer lines  typically weaken as the photosphere recedes into deeper layers of the ejecta devoid of H) are found to group among the SNe~Ib with much lower Amplitudes$_{3}$ values and higher Amplitude$_{2}$ values.
 
 Turning to SNe~Ic, its main grouping is generally located within the bottom-left quadrant of the Amplitude$_{2}$ versus Amplitude$_{3}$ diagram, reflecting  the traditional SE SN spectroscopic taxonomy where SNe~Ic lack both \ion{H}{i} and \ion{He}{i} features. However, both the SN~Ic and SN~Ib  subtypes do have several members that are mixed together along the Amplitude$_{2}$ (primarily \ion{He}{i}) axis. In addition, to the mixture between the SN~Ib and SN~Ic  subtypes in Amplitude$_{2}$, there are two early points located in the top-left quadrant among the peripheral grouping of young SNe~IIb spectra with positive Amplitude$_{3}$ values (i.e, $\sim 2$) that are associated with SN~2009ca. 
 SN~2009ca is  a superluminous SN~Ic that is more than 2 magnitudes brighter than the rest of the sample \citep{taddia2018}. Therefore, due to the expected spectral differences between SLSN-Ic and normal SNe~Ic, its position separated from the rest of the SNe~Ic is not surprising.

 Figure~\ref{fig:hel_vs_hyd} suggests that the SN~IIb, SN~Ib and SN~Ic subtypes can be differentiated based solely on PC$_2$, PC$_3$, and phase. The strongest dichotomy appears to exist among SNe~IIb and the SNe~Ibc, which is investigated more deeply in the next section. The difference in SNe~IIb is primarily from Amplitude$_{3}$ with a smaller contribution from Amplitude$_{2}$ and the phase, although at late times they do evolve toward lower Amplitude$_{3}$ and more closely resemble SNe~Ib. As discussed below, these findings are consistent with  pEW measurements of H$\alpha$ post maximum (see, e.g.,  \citealt{Liu2016}, and our  F5 panel in Fig.~\ref{fig:pew_vs_t}). 

\begin{figure}[t!]
 {\includegraphics[width=8cm]{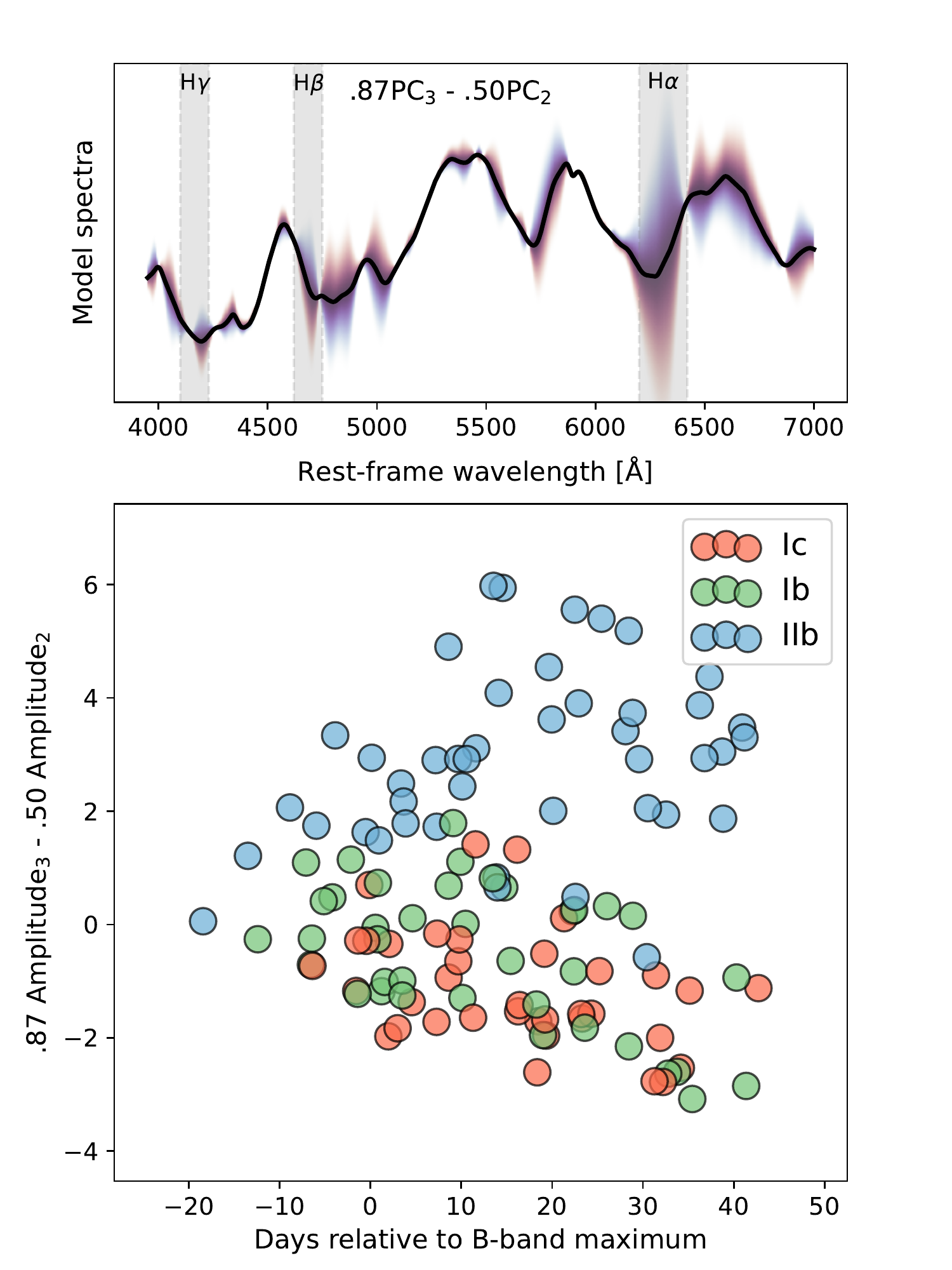}}
 \caption{Results from PCA. \textit{Top:} Full range of the linear combination of  Amplitude$_{2}$ and Amplitude$_{3}$ determined after applying a $\theta = 30^{\circ}$  rotation following Eq.~(\ref{eq:rotation}). Over-plotted the mean  spectrum (black line) determined from all spectra extending to $+45$~d and covering the wavelength range of 3950--7000~\AA. Blue and red shaded regions correspond to the full range of the first rotated PC ($0.87 \cdot \text{Amplitude}_3 -0.50 \cdot \text{Amplitude}_2$) with the blue corresponding  negative values and red to positive. Balmer series features are indicated with gray regions as in Fig.~\ref{fig:spectra_epoch1}. \textit{Bottom:} The corresponding linear combination of  Amplitude$_{2}$ and Amplitude$_{3}$, plotted versus phase. This particular linear combination separates particularly well between $+$20~d to $+$40~d the SNe~IIb from  the SNe~Ib and SNe~Ic.}
 \label{fig:pca_tbalmer}
\end{figure}

 \subsubsection{Linear combinations of PC$_{2}$ and  PC$_{3}$}
 \label{sect:linearcombos1}

 In order to explore whether PCA can be used to  differentiate SNe~Ibc from SNe~IIb, we use the fact that PCs are orthogonal, linear combinations of the input variables. A linear combination of a linear combination is another linear combination. Therefore, we search for a rotation of PC$_2$ and PC$_3$ (a rotation is just a linear combination) that creates a large  separation between SNe~IIb and the SNe~Ibc. As demonstrated below, by introducing linear combinations as a change of basis we achieve greater separation between the subtypes. This aids both in the visualizations and the interpretability of the PCA results. This is performed by applying the 2D rotation matrix to the basis defined by PC$_2$ and PC$_3$ following:
 
 \begin{align}
     \label{eq:rotation}
     \text{Rotated}~\text{PC}_{ij}(\theta) &= 
     \begin{bmatrix}
     cos(\theta) & -sin(\theta) \\
     sin(\theta) & cos(\theta)
     \end{bmatrix}
     \begin{bmatrix}
     \text{PC}_i \\
     \text{PC}_j
     \end{bmatrix} \\
     &= 
     \begin{bmatrix}
     cos(\theta)\text{Amplitude}_i -sin(\theta)\text{Amplitude}_j \\
     sin(\theta)\text{Amplitude}_i +cos(\theta)\text{Amplitude}_j
     \end{bmatrix} \nonumber .
 \end{align}
 
 \noindent We reiterate that the resulting Rotated PCs are  orthogonal, linear combinations of the original PCs. Instead of giving them new names, for clarity, we  explicitly refer to them by their  linear combination  expressions.  
 
Figure~\ref{fig:pca_tbalmer} presents the PCA results obtained for the first Rotated PC, ($0.87 \cdot \text{Amplitude}_3 -0.50 \cdot \text{Amplitude}_2$), corresponding to a Rotation of $\theta = 30^{\circ}$. As SNe~IIb and SNe~Ib generally evolve toward each other at late times (see Fig.~\ref{fig:hel_vs_hyd}), here PCA was  restricted to data obtained to $\sim +40$~d and the spectral range was limited to 3950--7000~\AA. Plotted in the top panel of  Fig.~\ref{fig:pca_tbalmer}  is the mean SE SN spectrum with the full range of the first rotated PC  ($0.87 \cdot \text{Amplitude}_3 -0.50 \cdot \text{Amplitude}_2$) over-plotted and indicated by  shaded regions, while the bottom panel contains the  linear combination of Amplitude$_{2}$ and Amplitude$_{3}$ versus phase. 
Inspection of the bottom panel reveals that  with a Rotation of  $\theta = 30^{\circ}$,  we obtain good separation between the SNe~IIb and SNe~Ibc. This  is attributed to the larger contribution of Amplitude$_{3}$, particularly during the post-maximum phases. This is also fully consistent with the results of Fig.~\ref{fig:hel_vs_hyd} where PC$_3$ was identified as primarily (but not entirely) associated with H Balmer features. 
Meanwhile Amplitude$_{2}$, which is primarily made up of He features, contributes less. 
Clearly, a linear combination of PC$_2$ and PC$_3$ can reliably identify  SNe~IIb and differentiate them from SNe~Ib between $\sim +20$ to $+40$~d. There is only a single interloping point that breaks this dichotomy, and we subsuquently return to the question of classifying and differentiating SNe~IIb from SNe~Ib (see Sect.~\ref{sec:discussion_classification}). Such interloping points also suggest that the differences among groupings are not just due to uncertainties (see Sect.~\ref{sec:interlopers}).

\begin{figure*}[t!]
 \centering
 {\includegraphics[width=12cm]{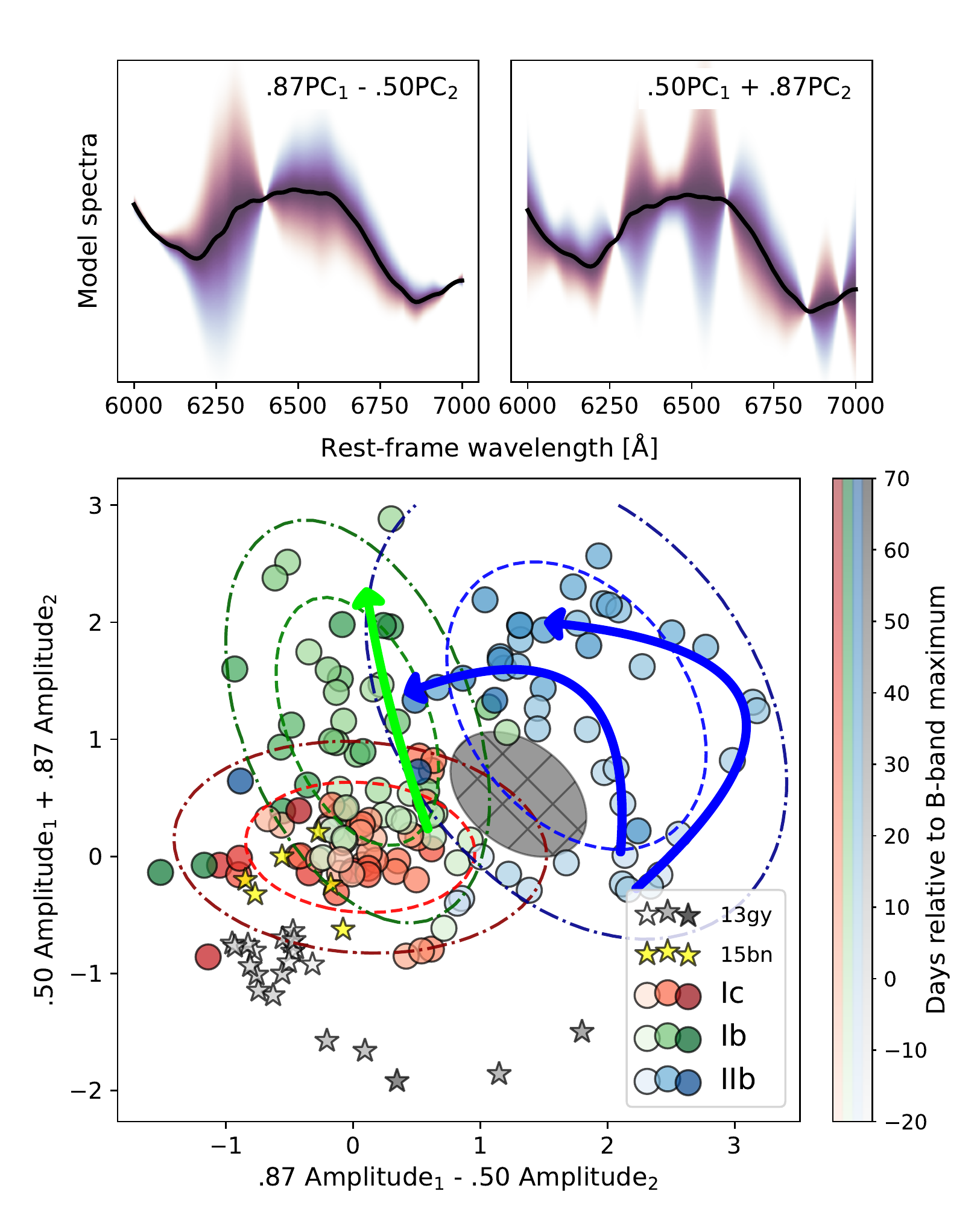}}
 \caption{Results from PCA. \textit{Top:} Linear combinations of the first two  PCs determined from PCA applied to the spectra extending to $+$70~d and between 6000--7000~\AA. 
 \textit{Bottom:} A linear combination of Amplitude$_{1}$ and Amplitude$_{2}$ plotted versus its orthogonal (30$^{\circ}$) rotation. 
 In addition to the CSP-I SE SN sample, PCA resutls are shown for the normal Type~Ia SN~2013gy \citep{Holmbo2019} and unpublished CSP-II spectra of the Type~Ic superluminous SN~2015bn \citep{Phillips2019}. 
 Solid curved lines highlight subtypes of \ion{He}{i}-rich SE SNe (see Sect.~\ref{sec:PCA6to7}).  GMM clusters are indicated with one and two sigma contours for each cluster using dashed and dot-dashed lines (see Sect.~\ref{sec:clusteringanalysis} and Table~\ref{tab:clusters}), respectively. To guide the eye, curved arrows are over-plotted highlighting the temporal evolution of SNe~IIb and SNe~Ib. }
 \label{fig:pca_hihei}
\end{figure*}

\subsection{PCA between 6000--7000~\AA}
\label{sec:PCA6to7}
As the rotated PCs in Sect.~\ref{sect:linearcombos1}  are able to separate SNe~IIb from the SNe~Ib and SNe~Ic, we now search for a manner to better separate the groupings of all three subtypes  using a similar methodology. Previously, the biggest uncertainty when attempting to do this with PC$_2$ and PC$_3$ was contamination from spurious spectral features associated with ions other than H and He, which tautologically should be the defining distinction between the SE SN subtypes. In order to ascertain whether the mixture among groupings of SNe~IIb, SNe~Ib and SNe~Ic as seen in Fig.~\ref{fig:hel_vs_hyd} are an effect of the  spectral wavelength range adopted in the PCA, we apply PCA to the even more restricted spectral range of 6000--7000~\AA. This spectral wavelength range should primarily contain H$\alpha$, and  if present, features associated with $\ion{He}{i}~\lambda6678$ and possibly $\ion{He}{i}~\lambda7065$. In addition, the narrow wavelength range ensures that the amplitude of any PCs primarily corresponding to the shape of the SED (like PC$_1$ encountered in Sect.~\ref{sec:color}) are limited.  

Following Sect.~\ref{section:PCAHeIHI}, we identified the linear combinations of PCs (i.e., PC$_1$ and PC$_2$) that represent a rotation creating separation among the different SE SN subtypes. Following Eq.~(\ref{eq:rotation}), we once again identified a rotation angle of $\theta = 30^{\circ}$ to be ideal. The results of this portion of our analysis are presented in Fig.~\ref{fig:pca_hihei}. 
The mean spectrum is plotted in the two upper panels as a black line,  while the left panel shows the first rotated PC $0.87 \cdot \text{Amplitude}_{1} - 0.50 \cdot \text{Amplitude}_{2}$ and  the right panel displays the (orthogonal) second rotated PC $0.50 \cdot \text{Amplitude}_{1} + 0.87 \cdot \text{Amplitude}_{2}$. The two rotated PCs are compared in the bottom panel of Fig.~\ref{fig:pca_hihei}, with each point color-coded by spectral subtype and with the intensity of the shading indicating the temporal phase.

Figure~\ref{fig:pca_hihei} reveals a separation between the groupings (similar to Fig.~\ref{fig:hel_vs_hyd}).  SNe~IIb clearly occupy the right half side of the figure with rotated PC $0.87 \cdot \text{Amplitude}_{1} - 0.50 \cdot \text{Amplitude}_{2}$ $\gtrsim$ 1, while the SNe~Ib and Ic occupy the left-half of the panel, with the SNe~Ib in the upper-quadrant and the SNe~Ic in the lower-quadrant. Between the right and left sides a clear separation appears within the plotted parameter space devoid of objects and 
which is highlighted in gray.
This region suggests that, at least for this sample, there is not a clear continuum over the parameter space covered by these PCs.  

Considering the evolution of SE SN subtypes with phase, we first notice  in Fig.~\ref{fig:pca_hihei} that the young SNe~IIb located toward the lower/middle right of the figure  evolve in time toward the right side of the quadrant reflected  by increasing values of the second rotated PC $0.50 \cdot \text{Amplitude}_{1} + 0.87 \cdot \text{Amplitude}_{2}$.
Beginning around maximum and extending over a month, the SNe~IIb then  follow one of two distinct tracks, which are highlighted in Fig.~\ref{fig:pca_hihei} with blue-curved arrows. 
One track is populated with objects exhibiting PC $0.87 \cdot \text{Amplitude}_{1} - 0.50 \cdot \text{Amplitude}_{2}$ $\gtrsim$  2.1--3.0 and the other track exhibits values $\lesssim 2.1$. However, after a month past maximum the two tracks begin to overlap, and as the SNe~IIb continue to evolve with decreasing values of the rotated PCs they are found to populate the same region of parameter space as old SNe~Ib. On the other hand, the SNe~Ib  located on the left quadrant of the figure (except for a few interlopers) first appear with  $0.50 \cdot \text{Amplitude}_{1} + 0.87 \cdot \text{Amplitude}_{2}$ $\lesssim 0.5$.  As time evolves through maximum and beyond, these objects  migrate upward with ever increasing  $0.50 \cdot \text{Amplitude}_{1} + 0.87 \cdot \text{Amplitude}_{2}$ of $\gtrsim 1.6$,  and then later exhibit lower  PCs values.
 Turning to the SNe~Ic, they occupy the lower left quadrant of Fig.~\ref{fig:pca_hihei}, and evolve further left with phase (although there are a handful of SNe~Ib mixed in at early and late times). 
 
 Based on the groupings reflected in Fig.~\ref{fig:pca_hihei}, in the case of the SNe~IIb and SNe~Ib, the first rotated PC $0.87 \cdot \text{Amplitude}_{1} - 0.50 \cdot \text{Amplitude}_{2}$ mainly traces H, while the second rotated PC $0.50 \cdot \text{Amplitude}_{1} + 0.87 \cdot \text{Amplitude}_{2}$ mainly traces He. As SNe~Ic by definition lack both H and He features, they occupy a relatively narrow region of the parameter space, especially among the second rotated PC (He). The evolution along the first rotated PC, tracing mainly H, can be explained by contamination from an additional feature. The origin of this feature in SNe~Ic spectra remains under debate (see discussion in  Sect.~\ref{sec:lineIDS-Ib-Ic}).   

\subsubsection{Clustering analysis}
\label{sec:clusteringanalysis}

In order to explore the inherent clustering in our PCA data and to statistically test our qualitative identification of the three main groupings, we used unsupervised learning making use of the \texttt{scikit-learn} package \citep{scikit-learn} to perform clustering analysis using K-means and Gaussian Mixture Modeling (GMM). In other words, rather than adopting a by-eye approach, we wish to   quantitatively assess how well  PCA can distinguish between the SE SN subtypes using the original (single template) PCs.  

Using three component models and as described with some detail at the end of Appendix~\ref{AppendixC}, both K-means and GMM readily identified three clusters centered at the qualitatively identified location of each grouping from the previous section. We tabulate the completeness of each algorithmically derived cluster compared to the known labels (i.e., IIb, Ib, Ic) in Table~\ref{tab:clusters}, and which are IIb $\sim 80\%$, Ib $\sim 50\%$, and Ic $\sim 95\%$, respectively. Results from this analysis consisting of the one and two sigma contours for each Gaussian component of the GMM fit are included in Fig.~\ref{fig:pca_hihei}.

\begin{figure}
 \resizebox{\hsize}{!}
 {\includegraphics[]{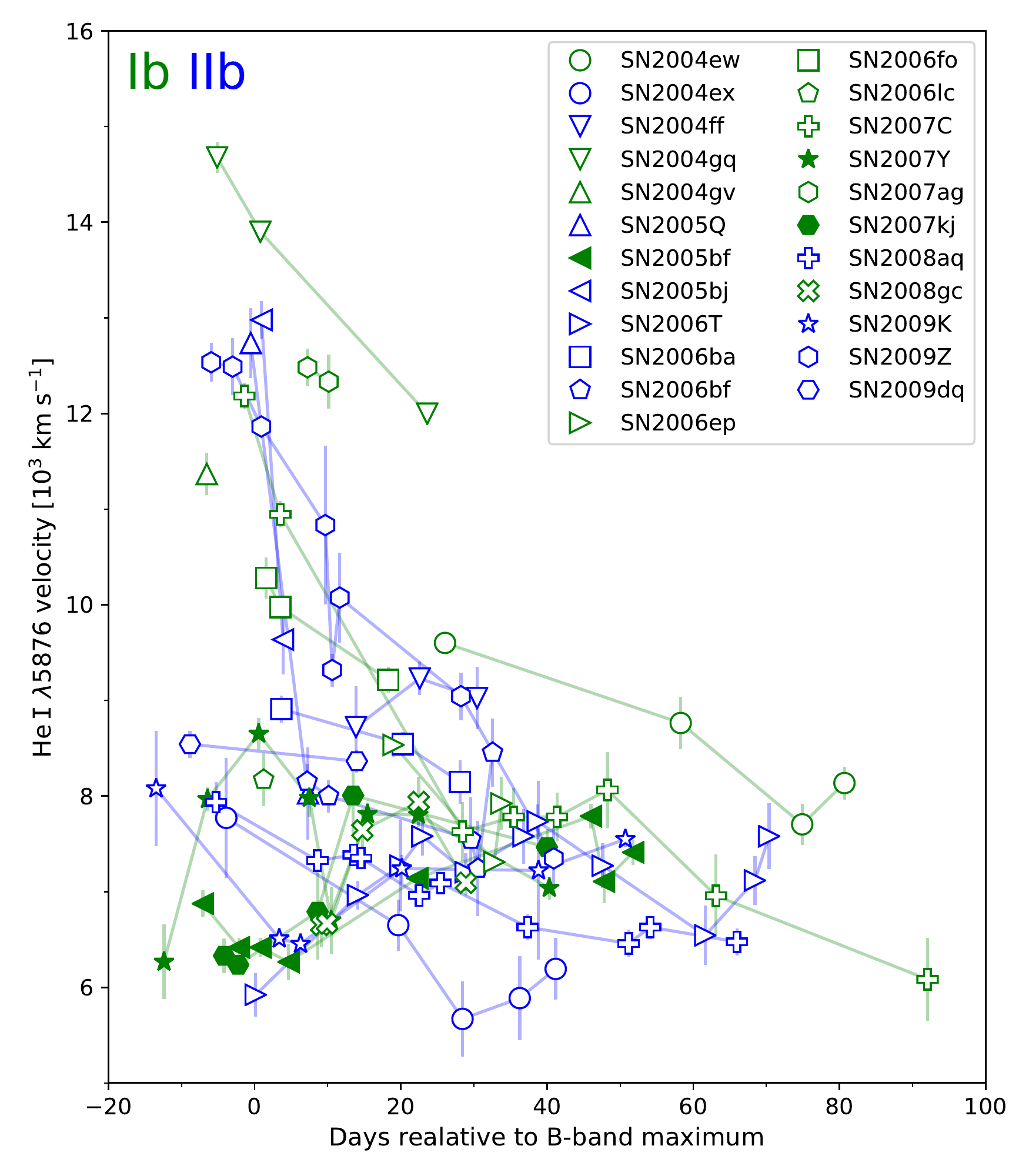}}
 \caption{Time-evolution of the Doppler velocity measured from the minimum of the \ion{He}{i} $\lambda5876$ feature.
 Each object is plotted with a unique combination of symbol and color where the color indicate subtype, with SN ~Ib plotted in blue and SN~Ib in green.
 The three SN~Ib plotted with filled symbols are SNe~2005bf, 2007Y, and 2007kj, and are identified as `flat-velocity' SNe~IIb by \citet{folatelli2014}, while others have referred to such objects as transitional SNe~Ib/c \citep[e.g.,][]{hamuy2002,stritzinger2009}.
 Using classification criteria based on the pEW of Features 4 and 5 as discussed in this paper these objects are consistent with a Type~Ib classification (see also \citealt{Liu2016} and \citealt{prentice2017} for additional discussion).
 }
\label{fig:gaston}
\end{figure}

\subsubsection{Interlopers among SNe Ib, IIb, and Ic groupings \protect\label{sec:interlopers}}

Upon examination of interlopers appearing among the groupings revealed in the bottom panel of Fig.~\ref{fig:pca_hihei}, we notice some interesting patterns. Most noticeably, a handful of SNe~Ib at early times intrude within the SNe~IIb grouping and have rotated PC $0.87 \cdot \text{Amplitude}_{1} - 0.50 \cdot \text{Amplitude}_{2}$ $\gtrsim$ 1. These are SN~2004ew (single spectrum at $+26$~d) and SN~2007kj (spectra at $+9$~d and +14~d). Two more  SNe~Ib including SN~2005bf and SN~2007Y also appear close, but do not violate our arbitrary limit. With the exception of SN~2004ew, the rest of these objects are part of a sub-group of  SNe~Ib that exhibit a flat-velocity evolution of the \ion{He}{i}~$\lambda6678$ line, as illustrated in Fig~\ref{fig:gaston}. This sub-group appears to show evidence of a weak feature at around 6250~\AA, which could be high-velocity H$\alpha$ \citep[see, e.g.,][]{folatelli2014}. Although SN~2004ew does not belong to this flat-velocity SN~Ib sub-group, it also initially shows the same absorption feature at $\sim 6250$~\AA. By the second spectrum at $+40$~d, this feature disappears and SN~2004ew returns to being firmly among the other SNe~Ib in Fig.~\ref{fig:pca_hihei}. The absorption related to this feature seems to drive the positive evolution along the x-axis (first rotated PC) in Fig.~\ref{fig:pca_hihei}. As can be seen in the top left panel of the Fig.~\ref{fig:pca_hihei}, there is a deep red trough at around this position, which is positively correlated with the first rotated principal component $0.87 \cdot \text{Amplitude}_{1} - 0.50 \cdot \text{Amplitude}_{2}$ and explains the interlopers. It is important to point out that, where we do have spectral coverage, they are found to evolve back over time toward the SNe~Ib grouping, and do not remain among the SNe~IIb. 

\begin{figure}
 \centering
\includegraphics[width=9cm]{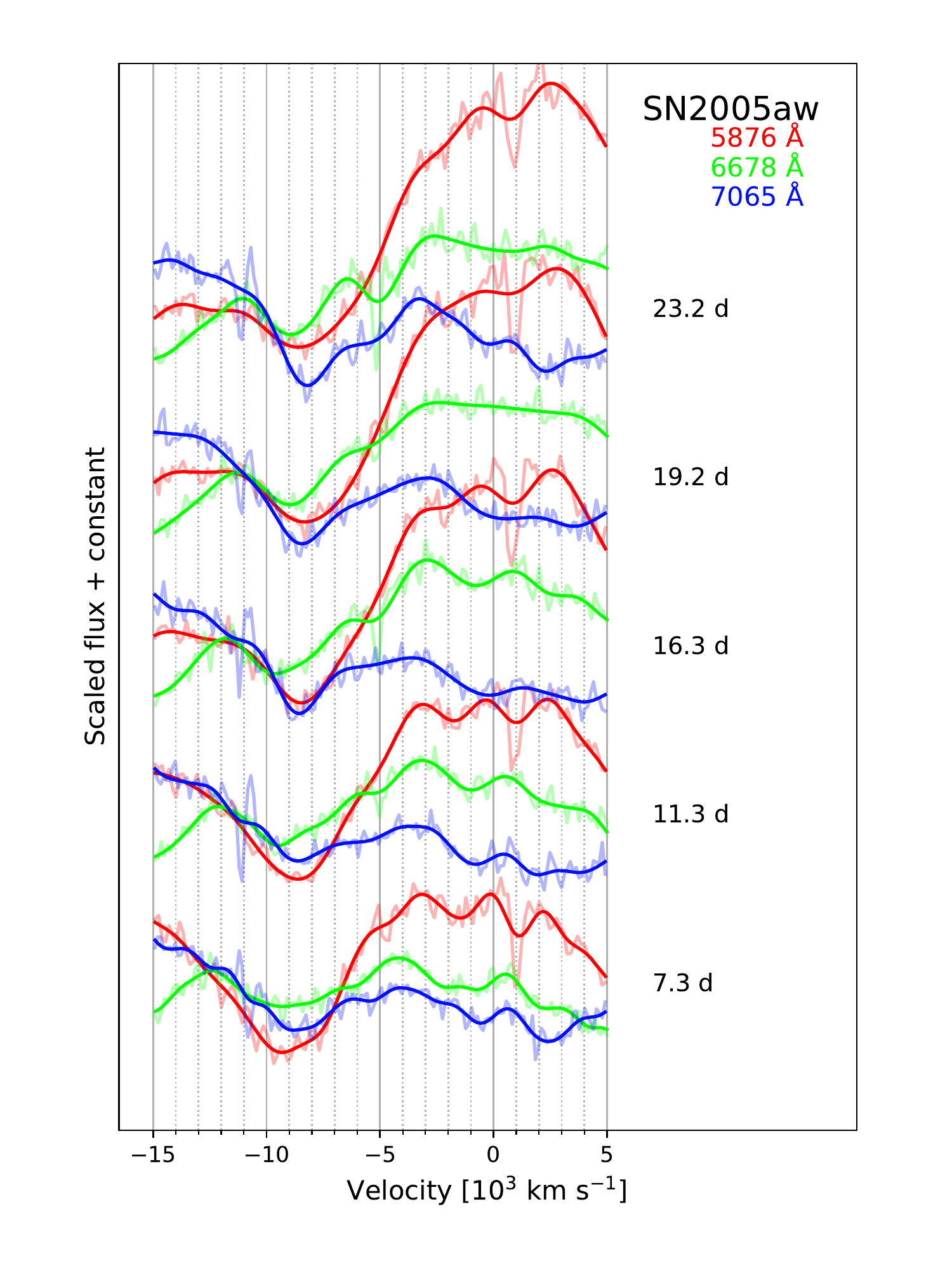}
 \caption{Spectroscopic time-series from $+$7~d to $+$23.2~d of the Type~Ic SN~2005aw. The spectra are plotted in velocity space with the segments of data containing the \ion{He}{I} $\lambda\lambda$5876, 6678, and 7065 features plotted on top of each other. As revealed from the PCA results and consistent the data, SN~2005aw exhibits discernible features likely formed by \ion{He}{i} with $-v_{abs} \sim 8,500$~km~s$^{-1}$.}
 \label{fig:2005awHe}
\end{figure}

Looking at the interlopers belonging to the SNe~Ic subtype, the most noticeable are SN~2005aw and SN~2009ca 
having, respectively, the highest and most negative values of the second rotated PC, corresponding to \ion{He}{i}. As discussed previously, the origin of the features in this region is under debate, but 
an association with H and He is not expected, especially a strong one. Nevertheless, as can be seen in Fig.~\ref{fig:2005awHe}, SN~2005aw shows evidence of early He features. Therefore, it is not surprising to find it as an interloper among the SNe~Ib and having a high value of the second rotated PC. In the case of the other interloper, SN~2009ca, it is known to be an unusually bright SN~Ic with peculiar features \citep[see][Paper 3]{taddia2018}. Hence, its separation from the rest of the grouping is not surprising. 

Looking at the interlopers belonging to SNe~IIb, we primarily see a few early phase points near the early SNe~Ib and SNe~Ic groupings, which all belong to the type~IIb SN~2009K. The earliest spectrum of SN~2009K taken on $-18.4$~d  actually lacks Balmer features, however they do  emerge by $-13.5$~d, which explains the early divergence. 

Switching gears, we note that the two very late-phase SNe~IIb points appearing in the same region as the SNe~Ib are in fact expected as the former evolve to resemble the latter during the post-maximum epochs as the H features dissipate  in  SNe~IIb. A similar convergence between SNe~IIb and SNe~Ib was seen in the late-time pEW evolution of the Feature~5. For this reason, spectral epochs later than $+$40~d were not included in Fig.~\ref{fig:pca_tbalmer}. On the whole, SNe~IIb seem to be the most well separated, which we subsuquently return to in the discussion.

The fact that distinguishing spectral features can explain the interlopers suggests that the differences are real, and not merely driven by uncertainty of the spectra or PCA analysis. Therefore, we have not considered the uncertainties of the PC amplitudes in the discussion, which is a nontrivial undertaking.

\section{Discussion}
\label{sec:discussion}

\subsection{Comparison with literature findings}

Here we examine our    line diagnostic    results presented in Sect.~\ref{sec:pEWs} and Sect.~\ref{sec:vel} to  average behavior of key features reported in the literature by the Berkeley SN group \citep{Matheson2001,Shivvers2019}, the  CfA SN group \citep{Modjaz2014,Liu2014}, the (i)PTF survey \citep{Fremling2018}, and  also  \citet{prentice2017} who compiled a (mostly) literature-based sample. Considering features studied in common by the literature, we find  broadly consistent results which we now summarize. 

\subsubsection{$\ion{Fe}{ii}$ \& $\ion{Co}{ii}$:  Feature 3}

Mean $-v_{abs}$ velocities of Feature~3 reported by \citet{Liu2016}  are  systematically higher in SNe~Ib than in SNe~IIb. We find for the CSP-I sample  a similar trend up to maximum  (see Fig.~\ref{fig:vel_vs_t}), however during post maximum phases, the sample shows no  bifurcation between the two subtypes. Turning to SNe~Ic, we find that around maximum they exhibit similar mean  $-v_{abs}$ values as the SNe~IIb, while SNe~Ib mean values are $\gtrsim 2,000$ km~s$^{-1}$ higher. Due to the lack of early SN~Ic spectra in the CSP-I sample, we are unable to comment on premaximum phases. However beginning around a week past maximum,  the average values in our  SNe~Ic tend to be larger than in SNe~Ib. We remind the reader that Feature 3 in SNe~Ic may have a nonnegligible contribution from  \ion{Co}{ii}, and therefore this comparison should be approached with caution.  Finally, to our knowledge, no measurements on the pEW of Feature~3 are available in the literature for us to compare.

\subsubsection{\ion{He}{i}: Features 4, 6, 7}

\ion{He}{i} lines corresponding to Features 4, 6, and 7 are the most commonly studied SE SN spectral features, with broad consensus in the literature on their characterization. In general, and as demonstrated in Fig.~\ref{fig:pew_vs_t}, the \ion{He}{i} features  emerge earlier in SNe~Ib and  consistently exhibit higher pEW mean values. As highlighted in the Feature 4 panel of the figure with a horizontal dashed line, the differences between the pEW values of the two subtypes prior to maximum is significant, while  mean pEW values for Features 6, 7 (and 8) reveal somewhat less significant differences. 
In the weeks following maximum brightness and as the photospheres of the SNe~IIb recede into deeper  He-rich layers of ejecta,  prominent \ion{He}{i} features emerge with mean pEW values fully consistent with those of the SNe~Ib. Again, this is in agreement with findings in the literature \citep[see][]{Liu2016,prentice2017,Fremling2018,Shivvers2019}.

Turning to  Doppler velocity,  work by \citet{Liu2016}, \citet{prentice2017} and  \citet{Fremling2018}  indicated  SNe~Ib tend to exhibit higher values prior to maximum  than SNe~IIb, while during post-maximum phases their mean values are similar. Similarly, and as demonstrated in Fig.~\ref{fig:vel_vs_t}, we also find agreement between their mean values beginning from maximum and beyond. Unfortunately, do to a dearth of data, we are unable to  comment on  premaximum phases. 

\subsubsection{Feature 5: \ion{Si}{ii} and/or H$\alpha$}

\citet{Liu2016} and \citet{prentice2017} report distinct pEW values between the SNe~IIb and SNe~Ib at all phases. As shown in Fig.~\ref{fig:pew_vs_t} and highlighted by a horizontal dashed line, the  SNe~IIb in the CSP-I sample do exhibit consistently higher  pEW values than the SNe~Ib, however, given the large dispersion of values measured for the SNe~Ib there is some overlap between $-$5~d to $+$15~d. 
In the case of SNe~Ic, we computed mean pEW values significantly less than those of the SNe~IIb at all phases, while between $+$0~d to $+$10~d they appear somewhat less than SNe~Ib, and then between $+$10~d to $+$30~d they are found to overlap. 

\citet{Liu2016} and \citet{prentice2017} report higher $-v_{abs}$ values in their SNe~Ib samples relative to SNe~IIb. This is contrary to the results of Feature~5 presented in Fig.~\ref{fig:vel_vs_t}, which indicates quite consistent rolling mean values between the two subtypes. In fact, at the earliest days of our coverage, the SNe~IIb  exhibit somewhat higher mean values, though we note this is  based on smaller sample size than that considered by \citeauthor{Liu2016} and \citeauthor{prentice2017}. Beyond maximum, the SNe~IIb and SNe~Ib exhibit similar rolling mean $-v_{abs}$ values for around a month.

\subsubsection{Feature 9: \ion{O}{i} \label{sec:disc:Feature9}}

Between maximum and ${\sim}$two weeks post maximum,  Fig.~\ref{fig:pew_vs_t} reveals that the mean pEW values of Feature~9 are   $\gtrsim50~\AA$ higher in the SNe~Ic, followed by SNe~Ib and then SNe~IIb. Similarly, Fig.~\ref{fig:vel_vs_t} indicates Feature~9 systematically exhibits higher mean $-v_{abs}$ values in SNe~Ic by a few thousand km~s$^{-1}$, followed by SNe Ib and then SNe~IIb. These results are in line with findings presented in the literature \citep[cf.][]{Matheson2001, Liu2016, Fremling2018,Shivvers2019}. 

\subsubsection{Feature 10: \ion{Ca}{ii}}

Previous sample studies of SE SN spectra do not consider the  \ion{Ca}{ii} NIR triplet. 
Only \citet{Matheson2001}, who previously reported (based on a  handful of spectra taken at different, mostly post-maximum phases) a mean velocity in SNe~Ic of $10,800\pm800$ km~s$^{-1}$, which is significantly less than what we find for CSP-I SNe~Ic. 
Even upon removal of a single object in our sample with very high velocities, we find mean $-v_{abs}$ around maximum light on the order of $\gtrsim 14,000$~km~s$^{-1}$.

\subsection{Classification}
\label{sec:discussion_classification}

\subsubsection{Line diagnostics}
\label{sec:pEWlinediagnostics}

Here we consider a few rules of thumb to follow that others may find useful in their quest to subtype SE SN based on either premaximum and/or post-maximum phase spectra based on a few of the key findings presented in Sect.~\ref{sect:pEWs}. Considering the pEW trends of the CSP-I SE SN shown in Fig.~\ref{fig:pew_vs_t}, we suggest that the pEW values of Features 4 and 5 in premaximum spectra can reliably differentiate between SNe~IIb and SNe~Ib. As indicated by the dashed line in the panel of Feature~4, SNe~Ib consistently exhibit pEW values in excess of  $\gtrsim 75$~\AA, while the other subtypes typically exhibit pEW values below this cutoff. The panel of Feature~5 shows a clear bifurcation at all phases between SNe~IIb and SNe~Ic with the latter consistently exhibiting pEW values below 100~\AA, while the former exhibit values in excess of 120~\AA. We suggest Feature~5 can serve as a  reliable indicator to differentiate SNe~IIb from SNe~Ib, beginning from $+$20~d onward. This is contrary to the advice of \citet{Liu2016} who advocated that Feature~5 provides a clear indication between SNe~IIb and SNe~Ib at all phases, and this is due to those SNe~Ib that do exhibit (modest) H$\alpha$ at early times. These objects are responsible for the significant variance of the premaximum  rolling mean pEW values. Feature~5 being a robust discriminator during post-maximum phases is in agreement with our PCA analysis (see below). Finally, we note that Feature~9 (\ion{O}{i}) also provides some indication of the SE SN subtype as discussed in Sect.~\ref{sec:disc:Feature9}. In particular, the velocity of Feature~9  in  SNe~Ic is consistently higher than exhibited in the  SNe~Ib and IIb, and at early times, SNe~Ic show  larger pEW values as compared with the other subtypes. 

\subsubsection{PCA} 

The PCA analysis presented in Sect.~\ref{section:PCAHeIHI} and the results obtained in  Fig.~\ref{fig:pca_tbalmer} demonstrate  we are able to reliably classify SNe~IIb distinctly from SNe~Ib using a combination of phase, PC$_2$, and PC$_3$, with results being strongest at phases between $+$20~d to $+$40~d. This result mirrors previous findings of \citet{Liu2016} and \citet{prentice2017} whom pointed out that the pEW value of H$\alpha$ is a suitable diagnostic to distinguish between SNe~IIb and Ib.  Appendix~\ref{AppendixC} contains a practical guide on how to leverage PCA  for those wishing to distinguish between SNe~IIb and Ib using one or more spectra taken at any phase. 
Whether SNe~Ib discovered during these post-maximum epoch are SNe~Ib, or are simply SNe~IIb that evolved to become SNe~Ib has been questioned previously \citep[e.g.,][]{Milisavljevic2013}. 
We note that all but one SN~Ib in our sample (SN~2004ew) have early spectra indicating they are not traditional SNe~IIb with strong H features. This suggests there is a clear separation of these observables between the two subtypes. Indeed, as estimated in Sect.~\ref{sec:clusteringanalysis}, interlopers from the SNe~IIb into SNe~Ib subtype can be found with $\sim 80\%$ completeness.

Contrary to \citet{Modjaz2014} and \citet{Liu2016} who report a similar separation, we did not re-classify any of our SNe, nor did we exclude transitional or peculiar SNe from our PCA as was done by \citet{Williamson2019}.  Nevertheless, we obtain a similar result and find a clear separation between SNe~IIb and SNe~Ib, indicating that the re-classification in these papers can not account for the observed differences. 
 
As discussed in Sect.~\ref{sec:interlopers} and depicted in Fig.~\ref{fig:pca_hihei}, we naturally find more scatter and some outliers within group properties such as PCs, pEWs, and $-v_{abs}$ values, even though they are accountable by peculiar outlier SNe. Specifically, for separating between SNe~IIb and SNe~Ib, the main exceptions to the PCA based classification, which were transitory outliers are: 1) SNe~Ib with early and weak  high-velocity H features all but one of which exhibit \ion{He}{i} lines that slowly evolve over time, and 2) SNe~IIb with H features that are stronger than SNe~Ib, but relatively weak compared to the rest of the SNe~IIb. Aside from these two interloping \emph{sub}-subtypes, our results indicate that SNe Ib, IIb, and Ic are distinct. It can be argued that these interlopers represent a continuum from SN IIb with weak H to SN Ib with weak H, or they can be interpreted as distinct subtypes themselves. Future PCA results from larger datasets and detailed study of these interlopers are required to disentangle the two possibilities.

To serve as a comparison, rotated PCs of the normal Type~Ia SN~2013gy relative to the SE~SN mean spectrum are also plotted in Fig.~\ref{fig:pca_hihei} (gray stars). The PCs of this normal SN~Ia are clearly separated from the SE~SNe, and therefore, such objects are not expected to serve as a source of confusion. However, PCs of  superluminous SNe~Ic (gold stars), as exemplified by SN~2015bn, do appear within the same region as less luminous SNe~Ic.

\citet{Shahbandeh2022} showed that PCA of NIR SE~SN spectra can differentiate SNe Ib from Ic. While a good degree of differentiation was also seen in our un-tuned optical results (Sect.~\ref{sec:PCA6to7}), the difference between SNe Ib and Ic was not as robust as in the NIR, nor as the difference between SNe IIb and Ibc (although we specifically delved deeper on the latter). One reason for this could be that in the optical, variance among SNe Ib and Ic is not as high as variance among SNe IIb and Ibc. Since we did not consider higher order PCs, we may have missed PCs which better differentiate SNe Ib from SNe Ic. As \citet{Shahbandeh2022} notes, NIR PCA is very good at this task due to features at 1 and 2 micron produced by He and/or C, while the optical is better for H. Similar to our investigation of H and He when aiming to classify SNe~IIb, our results could be extended to focusing on optical regions containing He and C, and extending to higher order PCs. This may tease out to search for an optical counterpart to the NIR results of \citet{Shahbandeh2022}. 

The significance of our PCA based result is that it is not reliant upon the particular methodology used to construct pEWs measurements or infer velocities. As noted by \citet{Fremling2018}, the choice of how to make these measurements can be the most significant uncertainty in the analysis. Our PCA methodology is fully reproducible and agnostically applicable to any past or future SE SN samples. Unlike spectral template matching methods widely used in the literature, the groupings can be entirely re-constructed from a given sample of SE~SNe without relying on external data. Thus, it can be used as an independent verification for classification of SE~SNe, particularly for SNe~IIb and  Ib, and especially if the initial classification is done via another method such as spectral template matching.

Future efforts related to PCA and SE SNe spectroscopic samples could focus on applying a similar analysis using an expanded sample. A guide to using PCA for this purpose, including a discussion of the practical considerations, is provided in Appendix~\ref{AppendixC}.

\section{Summary}
\label{sec:summary}

We presented a detailed analysis of the CSP-I SE SN optical spectroscopic sample. Key completed tasks and highlights of the analysis include:

\begin{itemize}
    \item The construction of mean spectra for each SE SN subtype at distinct phases. Prevalent spectral line features were then identified in the mean spectra of each SE SN subtype. Spectral synthesis modeling using \texttt{SYNAPPS} enabled the identification of the parent ions associated with the designated features. This includes the potential inclusion of: \ion{Si}{II} $\lambda$6355 in some SNe~IIb,  H$\alpha$ in some SNe~Ib, and a contribution to the $\sim 6150$~\AA\ feature in SNe~Ic by an unknown species.
    
    \item Pseudo-equivalent width (pEW) width and Doppler absorption velocity ($-v_{abs}$) measurements were measured for the spectral features in all photospheric phase spectra. With these measurements rolling mean values for both spectral indicators were determined and Spearman's rank correlation coefficient matrices constructed.

    \item Adopting a PCA formalism, we devise a method to reliably classify SE SNe using a single spectrum taken during the photospheric phase.  Using linear combinations of key principle components, we identify distinct groupings between the different SE SN subtypes. Moreover, based on a single post maximum spectrum we demonstrate the ability of PCA to provide a robust means to disentangle SNe~IIb and Ib. This finding reflects  results already in the literature  suggesting the pEW of H$\alpha$ can be used as a proxy to distinguish between SNe~IIb and Ib \citep[see][]{Liu2016}. 
    
    \item Outliers among the distinct SE SN groupings  are  identified and considered in detail. In summary we find: (i) two SNe~IIb subtypes, (ii) two SNe~Ib subtypes and (iii) a single SN~Ic with  \ion{He}{i} features all exhibiting similar $-v_{abs}$ values.  The difference between the two SNe~IIb subtypes is the strength of the H$\alpha$ feature. In the case of the SNe~Ib, along with those objects that fall within the traditional classification, our PCA analysis indicates the existence of a second type of SNe~Ib exhibiting  weak, high-velocity H$\alpha$  along with \ion{He}{i} features characterized by a flat-velocity evolution. However, as indicated by our PCA results those SNe~Ib that exhibit H$\alpha$ eventually evolve to appear as a normal SNe~Ib weeks after maximum.
 
\end{itemize}

In this paper, we demonstrated PCA  provides a avenue to gain deeper insights into the different SE SN subtypes and a  means to classify SE SNe free of human bias. Further efforts should aim to study the full public SE SN sample and include NIR spectroscopy.

\begin{acknowledgements}

The Aarhus supernova group is funded by the Independent Research Fund Denmark (IRFD, grant numbers 8021-00170B and 10.46540/2032-00022B), and by the VILLUM FONDEN (grant number 28021). L.G. acknowledges financial support from the Spanish Ministerio de Ciencia e Innovaci\'on (MCIN), the Agencia Estatal de Investigaci\'on (AEI) 10.13039/501100011033, and the European Social Fund (ESF) "Investing in your future" under the 2019 Ram\'on y Cajal program RYC2019-027683-I and the PID2020-115253GA-I00 HOSTFLOWS project, from Centro Superior de Investigaciones Cient\'ificas (CSIC) under the project PIE 20215AT016 and LINKA20409, and the program Unidad de Excelencia Mar\'ia de Maeztu CEX2020-001058-M.
The CSP has received support from the National Science Foundation (USA) under grants AST--0306969, AST--0607438, AST--1008343, AST--1613426, AST--1613455, and AST-1613472.  
This research used resources from the National Energy Research Scientific Computing Center (NERSC), which is supported by the Office of Science of the U.S. Department of Energy under Contract No. DE-AC02-05CH11231.
This research has made use of the NASA/IPAC Extragalactic Database (NED), which is operated by the Jet Propulsion Laboratory, California Institute of Technology, under contract with the National Aeronautics and Space Administration.\\

\textit{Lead authors contribution statement:}
Maximilian Stritzinger (MS) wrote the abstract, the majority of Sections 1 to 4, significant portions of Sect. 5, all of Sect. 6.1, portions of 6.2, and all of Sect. 7. Emir Karamehmetoglu (EK) performed the clustering analysis, co-wrote Sect. 5 (PCA analysis) and the discussion in Sect. 6.2, and the entirety of Appendix C. The rest of the paper, including data preparation, data analysis, \texttt{SYNAPPS} calculations and the initial figures, were created by Simon Holmbo (SH). SH had input on the text and wrote several paragraphs. SH led all aspects of the paper during his PhD thesis, while iteratively working through data analysis, figures, and discussion in each section with MS and EK. EK and MS made changes to and re-plotted Figs. 3, 5, 6, 7, 8, and 9, as well as revising the draft following co-author comments.

\end{acknowledgements}

\bibliographystyle{aa}
\bibliography{main.bib}
\clearpage

\input{sample.tex}

\clearpage
\input{specregions.tex}

\input{GMMresults.tex}

\begin{appendix}

\section{Spectral line identification with \texttt{SYNAPPS}}
\label{appendixA}

Here we summarize the results presented by \citet{Holmbo2018}, who computed \texttt{SYNAPPS} synthetic spectra  to match the mean spectra discussed in Sect.~\ref{sec:meanspectra}, in order to obtain plausible line identifications. 
Following the standard protocols \citep[see][]{thomas2011}, \texttt{SYNAPPS} requires a number of basic input parameters including:  a velocity range encompassing the outer and lower bounds of the photosphere ($v_{ph}$), a black-body temperature ($T_{BB}$), and an input list of ions each with their own set of input parameters.
The input list of ions include: \ion{H}{i}, \ion{He}{i}, \ion{C}{ii}, \ion{O}{i}, \ion{Na}{i}, \ion{Mg}{ii}, \ion{Si}{ii}, \ion{S}{ii}, \ion{Ca}{ii}, \ion{Ti}{ii}, \ion{Fe}{ii}, \ion{Co}{ii} and \ion{Ni}{ii}. All but \ion{S}{ii} and \ion{C}{ii} 
are found to most likely contribute to the formation of at least one or more of Features 1--10 (see Fig.~\ref{fig:median}) for at least one of the SE SN subtypes (see below). We note that there is also evidence for \ion{C}{ii} appearing as a notch in the mean SN~Ic spectra. This is highlighted with a vertical dashed line in Fig.~\ref{fig:spectra_epoch3}, though the line identification in our opinion is not significant enough to warrant a Feature designation in the present analysis. 

As for an initial $v_{ph}$ value we adopted the Doppler velocity at maximum absorption ($-v_{abs}$) measured from the \ion{Fe}{ii} $\lambda5169$ (i.e., Feature 3) in the maximum light mean spectra, while the lower bound was taken to be 5,000 km~s$^{-1}$ and the upper bound was set to 30,000 km~s$^{-1}$.
\ion{Fe}{ii} $\lambda5169$ provides a reasonable proxy for the velocity of the bulk of the SN ejecta (see Paper 3, and references therein).
{\tt SYNAPPS} tunes the $v_{ph}$ value in the fitting process and overall found values between 7,000--10,000~km~s$^{-1}$ for all of the mean spectra, except the $+$14~d SN~Ic mean spectrum, which has a best-fit spectrum characterized by $v_{ph}\approx12,600$~km~s$^{-1}$.
$T_{BB}$ was initially set to 7,000 K and then it was fine tuned by {\tt SYNAPPS} to values between 6,000 K to 8,000 K for all of the best-fit mean spectra, except the $+$21 d SN~Ic spectrum, which best-fit model has $T_{BB}$ $\approx$ 5,000~K.

Each input ion also has an accompanying set of input parameters.
These include the line opacity ($\tau$) at a specified reference velocity ($v_{ref}$), upper and lower velocity limits ($v_{max}$ and $v_{min}$), a value for the parameterization of the opacity profile (here with an exponential e-folding length $v_e$), and an excitation temperature ($T_{exc}$).
Values of these parameters can vary significantly between the various ions within a single synthetic spectrum, as well as significant variation of any given ion over the range of the spectral evolution covered by our set of mean spectra. 
\texttt{SYNOW} was used to obtain initial values for the input parameter set used in our \texttt{SYNAPPS} calculations.

 We found the most efficient manner to perform the calculations was to initially begin with the $+$21~d mean spectrum of each subtype and then work backwards to the earlier phases, where for each successive spectrum the results from the previous spectrum were used to guide the range of the various input parameters. 
Once the SNe~IIb spectra were modeled we continued with the SNe~Ib mean spectra, which were modeled including all of the same ions as in the case of the SNe~IIb. This was followed by modeling of the mean SNe~Ic spectra omitting both \ion{H}{i} and \ion{He}{i}.

\begin{figure*}
 {\includegraphics[width=14cm]{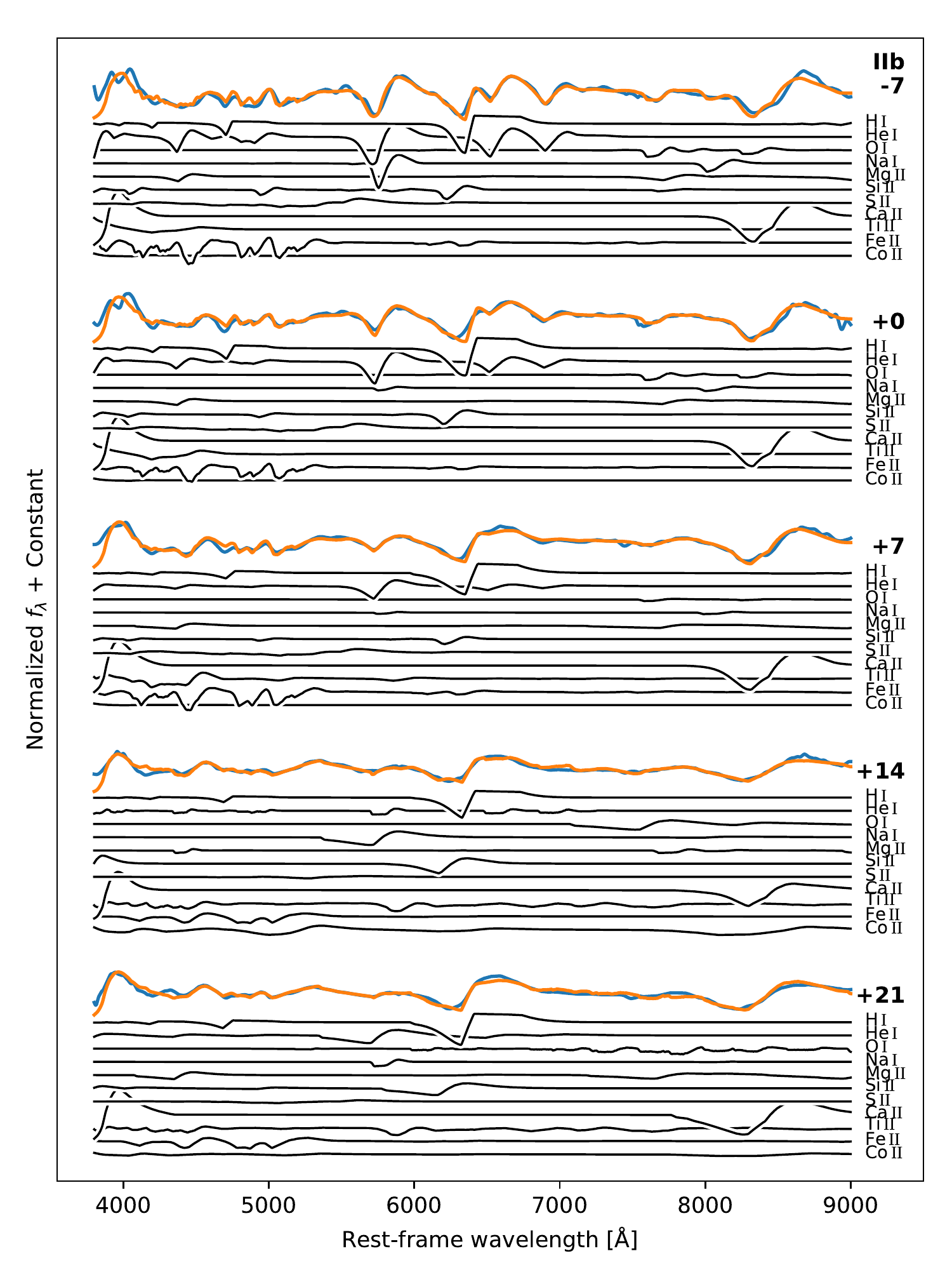}}
 \caption{{\tt SYNAPPS} fits (red lines) computed for mean template spectra (black lines) representing epochs of $-7$~d, $+0$~d, $+7$~d, $+14$~d and $+21$d. Each mean template spectrum was computed using data obtained within $\pm3.5$ days relative to its specific epoch. Spectral features attributed to each ion are also plotted in black.}
\label{fig:synapps1}
\end{figure*}

\begin{figure*}
 {\includegraphics[width=14cm]{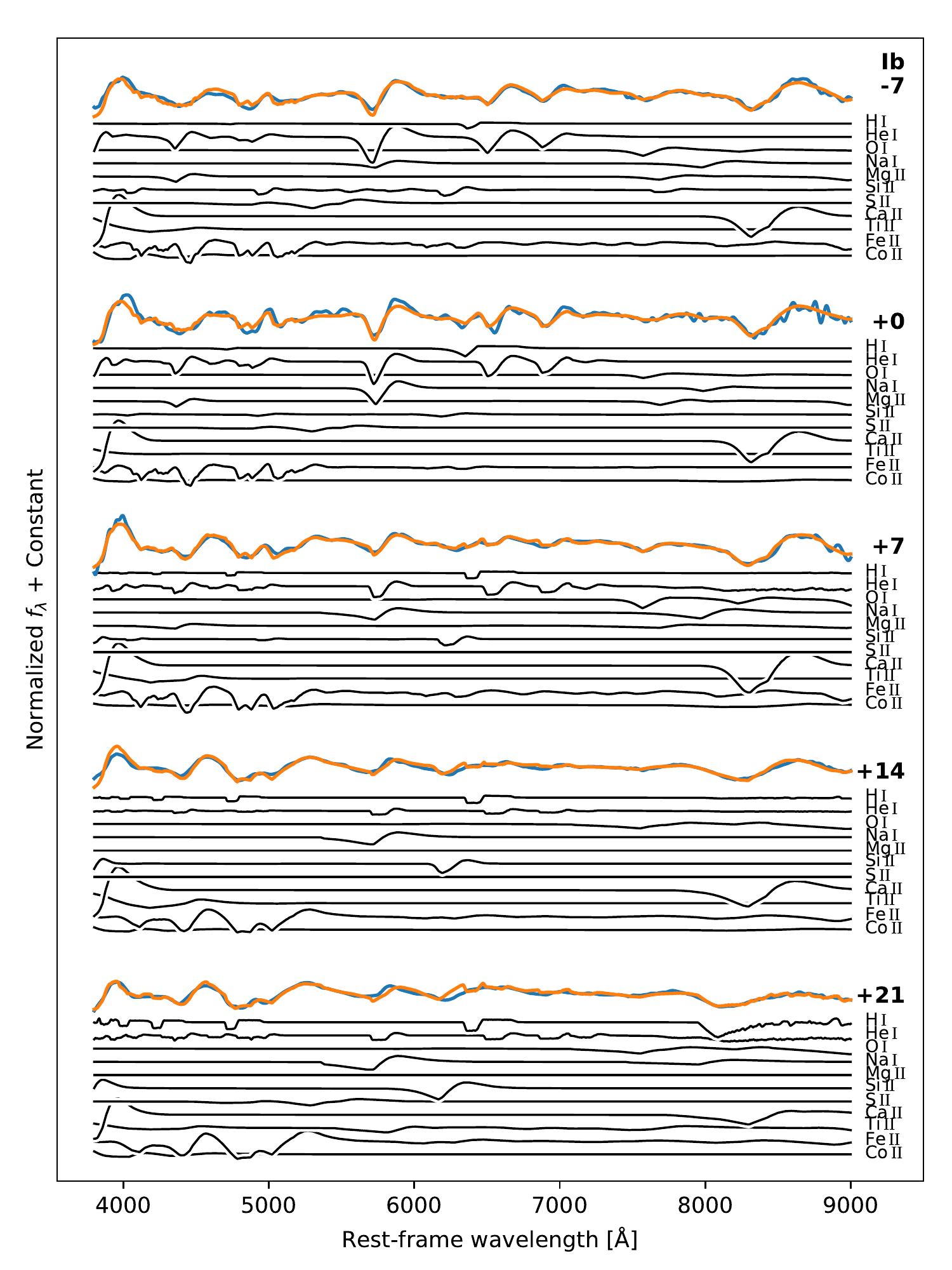}}
 \caption{{\tt SYNAPPS} fits (red lines) computed for SE SN template spectra (black lines) representing epochs of $-7$~d, $+0$~d, $+7$~d, $+14$~d and $+21$d. Each mean template spectrum was computed using data obtained within $\pm3.5$ days relative to its specific epoch. Spectral features attributed to each ion are also plotted in black.}
 \label{fig:synapps2}
\end{figure*}

\begin{figure*}
 {\includegraphics[width=14cm]{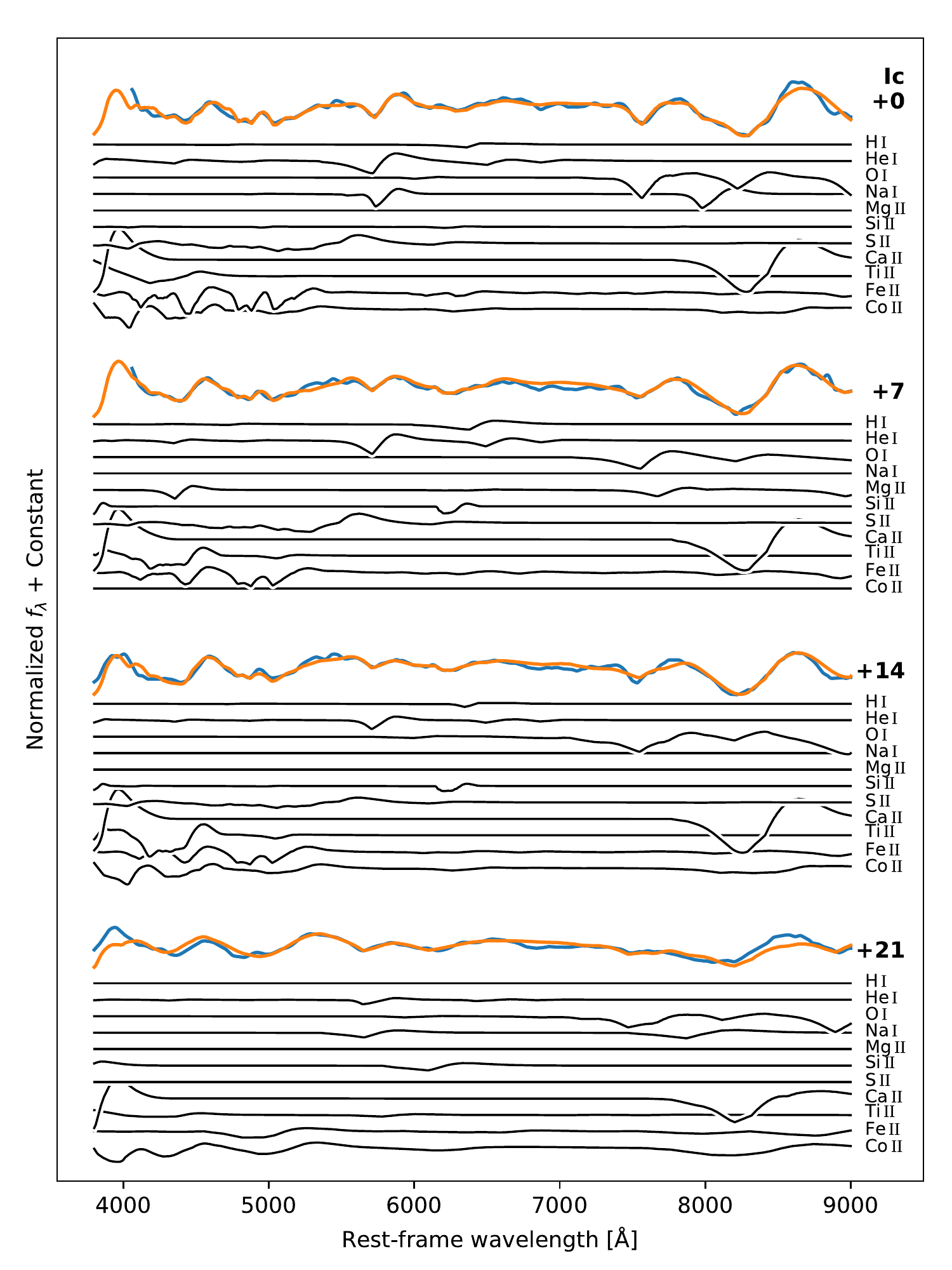}}
 \caption{{\tt SYNAPPS} fits (red lines) computed for SE SN template spectra (black lines) representing epochs of $+0$~d, $+7$d, $+14$~d and $+21$~d. Each mean template spectrum was computed using data obtained within $\pm3.5$ days relative to its specific epoch. Spectral features attributed to each ion are also plotted in black.}
 \label{fig:synapps3}
 \end{figure*}

The \texttt{SYNAPPS} spectra for each of the SE SN subtypes are plotted in Figs.~\ref{fig:synapps1}--\ref{fig:synapps3}.
This includes the \texttt{SYNAPPS} fit to  each of the template spectra, as well as the model spectrum of each of the individual ions. Comparison of the templates to the synthetic spectra reveals reasonable matches for the majority of Features 1--10, though in some cases there is some ambiguity. The ions identified contributing in whole or partially to Features 1--10 are listed in Table~\ref{tab:specfeatures}.

\begin{table}
\caption{Spectral features 1 through 10.\label{tab:specfeatures}}           
\begin{tabular}{ccc}          
\hline
\hline\\                        
Feature ID & Ion &  $\lambda_{rest}$ [\AA] \\    
\hline \\
\multicolumn{3}{c}{SNe~IIb}\\
\hline\\
   F1  & H$\gamma$    & $\lambda$4340 \\
    \ & \ion{Fe}{II} & \textit{forest of lines} \\
   \   & \ion{Ti}{II} & \textit{forest of lines} \\
   \   & \ion{He}{I} & $\lambda$4471 \\
   F2  & H$\beta$     & $\lambda$4861 \\ 
   \   & \ion{Fe}{II} & $\lambda$4924, $\lambda$5018 \\
   \   & \ion{He}{I} & $\lambda$4922\\
   F3  & \ion{Fe}{II} & $\lambda$5169             \\
   F4  & \ion{He}{I}  & $\lambda$5876             \\
   \   & \ion{Na}{I}  & $\lambda\lambda$5890, 5896 \\
   F5  & \ion{Si}{II}(?) & $\lambda$6355 \\
   \   & H$\alpha$    & $\lambda$6563 \\
   F6  & \ion{He}{I}  & $\lambda$6678 \\
   F7  & \ion{He}{I}  & $\lambda$7065     \\
   F8  & \ion{He}{I}  & $\lambda$7281     \\
   F9  & \ion{O}{I}   & $\lambda$7774     \\
   F10 & \ion{Ca}{II} & $\lambda\lambda$8498, 8542, 8662  \\
\hline\\                                             
\multicolumn{3}{c}{SNe~Ib}\\
\hline\\
   F1  & \ion{Fe}{II} & \textit{forest of lines} \\
   \   & \ion{Ti}{II} & \textit{forest of lines} \\
   \   & \ion{He}{I} & $\lambda$4471 \\
   F2  & \ion{Fe}{II} & $\lambda$4924, $\lambda$5018 \\
   \   & \ion{He}{I}  & $\lambda$4922 \\
   F3  & \ion{Fe}{II}                             & $\lambda$5169                                       \\
   F4  & \ion{He}{I} & $\lambda$5876             \\
   \   & \ion{Na}{I} & $\lambda\lambda$5890, 5896 \\
   F5  & \ion{Si}{II} & $\lambda$6355                   \\
   \   & H$\alpha$(?)    & $\lambda$6563 \\
   F6  & \ion{He}{I}                              & $\lambda$6678                                       \\
   F7  & \ion{He}{I}                              & $\lambda$7065                                       \\
   F8  & \ion{He}{I}                              & $\lambda$7281                                       \\
   F9  & \ion{O}{I}                               & $\lambda$7774                                       \\
   F10 & \ion{Ca}{II}                             & $\lambda\lambda$8498, 8542, 8662                           \\
\hline\\
\multicolumn{3}{c}{SNe~Ic}\\
\hline\\
   F1  & \ion{Fe}{II} & \textit{forest of lines} \\
   \   & \ion{Ti}{II} & \textit{forest of lines} \\ 
   F2  & \ion{Fe}{II} & $\lambda$4924, $\lambda$5018\\
   F3  & \ion{Fe}{II}, \ion{Co}{ii}      & $\lambda$5169, $\lambda$5526             \\
   F4  & \ion{Na}{I} & $\lambda\lambda$5890, 5896 \\
   F5  & \ion{Si}{II} & $\lambda$6355   \\
   F9  & \ion{O}{I}  & $\lambda$7774\\
   F10 & \ion{Ca}{II}   & $\lambda\lambda$8498, 8542, 8662   \\
   \hline 
\end{tabular}
\end{table}

\clearpage

\section{Spearman's rank pEW and Doppler velocity correlation coefficients}
\label{appendixB}

\begin{figure*}[!t]
 \centering
 \includegraphics[width=18cm]{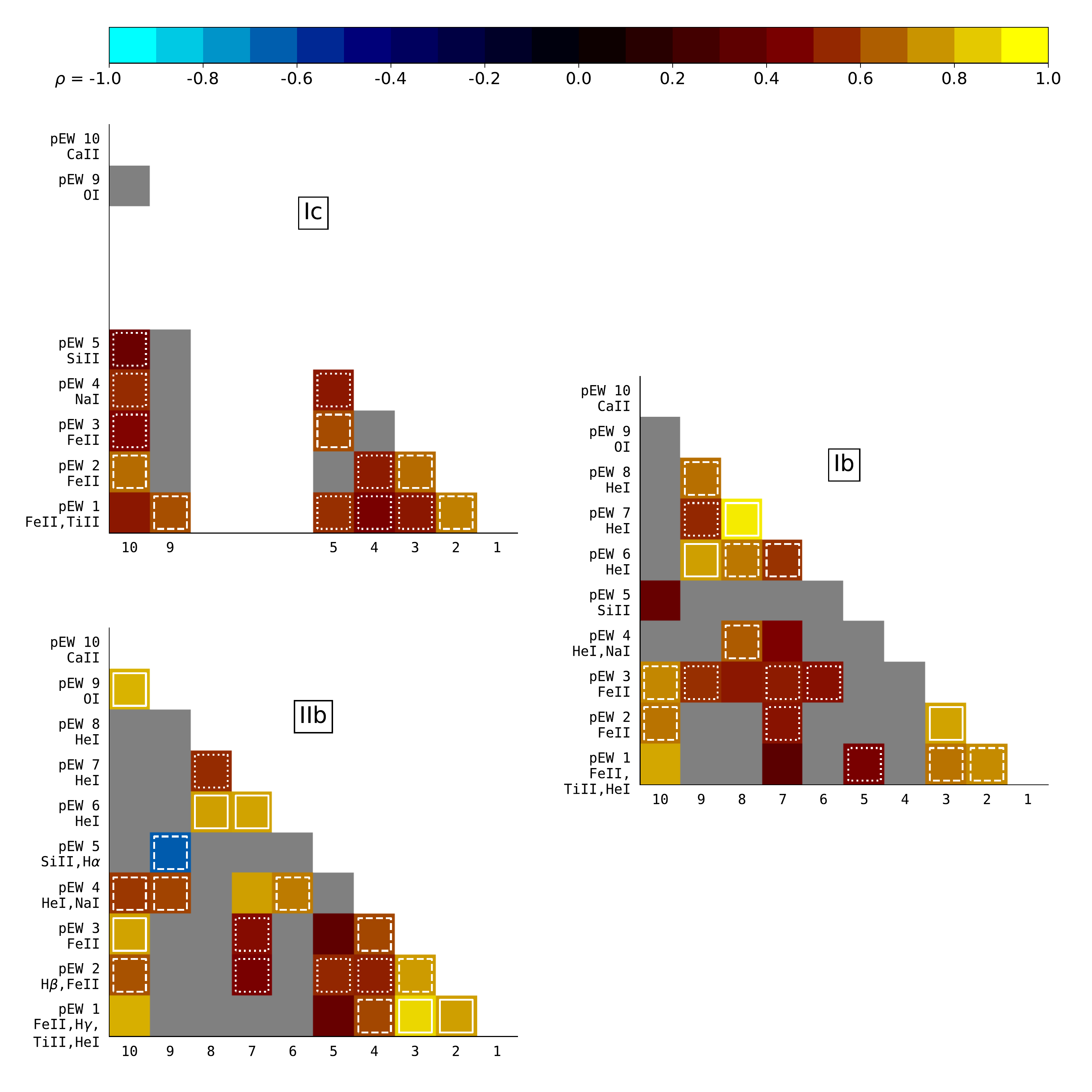}
 \caption{Spearman's rank correlation coefficients ($\rho$) between pairs of pEW measurements. The colors of the boxes is associated with the correlation coefficient with numeric values indicated by the colorbar at the top of the figure. The lighter the color, the stronger the correlation. Contained within the triangle of each panel are the correlation coefficients between pairs of pEW measurements computed using spectra obtained up to three weeks post the epoch of $B$-band maximum.  The relative significance of $\rho$ for each pair is indicated through the use of white lines. Here white-dotted, white-dashed and white-solid lines correspond to pEW pairs having low ($0.4 <$ $\vert\rho\vert$ $\leq 0.6$), moderate ($0.6 <$ $\vert\rho\vert$ $\leq 0.8$), or high ($0.8 <$ $\vert\rho\vert$) correlation coefficients, respectively. Colored boxes with no white lines exhibit minimal correlation. Gray squares have low statistical significance and are characterized by probability (p) values $p>0.1$.}
 \label{fig:pew_vs_pew}
\end{figure*}

With our pEW measurements in hand, we examined the extent of the correlations between different pairs of Features 1--10. To visualize the large amount of information encoded within the pEW measurements and to obtain quantitative measurements of the strength of correlation among various pairs of pEW parameters,  Spearman's rank correlation coefficients ($\rho$) were computed. The results of this analysis are summarized in Fig.~\ref{fig:pew_vs_pew}. In this case, each SE SN subtype has its own panel. Within the off-diagonal triangle containing the $\rho$ values determined from spectra obtained up to the first three weeks relative to the epoch of $B$-band maximum. 
Color-coding provides an indication of the degree of correlation with lighter colors indicating higher degrees of correlation or anticorrelation.
Quantitatively, pairs with $\rho$ values greater than $0.4$ are considered to be moderately to highly correlated, while those with $\rho$ values less than $-0.4$ are considered to be moderately to highly anticorrelated. 
Furthermore, pairs with $\rho$ values between $-0.4$ to $0.4$ are of low correlation, while pairs with $\rho$ values characterized by p-values below 0.1 are considered to be of low statistical significance and are shown in Fig.~\ref{fig:pew_vs_pew} by gray. 

Examination of the three panels indicates that a handful of the features are correlated, however determining whether or not any  
correlation is due to a particular physical relationship between a given pair of features is difficult. We note the following findings:

\begin{itemize}

 \item In SNe~Ic the pair of features that are most correlated are Features 1 (\ion{Fe}{ii}, \ion{Ti}{II}) and Feature 2 (\ion{Fe}{ii}), in SNe Ib they are Feature 7 (\ion{He}{i}) and Feature 8 (\ion{He}{i}), and in SNe IIb, Feature 1 (\ion{Fe}{ii}, H$\gamma$, \ion{Ti}{ii}, \ion{He}{i}) and Feature 3 (\ion{Fe}{ii}) display the highest correlation.
 
 \item The  diagonal of the SNe~IIb correlation coefficients panel indicates that the Feature~6 (\ion{He}{i}) presents a high degree of correlation with Features 7 and 8. A similar trend is also seen in the SNe~Ib, but to a lesser degree, while as mentioned previously, Features 7 and 8 are highly correlated.
 
 \item Feature~4 (\ion{He}{i} $\lambda$5876 and \ion{Na}{i} $\lambda\lambda$5890, 5896) and Feature~7 (\ion{He}{i} $\lambda$7065) are highly correlated with high statistical significance in SNe~IIb at all phases, while SNe~Ib exhibit moderate to low correlation. 
 The high degree of correlation between these two features therefore suggests \ion{He}{i} is a significant contributor to Feature~4 rather than \ion{Na}{i} in SNe~IIb, since \ion{Na}{i} does not, at all, contribute to Feature 7.

\item The pEW measurements of Feature~4 versus those of Feature~5 (H$\alpha$, \ion{Si}{ii}) at early times show a positive (though low) correlation for the SNe~Ic and no correlation among the SNe~IIb and SNe~Ib. This could indicate that the correlation in the SNe~Ic is driven by \ion{Na}{i} and \ion{Si}{ii}, where both features become somewhat more prevalent over time (see Fig.~\ref{fig:pew_vs_t}). The lack of correlation in the SNe~IIb and SNe~Ib may be due to \ion{He}{i} and H$\alpha$ being susceptible to nonthermal affects, and \ion{Na}{i} contributing less to the formation of Feature~4. 

\item SNe~Ic show only a handful of correlations with statistical significance. Most notable are Feature~1 and Feature~2 which are moderately correlated at early phases in all 3 subtypes. 

\item There is little evidence of anticorrelations among the various pEW pairs. However, the diagonal in the SNe~IIb panel of Fig.~\ref{fig:pew_vs_pew} does reveal some anticorrelation between Feature 5 and Feature 9 in SNe~IIb.  The physical causes for this anticorrelation could be related to the fact that SNe~IIb are less stripped than the other SE SN subtypes, which may then result in low pEW values inferred for Feature 9 from spectra obtained around maximum light.
\end{itemize}

\begin{figure*}[htb]
\centering
  {\includegraphics[width=18cm]{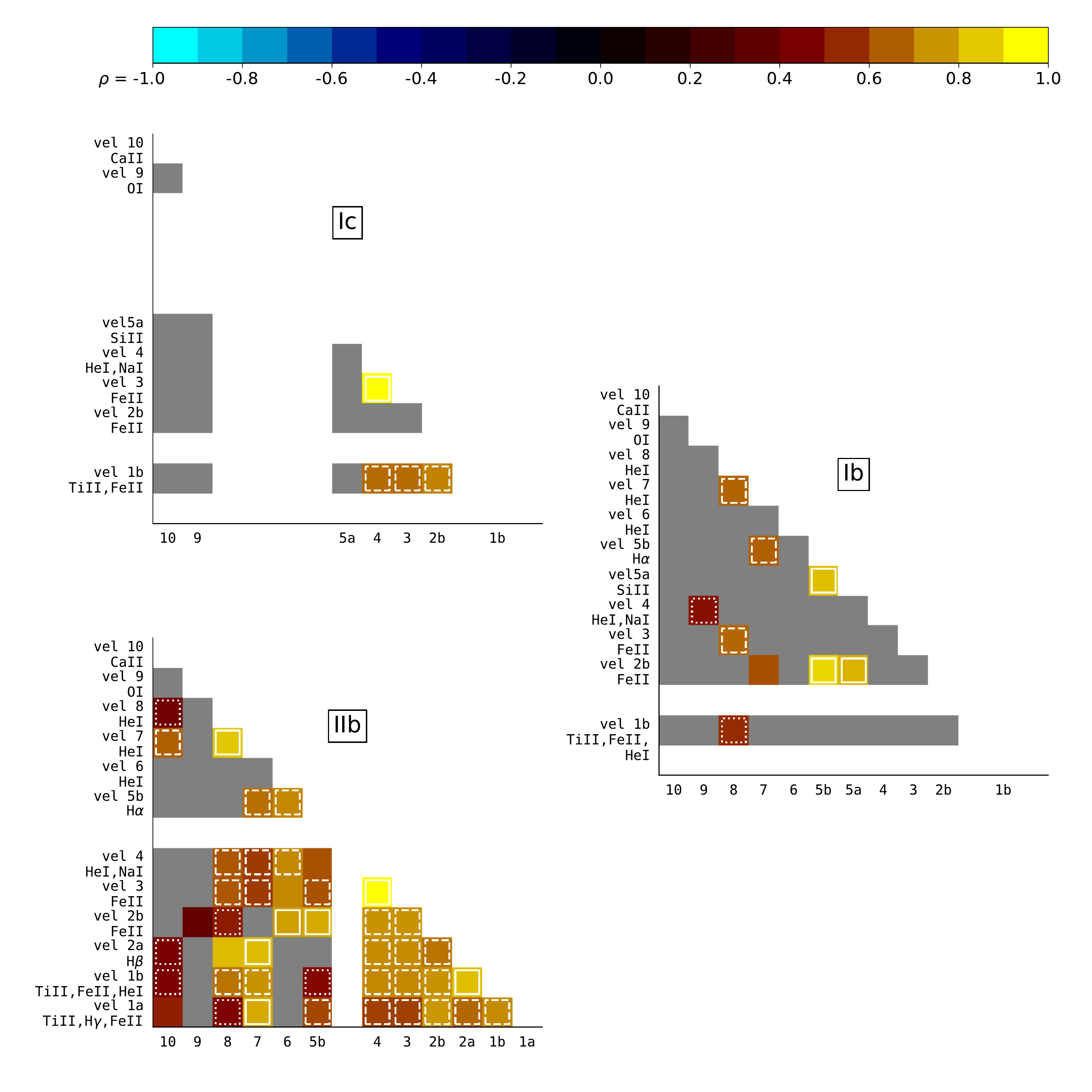}}
\caption{Spearman's rank correlation coefficients ($\rho$) between pairs of velocity ($-v_{abs}$) measurements. The color indicates the correlation coefficients with numeric values connected to colors as indicated by the mapping of the colorbar at the top of the figure. The lighter the color, the stronger the correlation. Contained within the triangle of each panel are the correlation coefficients between $-v_{abs}$ pairs computed from spectra obtained within the first three weeks relative to the time of $B$-band maximum. To highlight the relative significance of the Spearman's rank correlation coefficient each of the squares are marked with a different white line where the white dotted, white dashed and white solid lines indicate $-v_{abs}$ pairs of low ($0.4 <$ $\vert\rho\vert$ $\leq 0.6$), moderate ($0.6 <$ $\vert\rho\vert$ $\leq 0.8$), and high ($0.8 <$ $\vert\rho\vert$) correlation coefficients, respectively. Gray squares are correlations of low statistical significance characterized by probability values ($p$-values) $> 0.1$.}
  \label{fig:vel_vs_vel}
\end{figure*}

In the spirit of completeness, the Spearman's rank correlation coefficients ($\rho$) between pairs of velocity ($-v_{abs}$) measurements for the different SE SN subtypes are given in Fig.~\ref{fig:vel_vs_vel}.

\clearpage 

\clearpage 

\clearpage

\section{Classification and distinguishing SNe~IIb and Ib several weeks past maximum}
\label{AppendixC}
\FloatBarrier

As seen in Fig.~\ref{fig:pca_hihei} and discussed in Sect.~\ref{sec:discussion_classification}, our results indicate that it is possible to distinguish between SNe~IIb and SNe~Ib, even if the spectra are obtained a few weeks after maximum light. Previously, in the literature and in practice, there has been an appreciation that SNe~Ib classified with spectra taken after maximum could be SNe~IIb in disguise. Essentially, if the prevalent H features have disappeared, it is difficult to tell whether the objects in question are SNe~IIb or  SNe~Ib. This is an important consideration for population level studies of SE SNe, or their progenitors, since many SNe~Ib may either be excluded or contribute significantly to the final error budget. 

Our PCA results provide a blueprint for overcoming this practical difficulty. Therefore, we outline an algorithm for applying our PCA results to past and future SE~SN samples. Fundamentally, there are two different approaches. The first is seemingly harder one of creating and using a new template spectrum from a new or expanded sample, and determining appropriate PCs to use in the PCA. Second, using an established template from a literature sample (such as the template in this work), adding the new SNe only during the PCA phase (not into the template), and using the previously determined PCs of importance. Although the latter seems simpler, care must be taken to verify and keep the same PCs by not ``overloading'' the template sample, but also not washing out the new spectra. The simplest way to achieve this is by calculating the PCs for each new spectrum or SN, or small sets of spectra/SNe (a good rule of thumb is $\lesssim10\%$ of the template sample), and by verifying that the PCs can appropriately recreate the new input spectrum when added onto the template. Since both approaches have parts of the same basic steps, we now provide practical guidance on each step.

\textit{Create a template (mean) spectrum.} Using spectra of similar wavelength range and not too dissimilar resolving power (i.e., not mixing high-, medium-, and low-resolution spectra), a common template for all SE~SNe should be created. SNe with uncertain classification should be excluded if contamination by SNe~Ia or SLSNe is possible, or if they make balancing the classes difficult. The template is  a mean spectrum, defined by time and wavelength bins. We show that the wavelength range between 6000 to 7000~\AA\ is an appropriate choice, and the limited range has the benefit of significantly aiding in including the maximum number of SNe for consideration, while excluding PCs borne of the SED shape instead of spectral features. Under ideal conditions, the time bins would be small, but our results show that even binning the entire sample together, from $-20$ to $+40$ days, is adequate.

Most importantly, care should be taken to not over-represent individual SNe with dense spectral coverage by ensuring that one or two SNe do not dominate the template at any time or wavelength bin. Similarly, bin sizes should be chosen to provide a relatively even mix of SN subtypes and a relatively uniform distribution in time, if a large time bin is chosen. The most appropriate choice of bin size is an investigation we leave to future work, but our results can be replicated using the previously mentioned 6000 to 7000~\AA\ bin in the optical with a time bin covering the typical bell shaped light curves of SE~SN up to 5-6 weeks past maximum. All spectra must be normalized to a common scale, where we have chosen this to be such that the mean flux is zero, which is a mathematically advantageous choice. We note that well sampled, diverse, high N literature template(s) could be constructed and used in the future.
    
\textit{Perform PCA on \textit{all} of the spectra.} The spectra to perform PCA on can be included in the template sample, or an appropriately robust literature template could be used. Our analysis in Sect.~\ref{sec:PCA} already contains a detailed practical guide and software recommendation. The main point is that for each spectrum, a set of PCs is obtained, where the top few PCs correspond to the majority of the observed variation in the input spectrum compared to the template. A good check of having used enough PCs, and the appropriateness of PCA, is if the input spectrum can be approximately recreated by summing the PCs and adding them to the template spectrum. This verification gains increased importance if the input spectrum is not a part of the sample used to make the template. As discussed, it can be desirable to leave out the new input spectrum from the template sample to avoid the next step of selecting appropriate PCs. 

It is important to highlight that, even when using a previously calculated template, \emph{all} spectra, including the ones that went into the template, should be included in the PCA! Otherwise, the outcome will live in an entirely different PC space each time. However, this is not a requirement if the new spectra are numerous enough, and contain many secure classifications. Moreover, the most appropriate PCs will have to be re-derived, and systematic differences which make this more difficult are possible. 

\textit{Select the strongest PCs corresponding to spectral features.}
Because the variability among SE~SN subtypes is not random, the top PCs likely represent independent orthogonal processes. In our work, we found that in the limited range between 6000~\AA\ to 7000~\AA, the top PCs roughly corresponded to H and He features, while in the wider wavelength range the top PC represented SED color. However, the correlated nature of SN spectra might mean that the PCs constructed from a different template or sample can correspond to another feature, a mix of characteristics, or something even more difficult to interpret physically. To replicate our results and differentiate SNe~IIb from SNe~Ib, care should be taken to exclude the PC, if any, that corresponds to the overall SED shape/color, and focus on PCs which seem to correspond to spectral line features, even if the precise line cannot be identified. A plot like the top panels of our PCA Figs.~\ref{fig:pca_tbalmer} and \ref{fig:pca_hihei} can be helpful, showing the contribution of each PC to the observed variability about the template.

In theory, the number of PCs can be anywhere from two to N. However, we show that two are enough for this task and allow a visually verifiable PC space to investigate. Multidimensional clustering and classification is difficult to visually interpret, but if employed, should be able to boost these results.

Optionally, using the strongest PCs corresponding to spectral features, the PC space which creates the greatest variation among known (template) SNe~IIb, Ib, and Ic  groupings should be created. One way of doing this is using rotated PCs as we have shown in Fig.~\ref{fig:pca_hihei}. However, simple relations between the strongest PCs, or between PCs and epoch as in Fig.~\ref{fig:pca_tbalmer} can also work, as long as the separation between subtypes in observed.

\textit{Classification with clustering.} Using the newly created or existing PC space, SNe~Ib with spectra only at late time should be compared to the SN~IIb and SN~Ib groupings. In our work presented in Sect.~\ref{sec:clusteringanalysis}, interlopers from the SNe~IIb into SNe~Ib class can be found with more than 80\% completeness. The essential task is to create a decision boundary for classification in the PC space. This can be performed either by eye, or by using clustering analysis. Ultimately, this step will improve with increasing data size, allowing for more robust decision boundaries which take into account not just clustering, but the paths that various subtypes and sub-groups among the SN subtypes take in the PC space as they evolve. If using our template, a reasonable decision boundary is $>1$ in the first rotated PC space represented by the x-axis of Fig.~\ref{fig:pca_tbalmer}.

\textit{Technical details related to the clustering analysis.} The K-means fitting was done robustly using 50 random restarts, and identifies the three most tightly bound clusters in the dataset without assuming any distribution. Following standard procedure by using the K-means results to initialize our GMM, we used the Expectation Maximization algorithm to find the best fit maximum log-likelihood for our GMM model. Unlike K-means which forces a single label on every item, GMM is a soft clustering algorithm which assigns a probability for each item being inside a cluster or not based on well known properties of Gaussian distributions. Therefore, the results in Table~\ref{tab:clusters} assumes the label with the highest probability to be the correct one for calculating completeness of our GMM. While the means (centers) of our clusters are well separated, there is significant overlap at the two-sigma level for every cluster. This can be due to noise from the relatively low number of points or it could represent the fact that there is a continuum of classification between SE SN subtypes. 
However, the center of each cluster is well separated from the other. Only a minority of objects overlap between groupings or are interlopers. We note that our findings should be regarded with caution as (i) the sample is limited in size and wavelength range ($\sim 1000$~\AA), and (ii) assumes no time-dependency among the sample, which is incorrect given the time-dependent nature of the spectral energy distributions of SE SNe. 

\end{appendix}

\end{document}

%% file: sample.tex
\begin{deluxetable}{lcccc}
\tabletypesize{\scriptsize}
\tablecolumns{5}
\tablewidth{0pt}
\tablecaption{Spectroscopic sample of CSP-I SE SNe.\label{tab:sample.tex}}
\tablehead{
\colhead{SN} &
\colhead{Redshift$^{a}$} &
\colhead{Spectral} &
\colhead{No. of} &
\colhead{Phases}\\
\colhead{} &
\colhead{$z$} &
\colhead{Type} &
\colhead{Spectra} &
\colhead{}}
\startdata
  2004ew &  0.0218 & Ib     &  4 & $+$26.1, $+$58.3, $+$74.9, $+$80.7          \\ 
  2004ex &  0.0176 & IIb    &  5 & $-$3.9, $+$19.7, $+$28.4, $+$36.3, $+$41.2  \\
  2004fe &  0.0179 & Ic     &  4 & $+$8.6, $+$9.7, $+$18.5, $+$35.1           \\
  2004ff &  0.0227 & IIb    &  3 & $+$13.9, $+$22.6, $+$30.4                   \\
  2004gq &  0.0065 & Ib     &  3 & $-$5.1, $+$0.8, $+$23.6                     \\
  2004gt &  0.0055 & Ic     &  4 & $-$1.6, $+$21.3, $+$45.2, $+$49.2           \\
  2004gv &  0.0199 & Ib     &  1 & $-$6.6                                      \\
  2005Q  &  0.0224 & IIb    &  2 & $-$0.5, $+$7.3                              \\
  2005aw &  0.0095 & Ic     &  5 & $+$7.3, $+$11.3, $+$16.3, $+$19.2, $+$23.2  \\
  2005bf &  0.0189 & Ib     &  9 & $-$7.1, $-$2.2, $+$0.8, $+$4.6, $+$22.3, $+$27.2, $+$46.0, $+$47.8, $+$51.8  \\ 
  2005bj &  0.0222 & IIb    &  2 & $+$0.9, $+$3.9                             \\
  2005em &  0.0260 & Ic     &  1 & $-$0.1                                     \\
  2006T  &  0.0081 & IIb    & 11 & $+$0.2, $+$14.1, $+$20.0, $+$22.9, $+$28.9, $+$36.8, $+$38.7, $+$47.6, $+$61.6, $+$68.4, $+$70.3  \\
  2006ba &  0.0191 & IIb    &  3 & $+$3.7, $+$20.3, $+$28.1                    \\
  2006bf &  0.0239 & IIb    &  5 & $+$7.2, $+$10.1, $+$29.6, $+$30.5, $+$32.6  \\
  2006ep &  0.0151 & Ib     &  3 & $+$19.0, $+$32.8, $+$33.8                  \\
  2006fo &  0.0207 & Ib     &  3 & $+$1.6, $+$3.5, $+$18.3                     \\
  2006ir &  0.0200 & Ic     &  2 & $+$19.1, $+$42.7                            \\
  2006lc &  0.0162 & Ib     &  1 & $+$1.2                                      \\
  2007C  &  0.0056 & Ib     &  9 & $-$1.4, $+$3.5, $+$16.5, $+$28.5, $+$35.4, $+$41.4, $+$48.3, $+$63.1, $+$92.0  \\
  2007Y  &  0.0046 & Ib     &  12 & $-$12.4, $-$6.4, $+$0.6, $+$7.5, $+$10.5, $+$15.4, $+$22.4, $+$40.3, $+$230.8, $+$232.7, $+$257.6, $+$271.5  \\
  2007ag &  0.0207 & Ib     &  2 & $+$7.23, $+$10.1                            \\
  2007hn &  0.0273 & Ic     &  4 & $+$9.8, $+$25.2, $+$26.2, $+$56.3          \\
  2007kj &  0.0179 & Ib     &  5 & $-$4.2, $-$2.2, $+$8.6, $+$13.5, $+$40.0   \\
  2007rz &  0.0130 & Ic     &  2 & $+$7.4, $+$31.9                             \\
  2008aq &  0.0080 & IIb    & 12 & $-$5.3, $+$8.6, $+$13.6, $+$14.6, $+$22.5, $+$25.4, $+$37.3, $+$51.2, $+$54.1, $+$66.0, $+$76.9, $+$308.2 \\
  2008gc &  0.0492 & Ib     &  6 & $+$9.1, $+$9.9, $+$14.7, $+$22.4, $+$28.9, $+$117.3  \\
  2008hh &  0.0194 & Ic     &  3 & $+$2.1, $+$16.8, $+$31.4                    \\
  2009K  &  0.0117 & IIb    &  8 & $-$18.4, $-$13.5, $+$3.4, $+$6.3, $+$20.1, $+$38.8, $+$50.7, $+$282.3  \\
  2009Z  &  0.0248 & IIb    &  8 & $-$6.0, $-$3.0, $+$0.9, $+$9.7, $+$10.6, $+$11.7, $+$28.2, $+$40.9     \\
  2009bb &  0.0099 & Ic-BL  & 18 & $-$1.4, $-$0.5, $+$4.6, $+$8.4, $+$16.6, $+$18.4, $+$19.3, $+$23.3, $+$24.3, $+$27.5, $+$31.3, $+$32.2, $+$45.0, $+$53.9, $+$61.9, $+$282.9, $+$283.9, $+$307.5  \\
  2009ca &  0.0957 & Ic-BL  &  3 & $+$2.5, $+$11.6, $+$16.2                    \\
  2009dp &  0.0232 & Ic     &  3 & $+$2.0, $+$3.0, $+$34.2                     \\
  2009dq &  0.0047 & IIb    &  2 & $-$8.9, $+$14.0                             \\ 
  2009dt &  0.0104 & Ic     &  2 & $-$6.4, $+$16.4                             \\ 
\enddata
\tablenotetext{a}{Redshifts were retrieved from NED, or as determined from host-galaxy emission lines in the optical spectra of SNe~2007hn, 2008gc, 2009ca, and 2009dq.}
\end{deluxetable}

%% file: specregions.tex
\begin{deluxetable}{lccc}
\tabletypesize{\tiny}
\tablecolumns{4}
\tablewidth{0pt}
\tablecaption{Refined spectral feature IDs and line diagnostic fitting parameters.\label{tab:specregions}}
\tablehead{
\colhead{Feature ID} &
\colhead{Ion(s)} &
\colhead{Average $\lambda_{rest}$} &
\colhead{Detection boundaries}\\
\colhead{} &
\colhead{} &
\colhead{[\AA]} &
\colhead{intervals [\AA]}}
\startdata
\multicolumn{4}{c}{SNe~IIb}\\ 
\hline
   F1a & \ion{Ti}{II}, H$\gamma$, \ion{Fe}{II}   & 4540 & 3850-4100, 4400-4700 \\
   F1b & \ion{Ti}{II}, \ion{Fe}{II}, \ion{He}{I} & 4472 & 3850-4100, 4400-4700 \\
   F2a & H$\beta$                                & 4861 & 4400-4700, 4800-5100 \\
   F2b & \ion{He}{I}, \ion{Fe}{II}               & 5018 & 4400-4700, 4800-5100 \\
   F3  & \ion{Fe}{II}                            & 5169 & 4800-5100, 5100-5700 \\
   F4  & \ion{He}{I}, \ion{Na}{I}                & 5887 & 5000-5700, 5650-6050 \\
   F5a & \ion{Si}{II}(?)                         & 6355 & 5650-6050, 6050-6750 \\
   F5b & H$\alpha$                               & 6563 & 5650-6050, 6050-6750 \\
   F6  & \ion{He}{I}                             & 6678 & 6250-6520, 6520-6750 \\
   F7  & \ion{He}{I}                             & 7065 & 6350-6750, 6850-7100 \\
   F8  & \ion{He}{I}                             & 7281 & 6850-7100, 7100-7450 \\
   F9  & \ion{O}{I}                              & 7774 & 7100-7450, 7500-7900 \\
   F10 & \ion{Ca}{II}                            & 8567 & 7500-7900, 8450-8800 \\
   \hline
\multicolumn{4}{c}{SNe~Ib}\\
\hline
   F1b & \ion{Ti}{II}, \ion{Fe}{II}, \ion{He}{i} & 4472 & 3850-4100, 4400-4700 \\
   F2b &  \ion{He}{i}, \ion{Fe}{II}              & 5018 & 4400-4700, 4800-5100 \\
   F3  & \ion{Fe}{II}                            & 5169 & 4800-5100, 5100-5700 \\
   F4  & \ion{He}{I}, \ion{Na}{I}                & 5887 & 5250-5610, 5700-6050 \\
   F5a & \ion{Si}{II}                            & 6355 & 5650-6050, 6050-6750 \\
   F5b & H$\alpha$(?)                            & 6563 & 5650-6050, 6050-6750 \\
   F6  & \ion{He}{I}                             & 6678 & 6250-6520, 6520-6750 \\
   F7  & \ion{He}{I}                             & 7065 & 6350-6750, 6850-7100 \\
   F8  & \ion{He}{I}                             & 7281 & 6850-7100, 7100-7450 \\
   F9  & \ion{O}{I}                              & 7774 & 7100-7450, 7500-7900 \\
   F10 & \ion{Ca}{II}                            & 8567 & 7500-7900, 8450-8800 \\
\hline
\multicolumn{4}{c}{SNe~Ic}\\
\hline
   F1b & \ion{Ti}{II}, \ion{Fe}{II}              & 4472 & 3850-4100, 4400-4700 \\
   F2b &  \ion{Fe}{II}                           & 5018 & 4400-4700, 4800-5100 \\
   F3  & \ion{Fe}{II}, \ion{Co}{ii}              & 5169 & 4800-5100, 5100-5700 \\
   F4  &  \ion{Na}{I}                            & 5887 & 5250-5610, 5700-6050 \\
   F5a & \ion{Si}{II}                            & 6355 & 5650-6050, 6050-6750 \\
   F9  & \ion{O}{I}                              & 7774 & 7100-7450, 7500-7900 \\
   F10 & \ion{Ca}{II}                            & 8567 & 7500-7900, 8450-8800 \\
\enddata
\end{deluxetable}

%% file: GMMresults.tex
\begin{deluxetable}{lccccc}
\tabletypesize{\tiny}
\tablecolumns{6}
\tablewidth{0pt}
\tablecaption{Results of K-means and GMM clustering analysis.\label{tab:clusters}}
\tablehead{
\colhead{Subtype} &
\colhead{Total} &
\colhead{K-means} &
\colhead{GMM} &
\colhead{K-means} &
\colhead{GMM} \\
\colhead{} &
\colhead{counts} &
\colhead{matches} &
\colhead{matches} &
\colhead{completeness} &
\colhead{completeness} }
\startdata
    IIb & 49 &  37 & 40 & 0.76 & 0.82 \\
    Ib  & 46 &  21 & 20 & 0.46 & 0.43 \\
    Ic  & 43 &  41 & 40 & 0.95 & 0.93 \\
   \enddata
 \end{deluxetable}

%% file: main.bbl
\begin{thebibliography}{46}
\expandafter\ifx\csname natexlab\endcsname\relax\def\natexlab#1{#1}\fi

\bibitem[{{Blondin} {et~al.}(2006){Blondin}, {Dessart}, {Leibundgut}, {Branch},
  {H{\"o}flich}, {Tonry}, {Matheson}, {Foley}, {Chornock}, {Filippenko},
  {Sollerman}, {Spyromilio}, {Kirshner}, {Wood-Vasey}, {Clocchiatti},
  {Aguilera}, {Barris}, {Becker}, {Challis}, {Covarrubias}, {Davis},
  {Garnavich}, {Hicken}, {Jha}, {Krisciunas}, {Li}, {Miceli}, {Miknaitis},
  {Pignata}, {Prieto}, {Rest}, {Riess}, {Salvo}, {Schmidt}, {Smith}, {Stubbs},
  \& {Suntzeff}}]{Blondin2006}
{Blondin}, S., {Dessart}, L., {Leibundgut}, B., {et~al.} 2006, \aj, 131, 1648

\bibitem[{{Blondin} {et~al.}(2012){Blondin}, {Matheson}, {Kirshner}, {Mandel},
  {Berlind}, {Calkins}, {Challis}, {Garnavich}, {Jha}, {Modjaz}, {Riess}, \&
  {Schmidt}}]{2012AJ....143..126B}
{Blondin}, S., {Matheson}, T., {Kirshner}, R.~P., {et~al.} 2012, \aj, 143, 126

\bibitem[{{Branch} {et~al.}(2002){Branch}, {Benetti}, {Kasen}, {Baron},
  {Jeffery}, {Hatano}, {Stathakis}, {Filippenko}, {Matheson}, {Pastorello},
  {Altavilla}, {Cappellaro}, {Rizzi}, {Turatto}, {Li}, {Leonard}, \&
  {Shields}}]{2002ApJ...566.1005B}
{Branch}, D., {Benetti}, S., {Kasen}, D., {et~al.} 2002, \apj, 566, 1005

\bibitem[{{Branch} {et~al.}(2006){Branch}, {Dang}, {Hall}, {Ketchum},
  {Melakayil}, {Parrent}, {Troxel}, {Casebeer}, {Jeffery}, \&
  {Baron}}]{2006PASP..118..560B}
{Branch}, D., {Dang}, L.~C., {Hall}, N., {et~al.} 2006, \pasp, 118, 560

\bibitem[{{Branch} {et~al.}(2007){Branch}, {Troxel}, {Jeffery}, {Hatano},
  {Musco}, {Parrent}, {Baron}, {Dang}, {Casebeer}, {Hall}, \&
  {Ketchum}}]{branch2007}
{Branch}, D., {Troxel}, M.~A., {Jeffery}, D.~J., {et~al.} 2007, \pasp, 119, 709

\bibitem[{{Branch} \& {Wheeler}(2017)}]{Branch2017}
{Branch}, D. \& {Wheeler}, J.~C. 2017, {Supernova Explosions}

\bibitem[{{Cormier} \& {Davis}(2011)}]{Cormier2011}
{Cormier}, D. \& {Davis}, T.~M. 2011, \mnras, 410, 2137

\bibitem[{{Deng} {et~al.}(2000){Deng}, {Qiu}, {Hu}, {Hatano}, \&
  {Branch}}]{Deng2000}
{Deng}, J.~S., {Qiu}, Y.~L., {Hu}, J.~Y., {Hatano}, K., \& {Branch}, D. 2000,
  \apj, 540, 452

\bibitem[{{Fisher}(2000)}]{2000PhDT.........6F}
{Fisher}, A.~K. 2000, PhD thesis, The University of Oklahoma

\bibitem[{{Folatelli}(2004)}]{2004NewAR..48..623F}
{Folatelli}, G. 2004, \nar, 48, 623

\bibitem[{{Folatelli} {et~al.}(2014){Folatelli}, {Bersten}, {Kuncarayakti},
  {Olivares Estay}, {Anderson}, {Holmbo}, {Maeda}, {Morrell}, {Nomoto},
  {Pignata}, {Stritzinger}, {Contreras}, {F{\"o}rster}, {Hamuy}, {Phillips},
  {Prieto}, {Valenti}, {Afonso}, {Altenm{\"u}ller}, {Elliott}, {Greiner},
  {Updike}, {Haislip}, {LaCluyze}, {Moore}, \& {Reichart}}]{folatelli2014}
{Folatelli}, G., {Bersten}, M.~C., {Kuncarayakti}, H., {et~al.} 2014, \apj,
  792, 7

\bibitem[{{Folatelli} {et~al.}(2006){Folatelli}, {Contreras}, {Phillips},
  {Woosley}, {Blinnikov}, {Morrell}, {Suntzeff}, {Lee}, {Hamuy},
  {Gonz{\'a}lez}, {Krzeminski}, {Roth}, {Li}, {Filippenko}, {Foley},
  {Freedman}, {Madore}, {Persson}, {Murphy}, {Boissier}, {Galaz},
  {Gonz{\'a}lez}, {McCarthy}, {McWilliam}, \& {Pych}}]{folatelli2006}
{Folatelli}, G., {Contreras}, C., {Phillips}, M.~M., {et~al.} 2006, \apj, 641,
  1039

\bibitem[{{Folatelli} {et~al.}(2013){Folatelli}, {Morrell}, {Phillips},
  {Hsiao}, {Campillay}, {Contreras}, {Castell{\'o}n}, {Hamuy}, {Krzeminski},
  {Roth}, {Stritzinger}, {Burns}, {Freedman}, {Madore}, {Murphy}, {Persson},
  {Prieto}, {Suntzeff}, {Krisciunas}, {Anderson}, {F{\"o}rster}, {Maza},
  {Pignata}, {Rojas}, {Boldt}, {Salgado}, {Wyatt}, {Olivares E.}, {Gal-Yam}, \&
  {Sako}}]{2013ApJ...773...53F}
{Folatelli}, G., {Morrell}, N., {Phillips}, M.~M., {et~al.} 2013, \apj, 773, 53

\bibitem[{{Fremling} {et~al.}(2018){Fremling}, {Sollerman}, {Kasliwal},
  {Kulkarni}, {Barbarino}, {Ergon}, {Karamehmetoglu}, {Taddia}, {Arcavi},
  {Cenko}, {Clubb}, {De Cia}, {Duggan}, {Filippenko}, {Gal-Yam}, {Graham},
  {Horesh}, {Hosseinzadeh}, {Howell}, {Kuesters}, {Lunnan}, {Matheson},
  {Nugent}, {Perley}, {Quimby}, \& {Saunders}}]{Fremling2018}
{Fremling}, C., {Sollerman}, J., {Kasliwal}, M.~M., {et~al.} 2018, \aap, 618,
  A37

\bibitem[{{Gal-Yam}(2017)}]{galyam2017}
{Gal-Yam}, A. 2017, {Observational and Physical Classification of Supernovae},
  ed. A.~W. {Alsabti} \& P.~{Murdin}, 195

\bibitem[{{Garavini} {et~al.}(2007){Garavini}, {Folatelli}, {Nobili},
  {Aldering}, {Amanullah}, {Antilogus}, {Astier}, {Blanc}, {Bronder}, {Burns},
  {Conley}, {Deustua}, {Doi}, {Fabbro}, {Fadeyev}, {Gibbons}, {Goldhaber},
  {Goobar}, {Groom}, {Hook}, {Howell}, {Kashikawa}, {Kim}, {Kowalski},
  {Kuznetsova}, {Lee}, {Lidman}, {Mendez}, {Morokuma}, {Motohara}, {Nugent},
  {Pain}, {Perlmutter}, {Quimby}, {Raux}, {Regnault}, {Ruiz-Lapuente},
  {Sainton}, {Schahmaneche}, {Smith}, {Spadafora}, {Stanishev}, {Thomas},
  {Walton}, {Wang}, {Wood-Vasey}, \& {Yasuda}}]{2007A&A...470..411G}
{Garavini}, G., {Folatelli}, G., {Nobili}, S., {et~al.} 2007, \aap, 470, 411

\bibitem[{{Hachinger} {et~al.}(2012){Hachinger}, {Mazzali}, {Taubenberger},
  {Hillebrand t}, {Nomoto}, \& {Sauer}}]{hachinger2012}
{Hachinger}, S., {Mazzali}, P.~A., {Taubenberger}, S., {et~al.} 2012, \mnras,
  422, 70

\bibitem[{{Hamuy} {et~al.}(2006){Hamuy}, {Folatelli}, {Morrell}, {Phillips},
  {Suntzeff}, {Persson}, {Roth}, {Gonzalez}, {Krzeminski}, {Contreras},
  {Freedman}, {Murphy}, {Madore}, {Wyatt}, {Maza}, {Filippenko}, {Li}, \&
  {Pinto}}]{hamuy2006}
{Hamuy}, M., {Folatelli}, G., {Morrell}, N.~I., {et~al.} 2006, \pasp, 118, 2

\bibitem[{{Hamuy} {et~al.}(2002){Hamuy}, {Maza}, {Pinto}, {Phillips},
  {Suntzeff}, {Blum}, {Olsen}, {Pinfield}, {Ivanov}, {Augusteijn}, {Brillant},
  {Chadid}, {Cuby}, {Doublier}, {Hainaut}, {Le Floc'h}, {Lidman},
  {Petr-Gotzens}, {Pompei}, \& {Vanzi}}]{hamuy2002}
{Hamuy}, M., {Maza}, J., {Pinto}, P.~A., {et~al.} 2002, \aj, 124, 417

\bibitem[{{Harkness} {et~al.}(1987){Harkness}, {Wheeler}, {Margon}, {Downes},
  {Kirshner}, {Uomoto}, {Barker}, {Cochran}, {Dinerstein}, {Garnett}, \&
  {Levreault}}]{1987ApJ...317..355H}
{Harkness}, R.~P., {Wheeler}, J.~C., {Margon}, B., {et~al.} 1987, \apj, 317,
  355

\bibitem[{{Holmbo}({2018})}]{Holmbo2018}
{Holmbo}, S. {2018}, {Master Thesis}, {Aarhus University}

\bibitem[{{Holmbo}({2020})}]{Holmbo2020}
{Holmbo}, S. {2020}, {Phd Thesis}, {Aarhus University}

\bibitem[{{Holmbo} {et~al.}(2019){Holmbo}, {Stritzinger}, {Shappee}, {Tucker},
  {Zheng}, {Ashall}, {Phillips}, {Contreras}, {Filippenko}, {Hoeflich},
  {Huber}, {Piro}, {Wang}, {Zhang}, {Anais}, {Baron}, {Burns}, {Campillay},
  {Castell{\'o}n}, {Corco}, {Hsiao}, {Krisciunas}, {Morrell}, {Nielsen},
  {Persson}, {Taddia}, {Tomasella}, {Zhang}, \& {Zhao}}]{Holmbo2019}
{Holmbo}, S., {Stritzinger}, M.~D., {Shappee}, B.~J., {et~al.} 2019, \aap, 627,
  A174

\bibitem[{{Hsiao} {et~al.}(2015){Hsiao}, {Burns}, {Contreras}, {H{\"o}flich},
  {Sand}, {Marion}, {Phillips}, {Stritzinger}, {Gonz{\'a}lez-Gait{\'a}n},
  {Mason}, {Folatelli}, {Parent}, {Gall}, {Amanullah}, {Anupama}, {Arcavi},
  {Banerjee}, {Beletsky}, {Blanc}, {Bloom}, {Brown}, {Campillay}, {Cao}, {De
  Cia}, {Diamond}, {Freedman}, {Gonzalez}, {Goobar}, {Holmbo}, {Howell},
  {Johansson}, {Kasliwal}, {Kirshner}, {Krisciunas}, {Kulkarni}, {Maguire},
  {Milne}, {Morrell}, {Nugent}, {Ofek}, {Osip}, {Palunas}, {Perley}, {Persson},
  {Piro}, {Rabus}, {Roth}, {Schiefelbein}, {Srivastav}, {Sullivan}, {Suntzeff},
  {Surace}, {Wo{\'z}niak}, \& {Yaron}}]{Hsiao2015}
{Hsiao}, E.~Y., {Burns}, C.~R., {Contreras}, C., {et~al.} 2015, \aap, 578, A9

\bibitem[{{Hsiao} {et~al.}(2007){Hsiao}, {Conley}, {Howell}, {Sullivan},
  {Pritchet}, {Carlberg}, {Nugent}, \& {Phillips}}]{Hsiao2007}
{Hsiao}, E.~Y., {Conley}, A., {Howell}, D.~A., {et~al.} 2007, \apj, 663, 1187

\bibitem[{{Liu} \& {Modjaz}(2014)}]{Liu2014}
{Liu}, Y. \& {Modjaz}, M. 2014, ArXiv e-prints [\eprint[arXiv]{1405.1437}]

\bibitem[{{Liu} {et~al.}(2016){Liu}, {Modjaz}, {Bianco}, \& {Graur}}]{Liu2016}
{Liu}, Y.-Q., {Modjaz}, M., {Bianco}, F.~B., \& {Graur}, O. 2016, \apj, 827, 90

\bibitem[{{Marion} {et~al.}(2009){Marion}, {H{\"o}flich}, {Gerardy}, {Vacca},
  {Wheeler}, \& {Robinson}}]{marion2009}
{Marion}, G.~H., {H{\"o}flich}, P., {Gerardy}, C.~L., {et~al.} 2009, \aj, 138,
  727

\bibitem[{{Matheson} {et~al.}(2001){Matheson}, {Filippenko}, {Li}, {Leonard},
  \& {Shields}}]{Matheson2001}
{Matheson}, T., {Filippenko}, A.~V., {Li}, W., {Leonard}, D.~C., \& {Shields},
  J.~C. 2001, \aj, 121, 1648

\bibitem[{{Milisavljevic} {et~al.}(2013){Milisavljevic}, {Margutti},
  {Soderberg}, {Pignata}, {Chomiuk}, {Fesen}, {Bufano}, {Sanders}, {Parrent},
  {Parker}, {Mazzali}, {Pian}, {Pickering}, {Buckley}, {Crawford}, {Gulbis},
  {Hettlage}, {Hooper}, {Nordsieck}, {O'Donoghue}, {Husser}, {Potter},
  {Kniazev}, {Kotze}, {Romero-Colmenero}, {Vaisanen}, {Wolf}, {Bietenholz},
  {Bartel}, {Fransson}, {Walker}, {Brunthaler}, {Chakraborti}, {Levesque},
  {MacFadyen}, {Drescher}, {Bock}, {Marples}, {Anderson}, {Benetti},
  {Reichart}, \& {Ivarsen}}]{Milisavljevic2013}
{Milisavljevic}, D., {Margutti}, R., {Soderberg}, A.~M., {et~al.} 2013, \apj,
  767, 71

\bibitem[{{Modjaz} {et~al.}(2014){Modjaz}, {Blondin}, {Kirshner}, {Matheson},
  {Berlind}, {Bianco}, {Calkins}, {Challis}, {Garnavich}, {Hicken}, {Jha},
  {Liu}, \& {Marion}}]{Modjaz2014}
{Modjaz}, M., {Blondin}, S., {Kirshner}, R.~P., {et~al.} 2014, \aj, 147, 99

\bibitem[{{Parrent} {et~al.}(2016){Parrent}, {Milisavljevic}, {Soderberg}, \&
  {Parthasarathy}}]{parrent2016}
{Parrent}, J.~T., {Milisavljevic}, D., {Soderberg}, A.~M., \& {Parthasarathy},
  M. 2016, \apj, 820, 75

\bibitem[{{Pearson}(1901)}]{Pearson1901}
{Pearson}, K. 1901, The London, Edinburgh, and Dublin Philosophical Magazine
  and Journal of Science, 2, 559

\bibitem[{Pedregosa {et~al.}(2011)Pedregosa, Varoquaux, Gramfort, Michel,
  Thirion, Grisel, Blondel, Prettenhofer, Weiss, Dubourg, Vanderplas, Passos,
  Cournapeau, Brucher, Perrot, \& Duchesnay}]{scikit-learn}
Pedregosa, F., Varoquaux, G., Gramfort, A., {et~al.} 2011, Journal of Machine
  Learning Research, 12, 2825

\bibitem[{{Phillips} {et~al.}(2019){Phillips}, {Contreras}, {Hsiao}, {Morrell},
  {Burns}, {Stritzinger}, {Ashall}, {Freedman}, {Hoeflich}, {Persson}, {Piro},
  {Suntzeff}, {Uddin}, {Anais}, {Baron}, {Busta}, {Campillay}, {Castell{\'o}n},
  {Corco}, {Diamond}, {Gall}, {Gonzalez}, {Holmbo}, {Krisciunas}, {Roth},
  {Ser{\'o}n}, {Taddia}, {Torres}, {Anderson}, {Baltay}, {Folatelli},
  {Galbany}, {Goobar}, {Hadjiyska}, {Hamuy}, {Kasliwal}, {Lidman}, {Nugent},
  {Perlmutter}, {Rabinowitz}, {Ryder}, {Schmidt}, {Shappee}, \&
  {Walker}}]{Phillips2019}
{Phillips}, M.~M., {Contreras}, C., {Hsiao}, E.~Y., {et~al.} 2019, \pasp, 131,
  014001

\bibitem[{{Prentice} \& {Mazzali}(2017)}]{prentice2017}
{Prentice}, S.~J. \& {Mazzali}, P.~A. 2017, \mnras, 469, 2672

\bibitem[{{Shahbandeh} {et~al.}(2022){Shahbandeh}, {Hsiao}, {Ashall}, {Teffs},
  {Hoeflich}, {Morrell}, {Phillips}, {Anderson}, {Baron}, {Burns}, {Contreras},
  {Davis}, {Diamond}, {Folatelli}, {Galbany}, {Gall}, {Hachinger}, {Holmbo},
  {Karamehmetoglu}, {Kasliwal}, {Kirshner}, {Krisciunas}, {Kumar}, {Lu},
  {Marion}, {Mazzali}, {Piro}, {Sand}, {Stritzinger}, {Suntzeff}, {Taddia}, \&
  {Uddin}}]{Shahbandeh2022}
{Shahbandeh}, M., {Hsiao}, E.~Y., {Ashall}, C., {et~al.} 2022, \apj, 925, 175

\bibitem[{{Shivvers} {et~al.}(2019){Shivvers}, {Filippenko}, {Silverman},
  {Zheng}, {Foley}, {Chornock}, {Barth}, {Cenko}, {Clubb}, {Fox},
  {Ganeshalingam}, {Graham}, {Kelly}, {Kleiser}, {Leonard}, {Li}, {Matheson},
  {Mauerhan}, {Modjaz}, {Serduke}, {Shields}, {Steele}, {Swift}, {Wong}, \&
  {Yuk}}]{Shivvers2019}
{Shivvers}, I., {Filippenko}, A.~V., {Silverman}, J.~M., {et~al.} 2019, \mnras,
  482, 1545

\bibitem[{{Silverman} {et~al.}(2012){Silverman}, {Kong}, \&
  {Filippenko}}]{2012MNRAS.425.1819S}
{Silverman}, J.~M., {Kong}, J.~J., \& {Filippenko}, A.~V. 2012, \mnras, 425,
  1819

\bibitem[{{Stritzinger} {et~al.}(2009){Stritzinger}, {Mazzali}, {Phillips},
  {Immler}, {Soderberg}, {Sollerman}, {Boldt}, {Braithwaite}, {Brown}, {Burns},
  {Contreras}, {Covarrubias}, {Folatelli}, {Freedman}, {Gonz{\'a}lez}, {Hamuy},
  {Krzeminski}, {Madore}, {Milne}, {Morrell}, {Persson}, {Roth}, {Smith}, \&
  {Suntzeff}}]{stritzinger2009}
{Stritzinger}, M., {Mazzali}, P., {Phillips}, M.~M., {et~al.} 2009, \apj, 696,
  713

\bibitem[{{Stritzinger} {et~al.}(2018{\natexlab{a}}){Stritzinger}, {Anderson},
  {Contreras}, {Heinrich-Josties}, {Morrell}, {Phillips}, {Anais}, {Boldt},
  {Busta}, {Burns}, {Campillay}, {Corco}, {Castellon}, {Folatelli},
  {Gonz{\'a}lez}, {Holmbo}, {Hsiao}, {Krzeminski}, {Salgado}, {Ser{\'o}n},
  {Torres-Robledo}, {Freedman}, {Hamuy}, {Krisciunas}, {Madore}, {Persson},
  {Roth}, {Suntzeff}, {Taddia}, {Li}, \& {Filippenko}}]{stritzinger2018a}
{Stritzinger}, M.~D., {Anderson}, J.~P., {Contreras}, C., {et~al.}
  2018{\natexlab{a}}, \aap, 609, A134

\bibitem[{{Stritzinger} {et~al.}(2023){Stritzinger}, {Holmbo}, {Morrell},
  {Phillips}, {Burns}, {Castellon}, {Folatelli}, {Hamuy}, {Leloudas},
  {Suntzeff}, {Anderson}, {Ashall}, {Baron}, {Boissier}, {Hsiao},
  {Karamehmetoglu}, \& {Olivares}}]{Stritzinger2023}
{Stritzinger}, M.~D., {Holmbo}, S., {Morrell}, N., {et~al.} 2023, arXiv
  e-prints, arXiv:2302.11303

\bibitem[{{Stritzinger} {et~al.}(2018{\natexlab{b}}){Stritzinger}, {Taddia},
  {Burns}, {Phillips}, {Bersten}, {Contreras}, {Folatelli}, {Holmbo}, {Hsiao},
  {Hoeflich}, {Leloudas}, {Morrell}, {Sollerman}, \&
  {Suntzeff}}]{stritzinger2018b}
{Stritzinger}, M.~D., {Taddia}, F., {Burns}, C.~R., {et~al.}
  2018{\natexlab{b}}, \aap, 609, A135

\bibitem[{{Taddia} {et~al.}(2018){Taddia}, {Stritzinger}, {Bersten}, {Baron},
  {Burns}, {Contreras}, {Holmbo}, {Hsiao}, {Morrell}, {Phillips}, {Sollerman},
  \& {Suntzeff}}]{taddia2018}
{Taddia}, F., {Stritzinger}, M.~D., {Bersten}, M., {et~al.} 2018, \aap, 609,
  A136

\bibitem[{{Thomas} {et~al.}(2011){Thomas}, {Nugent}, \& {Meza}}]{thomas2011}
{Thomas}, R.~C., {Nugent}, P.~E., \& {Meza}, J.~C. 2011, \pasp, 123, 237

\bibitem[{{Williamson} {et~al.}(2019){Williamson}, {Modjaz}, \&
  {Bianco}}]{Williamson2019}
{Williamson}, M., {Modjaz}, M., \& {Bianco}, F.~B. 2019, \apjl, 880, L22

\end{thebibliography}
